\documentclass[twocolumn,showpacs,prb,citeautoscript,amsmath,amssymb,floatfix,eqsecnum]{revtex4-2}
\usepackage{amsmath,amssymb,color}
\usepackage[dvipsnames]{xcolor}
\usepackage{bm}
\usepackage{graphicx}
\usepackage{psfrag}
\usepackage{bbold}
\usepackage{bbm}
\newcommand{\bd}{\bm}

\begin{document}

\title{Dissipative spin dynamics  in hot
quantum paramagnets}

\author{Dmytro Tarasevych and Peter Kopietz}
  
\affiliation{Institut f\"{u}r Theoretische Physik, Universit\"{a}t
  Frankfurt,  Max-von-Laue Strasse 1, 60438 Frankfurt, Germany}

\date{July 16, 2021}

 \begin{abstract}
We use the functional renormalization approach
for quantum spin systems developed by 
Krieg and Kopietz [Phys. Rev. B {\bf{99}}, 060403(R) (2019)]
to calculate the
spin-spin correlation function $G ( \bd{k} , \omega )$ of
quantum Heisenberg magnets at infinite temperature.
For small wavevectors $\bd{k} $ and frequencies $\omega$ 
we  find that $G ( \bd{k} , \omega )$ assumes in dimensions $d > 2$
the diffusive form predicted by hydrodynamics.
In three dimensions
our result for the spin-diffusion coefficient  
${\cal{D}}$ is somewhat smaller than previous
theoretical predictions based on the extrapolation
of the short-time expansion, but is still about $30 \%$ larger
than the measured 
high-temperature value of ${\cal{D}}$ in
the Heisenberg ferromagnet 
Rb$_2$CuBr$_4\cdot 2$H$_2$O.
In reduced dimensions $d \leq 2$ we find superdiffusion
characterized  by a frequency-dependent  complex
spin-diffusion coefficient ${\cal{D}} ( \omega )$ which 
diverges logarithmically in $d=2$,
and as a power-law ${\cal{D}} ( \omega ) \propto \omega^{-1/3}$
in $d=1$.  Our result in one dimension implies scaling with dynamical
exponent $z =3/2$, in agreement with recent calculations for
integrable spin chains.
Our approach is not restricted to the hydrodynamic regime and
allows us to calculate the
dynamic structure factor $S ( \bd{k} , \omega )$ for all wavevectors. 
We show  how the short-wavelength behavior of
$S ( \bd{k} , \omega )$ at high temperatures reflects 
the relative sign and strength of competing
exchange interactions.

\end{abstract}


\maketitle
\tableofcontents

\section{Introduction}

Calculating  the dynamic spin-spin correlation function of quantum Heisenberg models in the paramagnetic regime is a challenging problem 
which requires advanced many-body techniques 
or large-scale numerical simulations.
Even in the limit of infinite temperature the spin dynamics remains 
non-trivial. Hydrodynamic arguments  suggest that
for sufficiently small wavevectors $\bd{k}$ and frequencies $\omega$ 
the Fourier transform $G ( \bd{k} , \omega )$ of the retarded
spin-spin correlation function of spin-rotationally invariant 
Heisenberg magnets has the diffusive form~\cite{Halperin69,Forster75}
 \begin{equation}
 G ( \bd{k} , \omega ) = G ( \bd{k} ) \frac{ {\cal{D}} k^2 }{ 
 {\cal{D}} k^2 - i \omega },
 \label{eq:Ghydro}
 \end{equation}
where $G ( \bd{k}  ) \equiv G ( \bd{k} , 0 )$ is the static (i.e., zero-frequency)
limit of the spin-spin correlation function, and ${\cal{D}}$ is the spin-diffusion coefficient. 
In the regime where the temperature $T$
is large compared with the exchange energy, the static correlation function $G ( \bd{k} )$
can be approximated by the  
susceptibility of an isolated spin $\bd{S}$,
 \begin{equation}
 G ( \bd{k} ) \approx   \frac{S ( S+1) }{ 3 T} ,
 \label{eq:G0stat}
 \end{equation}
where the spin operator is normalized such that
$\bd{S}^2 = S ( S+1)$.
On the other hand,
the calculation of the 
spin-diffusion coefficient ${\cal{D}}$ remains highly non-trivial even 
in the limit $T \rightarrow \infty$.
For three-dimensional Heisenberg magnets with nearest-neighbor exchange $J$
the spin-diffusion coefficient is expected to approach a constant of order 
$|J |$ at high temperatures. Note that at infinite temperature 
${\cal{D}}$ is independent of the sign of $J$, indicating that a simple expansion in powers of $J$ is not possible. 
In the 1960s and 1970s several approximate
calculations of the numerical value of ${\cal{D}}$ 
at high temperatures 
have been published \cite{DeGennes58,Mori62,Bennett65,Resibois66,Redfield68,TahirKheli69,Blume70,Morita72,Morita75}. 
Thereafter the interest in this problem has waned (see, however, Ref.~[\onlinecite{Kopietz93,Boehm94}]),
although a convergence of the results for ${\cal{D}}$ has 
not been achieved.
Surprisingly, an experiment~\cite{Labrujere82} 
measuring ${\cal{D}}$ in the three-dimensional 
Heisenberg ferromagnet
Rb$_2$CuBr$_4\cdot 2$H$_2$O at high temperatures produced a result which 
was consistently smaller (by a factor ranging between $0.5$ and $0.7$)
than theoretical predictions \cite{Mori62,Bennett65,Resibois66,TahirKheli69}.
As far as we know, this discrepancy between theory and experiment has never been resolved. The authors of Ref.~[\onlinecite{Labrujere82}] speculated that 
methods based on the extrapolation
of the short-time expansion
of the spin-spin correlation function to long 
times~\cite{DeGennes58,Mori62,Bennett65,Redfield68,Morita72,Morita75,Kopietz93,Boehm94}
tend to overestimate the magnitude of the spin-diffusion coefficient.
A numerical simulation \cite{Mueller88} for classical 
Heisenberg models at infinite temperatures revealed long-time tails in dimensions
$d =1,2,3$ which are incompatible with a frequency-independent spin-diffusion 
coefficient assumed by hydrodynamics. It is not clear, however, whether
in $d=3$ the simulated systems are large enough to eliminate finite-size effects.

In $d=1$ the problem of infinite-temperature spin-transport has recently
been studied by several authors~\cite{Ljubotina17,Gopalakrishnan19a,Gopalakrishnan19b,Nardis19,Nardis20,Bulchandani20,Dupont20,Nardis21,Bulchandani21} 
using insights from the Bethe ansatz for integrable chains and 
state-of-the-art numerical methods.
Most authors found that for $T = \infty$ the spin transport in isotropic spin chains is superdiffusive and can be described by a frequency-dependent diffusion coefficient ${\cal{D}} ( \omega ) \propto \omega^{-1/3}$.
However, contrary to that, recent numerical simulations~\cite{Dupont20} 
based on tensor network methods predict at $T = \infty$ normal
diffusion for non-integrable spin chains with $S > 1/2$.
At this point  
the conditions for the persistence of superdiffusive spin dynamics in non-integrable 
spin chains  are not completely understood~\cite{Nardis21,Bulchandani21}.

In this work we use recent advances in the application of
functional renormalization group (FRG) methods to
quantum spin systems~\cite{Krieg19,Tarasevych18,Goll19,Goll20} 
to calculate the dynamic spin-spin correlation function of 
Heisenberg magnets in the high-temperature limit.
By integrating the 
truncated FRG flow equation for the suitably defined irreducible
part of the spin-spin correlation function  $G ( \bd{k} , \omega )$
we derive an integral equation in momentum space 
which determines $G ( \bd{k} , \omega )$
in the entire paramagnetic phase 
of a Heisenberg magnet
on a $d$-dimensional  Bravais lattice with arbitrary exchange 
interaction.
We explicitly solve this equation in the limit of infinite temperature
in dimensions $d=1,2,3$. In three dimensions we find normal diffusion and explicitly calculate
the numerical value of the spin-diffusion coefficient ${\cal{D}}$ for Heisenberg magnets
with nearest- and next-nearest-neighbor coupling
on a simple cubic lattice.
We also calculate ${\cal{D}}$  for a
body-centered cubic lattice describing the material
Rb$_2$CuBr$_4\cdot 2$H$_2$O where 
experimental  high-temperature 
data for ${\cal{D}}$ are available~\cite{Labrujere82}.
It turns out that our result for ${\cal{D}}$ is somewhat closer 
to the experimental value than previous theoretical predictions
based on the extrapolation of the short-time expansion, 
although the measured value of ${\cal{D}}$ is still smaller than
predicted by theory. 
In two dimensions we find anomalous diffusion in the sense 
that the hydrodynamic form (\ref{eq:Ghydro}) should be generalized by replacing ${\cal{D}}$ with a frequency-dependent function $ {\cal{D}} ( \omega )$ which diverges logarithmically for $\omega \rightarrow 0$. In $d=1$ 
we find that the singularity is even stronger, $ {\cal{D}} ( \omega ) \propto \omega^{-1/3}$;
the usual dynamic scaling ${\cal{D}} ( \omega ) k^2 \propto \omega \propto k^z$
then implies the dynamical exponent $z = 3/2$,  in agreement with the 
established result for integrable spin chains~[\onlinecite{Ljubotina17,Gopalakrishnan19a,Gopalakrishnan19b,Nardis19,Nardis20,Bulchandani20,Dupont20,Nardis21,Bulchandani21}].

The spin functional renormalization group (SFRG)  approach developed in this work also
allows us to
calculate $G ( \bd{k} , \omega )$ for all wavevectors $\bd{k}$, 
including the short-wavelength regime which cannot be described by
hydrodynamics.
Therefore we parametrize  the retarded spin-spin correlation function in the form
 \begin{equation}
 G ( \bd{k} , \omega  ) = G ( \bd{k} ) \frac{ \Delta ( \bd{k} , \omega )  }{ 
  \Delta ( \bd{k}, \omega)  - i \omega },
 \label{eq:Gall}
 \end{equation}
and  explicitly calculate the dissipation energy $ \Delta ( \bd{k}  ,  \omega )$ 
at infinite temperature for all 
wavevectors $\bd{k}$ in the first Brillouin zone.
For a three-dimensional nearest-neighbor Heisenberg model
on a cubic lattice
we find that $ \Delta ( \bd{k}  , 0)$ assumes a global maximum at  the  
corners of the Brillouin zone.
For Heisenberg magnets  with interactions 
beyond nearest neighbors 
the momentum dependence of
$ \Delta ( \bd{k}  , \omega )$ 
in the first Brillouin zone leads to characteristic features 
in the dynamic structure factor which put constraints on the
relative sign and strength of
competing exchange interactions.

The rest of this article is organized as follows:
In Sec.~\ref{sec:SFRG} we present a specific variant of the SFRG approach \cite{Krieg19}
which enables us to calculate   the dynamic spin-spin correlation function
in the paramagnetic regime of quantum Heisenberg models. 
We also write down exact flow equations
for the suitably defined irreducible static self-energy $\Sigma ( \bd{k} )$ and
the irreducible dynamic susceptibility $\tilde{\Pi} ( \bd{k} , \omega )$
which is inversely proportional to the 
dissipation energy $\Delta ( \bd{k} ,  \omega )$ defined via
Eq.~(\ref{eq:Gall}).
In Sec.~\ref{sec:trunc} we use constraints  on the irreducible three-point and four-point vertices
imposed by Ward identities and a continuity condition due to ergodicity
to derive a truncated flow equation  for $\tilde{\Pi} ( \bd{k} , \omega )$. We then integrate this flow equation
to obtain an integral equation for the dissipation energy
${\Delta} ( \bd{k} , \omega )$ which depends on
the static spin-spin correlation function $ G ( \bd{k} )$.
Using the fact that in the high-temperature limit
$G ( \bd{k} )$ can be obtained from a controlled expansion in powers of $1/T$, 
in Sec.~\ref{sec:high} we explicitly solve the integral equation for 
$\Delta ( \bd{k} , \omega )$
in the limit of infinite temperature and calculate the resulting dynamic spin-spin correlation function
$G ( \bd{k} , \omega )$ in different dimensions.
In Sec.~\ref{sec:dissip} we discuss the behavior of the dissipation  energy
$\Delta ( \bd{k} , \omega )$ defined via Eq.~(\ref{eq:Gall}) 
and the corresponding dynamic structure factor
as a function of $\bd{k}$ in the first Brillouin zone.
In the concluding Sec.~\ref{sec:summary} we 
summarize our results and give an outlook on future applications of our method. 
Finally, in two appendices we present  technical details of the solution of the integral equation
for  $\Delta ( \bd{k} , \omega )$ on different lattices.

\section{SFRG with 
classical-quantum decomposition}
\label{sec:SFRG}

In this section we shall develop a variant of the SFRG approach proposed in Ref.~[\onlinecite{Krieg19}]
which  is specially tailored to the problem of calculating  
the dynamic spin-spin correlation
function in the paramagnetic phase of 
spin-rotationally invariant quantum Heisenberg models
with Hamiltonian
  \begin{equation}
 {\cal{H}} =  \frac{1}{2} \sum_{ij} {J}_{ij} \bd{S}_i \cdot \bd{S}_j.
 \label{eq:Heisenberg}
 \end{equation}
Here $\bd{S}_i$ are spin-$S$ operators localized at the sites
$\bd{R}_i$
of a $d$-dimensional Bravais lattice with lattice spacing $a$, where
the index $i = 1, \ldots , N$ labels the  lattice sites.
The exchange couplings 
$J_{ij}$ are assumed to depend only on the difference $\bd{R}_i - \bd{R}_j$ 
so that they can be expanded in a Fourier series,
 \begin{equation}
 J_{ij} = \frac{1}{N} \sum_{\bd{k} } e^{ i \bd{k} \cdot ( \bd{R}_i - \bd{R}_j ) }
 J ( \bd{k} ),
 \end{equation}
where the $\bd{k}$-sum is over the first Brillouin zone. 

\subsection{Subtracted exchange interaction and irreducible dynamic
susceptibility}

The basic idea of Ref.~[\onlinecite{Krieg19}] is to replace the
exchange couplings $J_{ij}$ in the original Heisenberg model (\ref{eq:Heisenberg})
by some continuous deformation
$J^{\Lambda}_{ i j}$ and to derive a formally exact flow equation
describing the evolution of the imaginary-time ordered connected
spin correlation functions under changes of the deformation 
parameter $\Lambda$. For classical spin models this strategy  has been implemented 
previously by Machado and Dupuis~\cite{Machado10}. 
A related  strategy has also been adopted for bosonic quantum lattice 
models~\cite{Rancon11a,Rancon11b, Rancon12a,Rancon12b,Rancon14}.
Starting point is the deformed generating functional 
of the imaginary-time ordered connected spin correlation functions of our deformed 
quantum spin model, 
 \begin{eqnarray}
 & & {\cal{G}}_{\Lambda} [ \bd{h} ] 
 \nonumber
 \\
 &  & =
\ln {\rm Tr}
 \left[  {\cal{T}} e^{ \int_0^{\beta} d \tau 
 \left[
\sum_i \bd{h}_i ( \tau )
 \cdot \bd{S}_i ( \tau ) - \frac{1}{2} \sum_{ij} J^{\Lambda}_{ ij}
 \bd{S}_i ( \tau ) \cdot \bd{S}_j ( \tau ) \right]} \right],
 \hspace{7mm}
 \label{eq:Gcondef}
 \end{eqnarray}
where $\beta = 1/T$ denotes the inverse temperature,
$\bd{h}_i ( \tau )$ is  fluctuating source magnetic field,  ${\cal{T}}$ denotes time-ordering
in imaginary time, and the imaginary-time label $\tau$ of the spin operators 
$ \bd{S}_i ( \tau )$ keeps track of the time-ordering. 
By simply differentiating both sides of Eq.~(\ref{eq:Gcondef}) 
with respect to the deformation parameter $\Lambda$ we obtain the exact flow 
equation~\cite{Krieg19},
 \begin{eqnarray}
\partial_{\Lambda}{\cal{G}}_{\Lambda} [ \bd{h} ]  
 &  = & -
\frac{1}{2} \int_0^{\beta} d \tau 
 \sum_{ij, \alpha} ( \partial_{\Lambda} J^{\Lambda}_{ij} ) 
 \Biggl[\frac{ \delta^2 {\cal{G}}_{\Lambda} [ \bd{h} ] }{\delta h_i^{\alpha} ( \tau )   
\delta h_j^{\alpha} ( \tau ) }
 \nonumber
 \\
 & & \hspace{16mm}
 + 
\frac{ \delta {\cal{G}}_{\Lambda} [ \bd{h} ] }{\delta h_i^{\alpha} ( \tau ) }
\frac{ \delta {\cal{G}}_{\Lambda} [ \bd{h} ] }{\delta h_j^{\alpha} ( \tau ) }
  \Biggr],
 \label{eq:flowW}
 \end{eqnarray}
where $\alpha = x,y,z$ labels the three Cartesian components of $\bd{h}_i ( \tau )$.
In principle, we can now introduce the (subtracted) Legendre transform
of ${\cal{G}}_\Lambda [ \bd{h} ]$ in the usual 
way~\cite{Berges02,Pawlowski07,Kopietz10,Metzner12,Dupuis21} and derive the corresponding Wetterich equation \cite{Wetterich93}.
The problem with this procedure is that in a deformation scheme
where at the initial value $\Lambda =0$ of the deformation parameter
 the deformed exchange interaction vanishes the Legendre transform
of ${\cal{G}}_{\Lambda =0} [ \bd{h}] $ does not exist \cite{Goll19,Rancon14} 
because for vanishing exchange couplings the
spins do not have any dynamics. 
As already noticed in
Refs.~[\onlinecite{Krieg19,Goll19,Goll20}], this problem can be avoided
by introducing a 
hybrid functional which generates amputated correlation functions where the
external interaction lines are removed.
In this work we further develop this idea by noting that
in the classical approximation where the time-dependence of all operators is simply neglected the Legendre transform of  ${\cal{G}}_\Lambda [ \bd{h} ]$ 
does exist. It is therefore useful to decompose 
the source field $\bd{h}_i ( \tau )$
into  classical and  quantum components. Technically, this can be achieved by
expanding $\bd{h}_i ( \tau )$ in frequency space and identifying the
zero-frequency component with the classical source $\bd{h}_i^c$,
 \begin{eqnarray}
 \bd{h}_i ( \tau )  =   T \sum_{\omega} e^{- i \omega \tau } \bd{h}_{i, \omega }
= \bd{h}_{ i }^c + \bd{h}_i^q ( \tau ),
 \end{eqnarray}
where
 \begin{subequations}
  \begin{eqnarray}
 \bd{h}_i^c & = &  T \bd{h}_{i , \omega =0},
 \label{eq:hcdef}
 \\
 \bd{h}_i^q ( \tau ) & = &   T \sum_{\omega \neq 0} e^{- i \omega \tau } 
\bd{h}_{i, \omega }.
 \label{eq:hqdef}
 \end{eqnarray}
 \end{subequations}
In frequency space this decomposition is equivalent with
 \begin{equation}
 \bd{h}_{i , \omega } = \beta \delta_{\omega,0} \bd{h}_i^c + ( 1 - \delta_{\omega , 0} ) \bd{h}_{i, \omega}^q.
 \end{equation} 
Since in the classical sector the Legendre transform of
${\cal{G}}_{\Lambda} [ \bd{h}^c ]$ is well defined even for vanishing 
exchange coupling, it is convenient to introduce a hybrid 
functional $\Gamma_{\Lambda} [ \bd{m}^c,
 \bd{\eta}^q ]$ which for vanishing quantum source $\bd{\eta}^q =0$ reduces to the Legendre transform of the generating functional 
${\cal{G}}_{\Lambda} [ \bd{h}^c ]$ with classical sources.
To construct such a functional, recall that in the paramagnetic phase
the imaginary-frequency spin-spin correlation 
function can be written as 
 \begin{equation}
  G ( \bd{k} , i \omega ) = \frac{ \Pi (\bd{k} , i \omega ) }{ 1 + J ( \bd{k} )
 \Pi ( \bd{k} , i \omega ) },
 \label{eq:Gpi}
 \end{equation}
where
$\Pi ( \bd{k} , i \omega )$  is the interaction-irreducible part of
$G ( \bd{k} , i \omega)$. We shall refer to  $\Pi ( \bd{k} , i \omega )$
as the irreducible dynamic susceptibility.
In the spin-diagram technique developed by Vaks, Larkin, and Pikin~\cite{Vaks68,Vaks68b}
(see also the textbook by Izyumov and Skryabin \cite{Izyumov88}) 
the  function $\Pi ( \bd{k} , i \omega )$ is the sum of all diagrams contributing to
$G ( \bd{k} , i \omega )$ which cannot be separated into two parts by 
cutting a single interaction line representing $J ( \bd{k} )$.
For our purpose, it is more convenient to parametrize the
spin-spin correlation function in a slightly different way,
\begin{equation}
  G ( \bd{k} , i \omega ) = \frac{ \tilde{\Pi} (\bd{k} , i \omega ) }{ 1 + \tilde{J} ( \bd{k} )
 \tilde{\Pi} ( \bd{k} , i \omega ) },
 \label{eq:Gtildepi}
 \end{equation}
where the  subtracted exchange interaction is defined by
 \begin{equation}
 \tilde{J} ( \bd{k} )  \equiv J ( \bd{k} ) + \Pi^{-1} ( \bd{k} , 0 ).
 \label{eq:tildeJdef}
 \end{equation}
Combining this with 
the definition  of $G ( \bd{k} , 0 )   \equiv  G ( \bd{k} )$
in  Eq.~(\ref{eq:Gpi}) we find
   \begin{equation}
  \frac{1}{\tilde{J} ( \bd{k} ) } =  \frac{1}{ J ( \bd{k} ) + \Pi^{-1} ( \bd{k} ,0 )} =
\frac{ {\Pi} (\bd{k} , 0 ) }{ 1 + {J} ( \bd{k} )
   {\Pi} ( \bd{k} , 0 ) } = G ( \bd{k} ),
 \label{eq:tildeJG}
  \end{equation}
i.e.,  our subtracted exchange interaction $\tilde{J} ( \bd{k})$ is the inverse of the
static spin-spin correlation function $G ( \bd{k} )$.
Using the definitions (\ref{eq:Gpi})--(\ref{eq:tildeJdef}), we conclude that 
for finite frequency the irreducible susceptibility $\Pi ( \bd{k} , i \omega )$ and its subtracted counterpart
are related  as follows
 \begin{equation}
 \tilde{\Pi}^{-1} ( \bd{k} , i \omega ) \equiv \Pi^{-1} ( \bd{k} , i \omega )
 - \Pi^{-1} ( \bd{k} , 0 ).
 \label{eq:Pisub}
 \end{equation}

For vanishing exchange interaction the imaginary frequency
spin-spin correlation function has a non-analytic frequency dependence,
 \begin{equation}
 G_0 ( \bd{k} , i \omega ) = \Pi_0 ( \bd{k} , i \omega ) =
\delta_{\omega ,0 } \frac{b_0^{\prime}}{T} ,
 \label{eq:nonana}
 \end{equation}
where
 \begin{equation}
 b_0^{\prime} = \frac{ S ( S+1)}{3}
 \label{eq:b0prime}
 \end{equation}
is the first order coefficient in the Taylor expansion  of the spin-$S$ Brillouin function 
 \begin{eqnarray}
 b ( y ) & = &  \left( S + \frac{1}{2} \right)
 \coth \left[ \left( S + \frac{1}{2} \right) y \right]
 - \frac{1}{2} \coth \left[ \frac{y}{2} \right]
 \nonumber
 \\
 & = & b_0^{\prime} y + {\cal{O}} ( y^3 ).
 \label{eq:bydef}
 \end{eqnarray}
Whether or not such a non-analytic contribution  proportional to 
$\delta_{\omega ,0}$ survives for finite exchange coupling 
is closely related to the 
ergodicity of the system
and the distinction between
the isolated (Kubo) susceptibility and the isothermal 
susceptibility~\cite{Kubo57,Wilcox68,Kwok69,Pirc74,Chiba20}.
Note that for $\bd{k} =0$ the zero-frequency limit of the finite-frequency
thermal  
spin-spin correlation function $ G ( \bd{k}=0, i \omega)$ gives the 
isolated (Kubo) susceptibility, which in general does not agree 
with the isothermal susceptibility defined via the derivative of the magnetization with respect to an external magnetic field
at constant temperature \cite{Kubo57,Wilcox68,Kwok69,Pirc74,Chiba20}.
However, as recently shown by Chiba {\it{et al.}}~\cite{Chiba20}, under  conditions
similar to the eigenstate thermalization hypothesis~\cite{Alessio16}, at finite momentum
$\bd{k} \neq 0$
all static susceptibilities agree. This rules out a non-analytic contribution
to the thermal spin-spin correlation function
$G ( \bd{k} , i \omega )$ 
similar to Eq.~(\ref{eq:nonana})
for finite exchange coupling and finite $\bd{k}$.
Consequently, in this case the irreducible susceptibility 
 $\Pi ( \bd{k} , i \omega )$ is expected to be a continuous function of $\omega $, so that for finite exchange coupling and finite
momentum we conclude from Eq.~(\ref{eq:Pisub}) that in the zero-frequency limit
the inverse
of the irreducible subtracted susceptibility  defined via Eqs.~(\ref{eq:Gtildepi}) and
 (\ref{eq:tildeJdef}) vanishes,
 \begin{equation}
 \tilde{\Pi}^{-1} ( \bd{k} \neq 0  , 0) \equiv \lim_{\omega \rightarrow 0}\tilde{\Pi}^{-1} ( \bd{k} \neq 0  , i\omega) = 0.
 \label{eq:Piinvnull}
 \end{equation}
We shall to refer to Eq.~(\ref{eq:Piinvnull})  as the {\it{continuity condition}}.

\subsection{Hybrid functional and generalized Wetterich equation}

Let us write the deformed exchange interaction in momentum
space in the form
 \begin{equation}
 J_{\Lambda} ( \bd{k} ) = J ( \bd{k} ) + R_\Lambda ( \bd{k} ),
 \end{equation}
where $R_{\Lambda} ( \bd{k} )$ is some momentum dependent  regulator
which vanishes at $\Lambda =1$ where we recover our original model.
The deformed spin-spin correlation function can then be written in two  equivalent ways,
 \begin{subequations}
\begin{eqnarray}
  G_{\Lambda} ( \bd{k} , i \omega ) & = &
 \frac{ {\Pi}_{\Lambda} (\bd{k} , i \omega ) }{ 
 1 + {J}_{\Lambda} ( \bd{k} )
 {\Pi}_{\Lambda} ( \bd{k} , i \omega ) }
 \label{eq:GLambdaPi}
 \\
 & = & 
\frac{ \tilde{\Pi}_{\Lambda} (\bd{k} , i \omega ) }{ 
1 + \tilde{J}_{\Lambda} ( \bd{k} )
 \tilde{\Pi}_{\Lambda} ( \bd{k} , i \omega ) },
 \label{eq:GLambda}
 \end{eqnarray}
 \end{subequations}
where the deformed subtracted exchange interaction is defined
analogously to $\tilde{J} ( \bd{k} )$ in Eq.~(\ref{eq:tildeJdef}),
 \begin{equation}
  \tilde{J}_{\Lambda} ( \bd{k} ) =  {J}_{\Lambda} ( \bd{k} ) + 
 \Pi^{-1}_{\Lambda} ( \bd{k} ,0 )   
= G^{-1}_{\Lambda} ( \bd{k} , 0 ) \equiv G^{-1}_{\Lambda}  ( \bd{k} ).
 \label{eq:tildeJLambda}
 \end{equation}
The subtracted irreducible susceptibility therefore satisfies by construction
the continuity condition
 \begin{equation}
  \tilde{\Pi}^{-1}_{\Lambda} ( \bd{k} \neq 0 , 0 ) =0,
 \label{eq:continuity}
 \end{equation}
which generalizes the condition (\ref{eq:Piinvnull}) for all values of the
deformation parameter $\Lambda$.

Our aim is to derive exact FRG flow equations for  the static self-energy \begin{equation}
 \Sigma_{\Lambda} ( \bd{k} )  \equiv \Pi^{-1}_{\Lambda} ( \bd{k} , 0 )
 \label{eq:Sigmastat}
 \end{equation}
and for the subtracted irreducible dynamic  
susceptibility $\tilde{\Pi}_{\Lambda} ( \bd{k} , i \omega )$
defined via Eq.~(\ref{eq:GLambda}).
To  construct the corresponding  generating functional, we 
first introduce the auxiliary functional
 \begin{equation}
 {\cal{F}}_\Lambda [ {\bd{h}}^c , {\bd{s}}^q ] = {\cal{G}}_{\Lambda} [  \bd{h}^c - \tilde{\mathbf{J}}_{\Lambda} \bd{s}^q ]
 - \frac{1}{2} ( \bd{s}^q, \tilde{\mathbf{J}}_{\Lambda} \bd{s}^q ),
 \label{eq:Fhybrid}
 \end{equation}
which depends on the classical component $\bd{h}_i^c$
of the source field $\bd{h}_i ( \tau )$ defined in Eq.~(\ref{eq:hcdef}) and on a
quantum field $\bd{s}_i^q ( \tau )$ which is introduced via the following substitution of  the quantum component $\bd{h}_i^q ( \tau )$
of the source field defined in Eq.~(\ref{eq:hqdef}),
 \begin{equation}
\bd{h}_i^q ( \tau ) =  - [ \tilde{\mathbf{J}}_{\Lambda} \bd{s}^q ]_{i \tau }  =
   - \sum_j \tilde{J}^{\Lambda}_{ ij} \bd{s}_j^q ( \tau ).
 \end{equation}
In Eq.~(\ref{eq:Fhybrid}) the symbol 
$\tilde{\mathbf{J}}_{\Lambda}$ 
represents an infinite matrix in the site label $i$  and imaginary time $\tau$,
 \begin{equation}
 [ \tilde{\mathbf{J}}_{\Lambda} ]_{ i \tau , j \tau^{\prime} }
 = \delta ( \tau - \tau^{\prime} ) \tilde{J}^{\Lambda}_{ij},
 \label{eq:tildeJmatrix}
 \end{equation}
and the last term in Eq.~(\ref{eq:Fhybrid}) is a short notation for
 \begin{equation}
 ( \bd{s}^q, \tilde{\mathbf{J}}_{\Lambda} \bd{s}^q ) = \int_0^{\beta} d \tau
 \sum_{ij}  \tilde{J}^{\Lambda}_{ij}  \bd{s}_i^{q} ( \tau ) 
 \cdot \bd{s}_j^q ( \tau ).
 \end{equation}
Differentiation of the auxiliary functional  ${\cal{F}}_\Lambda [ {\bd{h}}^c , {\bd{s}}^q ]$
defined in Eq.~(\ref{eq:Fhybrid})
with respect to the source fields generates connected correlation functions which are 
partially amputated in the quantum sector. The corresponding two-point function at finite frequencies can then be interpreted as an effective, subtracted exchange interaction, while higher order correlation functions can be obtained
from their connected counterparts by multiplying the quantum legs by factors of$-\tilde{\mathbf{J}}_{\Lambda}$. A related auxiliary functional has been
introduced in Ref.~[\onlinecite{Goll19}].
Our hybrid functional with the desired properties is now
given by the subtracted Legendre transform of the above auxiliary functional 
${\cal{F}}_\Lambda [ {\bd{h}}^c , {\bd{s}}^q ]$, 
 \begin{eqnarray}
 \Gamma_{\Lambda} [ {\bd{m}}^c , {\bd{\eta}}^q ] & = & ( {\bd{m}}^c, {\bd{h}}^c ) + 
 ( {\bd{\eta}}^q , {\bd{s}}^q ) 
 - {\cal{F}}_\Lambda [ {\bd{h}}^c , {\bd{s}}^q ] 
 \nonumber
 \\
 &- &  \frac{1}{2} ( {\bd{m}}^c , \mathbf{R}_\Lambda^c  {\bd{m}}^c ) - 
 \frac{1}{2} ( \bd{\eta}^q , \mathbf{R}_{\Lambda}^q  \bd{\eta}^q ),
 \hspace{7mm}
 \label{eq:Gammahybrid}
 \end{eqnarray}
where on  the right-hand side we should substitute
$\bd{h}^c = \bd{h}^c [   \bd{m}^c,  \bd{\eta}^q ]$ and 
$\bd{s}^q = \bd{s}^q [ \bd{m}^c, \bd{\eta}^q ]$ as functionals of $\bd{m}^c$ and $\bd{\eta}^q$ 
by inverting the relations
 \begin{eqnarray}
  \bd{m}^c &  = & \frac{ \delta {\cal{F}}_{\Lambda} [ \bd{h}^c , \bd{s}^q ]}{\delta \bd{h}^c },
 \\
 \bd{\eta}^q & = &  \frac{ \delta {\cal{F}}_{\Lambda} [ \bd{h}^c , \bd{s}^q ]}{\delta \bd{s}^q } .
 \end{eqnarray}
The regulator matrices in Eq.~(\ref{eq:Gammahybrid}) are in the momentum-time domain given by
 \begin{eqnarray}
 {[} \mathbf{R}^\alpha_{\Lambda} {]}_{ \bd{k} \tau , \bd{k}^{\prime} \tau^{\prime} }
 & = &  \delta ( \tau - \tau^{\prime} )\delta (  \bd{k} + \bd{k}^{\prime} )
 R^\alpha_{\Lambda} ( \bd{k} ), \; \; \; \alpha = c, q, \hspace{9mm}
 \end{eqnarray}
with
 \begin{subequations}
 \begin{eqnarray}
 R^c_{\Lambda} ( \bd{k} ) & = &  
{J}_{\Lambda} ( \bd{k} ) - {J} ( \bd{k} ) =  R_{\Lambda} ( \bd{k} )   , \\
  R^q_{\Lambda} ( \bd{k} ) & = &  - 
 \frac{1}{\tilde{J}_{\Lambda} ( \bd{k} )}  +  \frac{1}{\tilde{J} ( \bd{k} )}  .
 \end{eqnarray}
 \end{subequations}
Here the
$\delta$-symbol in wavevector space is defined by
$\delta ( \bd{k} ) = N \delta_{ \bd{k} , 0 }$.
 
After some standard manipulations similar to those outlined
in Ref.~[\onlinecite{Goll19}] we find
that the hybrid functional 
$
\Gamma_{\Lambda} [ {\bd{m}}^c , {\bd{\eta}}^q ]$  defined in 
Eq.~(\ref{eq:Gammahybrid})
satisfies the generalized Wetterich equation
 \begin{widetext}
 \begin{eqnarray}
 \partial_{\Lambda} \Gamma_{\Lambda} [ \bd{m}^c , \bd{\eta}^q  ] 
   & = &  
  \frac{1}{2} {\rm Tr} 
 \left\{ \left[
 \left(    \mathbf{\Gamma}^{\prime \prime} _{\Lambda} [ \bd{m}^c , \bd{\eta}^q  ] +
 \mathbf{R}_{\Lambda} \right)^{-1}   + {\mathbf{{J}}}_{\Lambda}^q \right]  \dot{\mathbf{R}}_{\Lambda}    \right\}
 \nonumber
 \\
 &-  &  \frac{1}{2} \sum_{ij} ( \partial_{\Lambda} 
 \Sigma^{\Lambda}_{ij} )  \int_0^{\beta} d \tau  d \tau'    \delta^{q} (\tau - \tau') \Big( \frac{ 
 \delta \Gamma_{\Lambda} }{\delta \bd{\eta}^q_{ i} ( \tau ) }
  + \big[ \mathbf{\tilde J}^{-1}\bd{\eta}^q\big]_{i \tau} \Big)
 \cdot  \Big( \frac{
\delta \Gamma_{\Lambda}}{\delta \bd{\eta}^q_{ j} ( \tau') } + \big[\mathbf{\tilde J}^{-1}\bd{\eta}^q\big]_{j \tau'} \Big),
 \label{eq:Wetterichhybrid}
 \hspace{7mm}
 \end{eqnarray}
 \end{widetext}
where 
 \begin{equation}
\Sigma^{\Lambda}_{ij}  = \frac{1}{N} \sum_{\bd{k}}
 e^{ i \bd{k} \cdot ( \bd{R}_i - \bd{R}_j ) } \Sigma_{\Lambda} ( \bd{k} ) \end{equation}
 is the real-space Fourier transform of the flowing static self-energy,
 the finite-frequency part of the periodic 
imaginary-time $\delta$-function $\delta ( \tau ) = T \sum_\omega e^{ i \omega \tau }$ is denoted by
 \begin{equation}
 \delta^q ( \tau  ) = T \sum_{\omega \neq 0 }  e^{ i \omega \tau } = \delta ( \tau ) - T,
 \end{equation}
and  the matrix 
$\mathbf{\Gamma}^{\prime \prime}_{\Lambda} [ \bd{m}^c , \bd{\eta}^q ]$
of second functional derivatives of 
$\Gamma_{\Lambda} [ \bd{m}^c , \bd{\eta}^q ]$ is explicitly given by
 \begin{equation}
 \left(    \mathbf{\Gamma}^{\prime \prime} _{\Lambda} [ \bd{m}^c , \bd{\eta}^q  ]
 \right)_{ i \tau \alpha , j \tau^{\prime} \alpha^{\prime}} =
 \frac{ \delta^2 \Gamma_{\Lambda} [ \bd{m}^c , \bd{\eta}^q  ] }{
 \delta \Phi^{\alpha}_i ( \tau ) \delta \Phi^{\alpha^{\prime}}_j ( \tau^{\prime} )}.
  \end{equation}
Here we have  combined the components of $\bd{m}_i^c$ and
$\bd{\eta}_i^q ( \tau )$ into a six-component field
 \begin{equation}  
 \left(   \begin{array}{c} \Phi_i^{m_x} ( \tau ) \\ \Phi_i^{m_y} ( \tau ) 
 \\ \Phi_i^{m_z} ( \tau )  \\
 \Phi_i^{\eta_x} ( \tau ) \\
  \Phi_i^{\eta_y} ( \tau ) \\
 \Phi_i^{\eta_z} ( \tau )
\end{array} \right)    = \left( \begin{array}{c} {m}^{c}_{x,i}  \\
   m^{c}_{y,i}  \\  m^{c}_{z,i} \\
 \eta^{q}_{x,i} ( \tau ) 
 \\
 \eta^{q}_{y,i} ( \tau ) \\ \eta^{q}_{z,i} ( \tau )
 \end{array}  \right)
 = \left( \begin{array}{c} \bd{m}^c_i \\ \bd{\eta}^q_i ( \tau ) \end{array}
 \right).
 \label{eq:collectivefield}
 \end{equation}
The regulator matrix
$\mathbf{R}_{\Lambda}$
and the matrix $\mathbf{J}^q_{\Lambda}$ in the generalized Wetterich 
equation~(\ref{eq:Wetterichhybrid}) have the following
block structure in the space of field components,
 \begin{equation}
 \mathbf{R}_{\Lambda} = \left( \begin{array}{cc} \mathbf{R}^c_{\Lambda} & 0 \\
 0 & \mathbf{R}^q_{\Lambda} \end{array} \right), \; \; \;
\mathbf{J}^q_{\Lambda} = \left( \begin{array}{cc} 0 & 0 \\
 0 & \tilde{\mathbf{J}}_{\Lambda} \end{array} \right),
 \end{equation}
and the matrix $\dot{\mathbf{R}}_{\Lambda}$ is defined by
 \begin{equation}
  \dot{\mathbf{R}}_{\Lambda} \equiv \left( \begin{array}{cc} {\partial}_\Lambda  
\mathbf{J}_{\Lambda} & 0 \\
 0 & \tilde{\mathbf{J}}^{-1}_{\Lambda}[\partial_\Lambda \mathbf{J}_\Lambda]  \tilde{\mathbf{J}}^{-1}_{\Lambda}  \end{array} \right),
 \label{eq:modifiedregderivative}
 \end{equation}
where the deformed exchange interaction matrix
  \begin{equation}
 [{\mathbf{J}}_{\Lambda} ]_{ i \tau , j \tau^{\prime} }
= \delta ( \tau - \tau^{\prime} ) J^{\Lambda}_{ij}
 \end{equation}
is defined analogously to its subtracted counterpart in Eq.~(\ref{eq:tildeJmatrix}).
Note that $\dot{\mathbf{R}}_{\Lambda} \neq \partial_{\Lambda} \mathbf{R}_{\Lambda}$,
because by taking the $\Lambda$-derivative of $\Gamma_{\Lambda} [ \bd{m}^c,
 \bd{\eta}^q ]$ in Eq.~(\ref{eq:Gammahybrid}) we generate only terms involving 
 derivatives  $\partial_{\Lambda} \mathbf{J}_\Lambda$ of the deformed bare coupling;
due to the $\Lambda$-dependent  subtraction $\Pi^{-1}_{\Lambda} ( \bd{k} , 0 )$
in the definition (\ref{eq:tildeJLambda}) of $\tilde{J}_{\Lambda} ( \bd{k} )$,
the quantum sector of the matrix  $\dot{\mathbf{R}}_{\Lambda}$
is therefore in general different from
${\partial}_\Lambda  \mathbf{R}^q_{\Lambda}$.
Diagrammatically,  the vertices generated by $\Gamma_{\Lambda} [ \bd{m}^c , \bd{\eta}^q ]$
are classical propagator irreducible, i.e. the diagrams contributing to the
vertices cannot be separated into two parts by cutting a single
classical propagator line representing $ {G}_\Lambda ( \bd{k} ,0  )$.
Moreover, for finite frequencies the  
vertices generated by $\Gamma_{\Lambda} [ \bd{m}^c , \bd{\eta}^q ]$
are also interaction-irreducible, which means
that diagrammatically  the vertices generated by expanding
$\Gamma_{\Lambda} [ \bd{m}^c , \bd{\eta}^q ]$ in powers of $\bd{\eta}^q$
cannot be separated into two parts by  cutting a single effective interaction 
line representing $\tilde{J}_{\Lambda} ( \bd{k} )$. 
Note that the last term on the right-hand side of Eq.~(\ref{eq:Wetterichhybrid}),
which is  absent in the usual Wetterich equation,\cite{Wetterich93}
is generated by the scale-dependent subtraction $\Pi^{-1}_{\Lambda} ( \bd{k} , 0 )
 = \Sigma_{\Lambda} ( \bd{k} )$ in the definition (\ref{eq:tildeJLambda})
of $\tilde{J}_{\Lambda} ( \bd{k} )$.
This term gives rise to local tree contributions to the flow equations for the vertex functions which do not contribute to the flow of $\Sigma_{\Lambda} ( \bd{k} )$ or any other static irreducible vertices, due to the subtraction of the time-independent contribution in the second line of Eq.~(\ref{eq:Wetterichhybrid}).

\subsection{Vertex expansion}

The generalized Wetterich equation (\ref{eq:Wetterichhybrid})
implies an infinite hierarchy of exact FRG flow equations 
for the irreducible vertices which can be obtained  by
expanding the functional $\Gamma_{\Lambda} [ \bd{m}^c , \bd{\eta}^q ]$ in powers of the fields.
Let us first consider the vertex expansion 
in the classical sector, which is obtained by setting 
$\bd{\eta}^q =0$.
To simplify our notation, let us rename $\bd{m}^c \rightarrow \bd{m}$.
In the paramagnetic regime the classical magnetization field $\bd{m}$ vanishes
for vanishing external magnetic fields, so that 
the first few terms of the vertex expansion  in the classical sector are
 \begin{widetext}
 \begin{eqnarray}
  {\Gamma}_{\Lambda} [  \bd{m} , 0  ]  & = &   
   {\Gamma}_{\Lambda} [ 0 , 0 ]
 +  \frac{\beta }{2!} \int_{\bd{k}} \left[  
  {J} ( \bd{k} ) +  {\Sigma}_{\Lambda} ( \bd{k} ) \right] 
 \bd{m}_{- \bd{k} } \cdot \bd{m}_{\bd{k}}
\nonumber
 \\
 &   +  & \beta  \int_{\bd{k}_1} \int_{\bd{k}_2} \int_{\bd{k}_3}  \int_{ \bd{k}_4} 
\delta ( \bd{k}_1 + \bd{k}_2 + \bd{k}_3 + \bd{k}_4) 
 \biggl\{  \frac{1}{ (2! )^2}
  \Gamma^{--++}_{\Lambda} ( \bd{k}_1, \bd{k}_2 ,  \bd{k}_3 ,  \bd{k}_4 )  
 m^-_{\bd{k}_1} m^-_{\bd{k}_2} m^+_{ \bd{k}_3 } m^+_{  \bd{k}_4} 
 \nonumber
 \\
 & +   & 
 \frac{1}{ 2! } 
  \Gamma^{-+zz}_{\Lambda} ( \bd{k}_1, \bd{k}_2 ,  \bd{k}_3 ,  \bd{k}_4 )  
 m^-_{\bd{k}_1} {m}^+_{\bd{k}_2} m^z_{ \bd{k}_3 } m^z_{  \bd{k}_4} 
+  \frac{1}{4!} 
\Gamma^{zzzz}_{\Lambda} ( \bd{k}_1 , \bd{k}_2 , \bd{k}_3 , \bd{k}_4 ) 
 m^z_{\bd{k}_1 } m^z_{\bd{k}_2} m^z_{ \bd{k}_3 } 
 m^z_{\bd{k}_4} \biggr\}
+ \ldots \; , \hspace{7mm}
 \label{eq:Gammahcomplete}
 \end{eqnarray}
 \end{widetext}
where $\int_{\bd{k}} = \frac{1}{N} \sum_{\bd{k}}$, 
the interaction vertices are given in the spherical basis with
$m^{\pm}_{\bd{k}} = ( m^x_{\bd{k}} \pm i m^y_{\bd{k}} ) / \sqrt{2}$,
and
we have omitted vertices with five and more external legs.
Note that by  adding to the coefficient of the
quadratic term in Eq.~(\ref{eq:Gammahcomplete})
the regulator $R_{\Lambda} ( \bd{k} )$ we obtain
 \begin{eqnarray}
 & &  {J} ( \bd{k} ) +  {\Sigma}_{\Lambda} ( \bd{k} ) + R_{\Lambda} ( \bd{k} ) 
  =   
J_{\Lambda} ( \bd{k} )  +  {\Sigma}_{\Lambda} ( \bd{k} ) 
 \nonumber
 \\  
& = & J_{\Lambda} ( \bd{k} )  +  {\Pi}^{-1}_{\Lambda} ( \bd{k} ,0 ) 
 =
 G^{-1}_{\Lambda} ( \bd{k} , 0 ) \equiv G^{-1}_{\Lambda} ( \bd{k} ),
 \hspace{7mm}
 \label{eq:Gclassreg}
 \end{eqnarray}
which can be identified with the inverse of the
deformed static spin-spin correlation function defined via
Eq.~(\ref{eq:GLambdaPi}).
In a cutoff scheme where for $\Lambda =0$ the deformed exchange coupling vanishes,
$J_{ \Lambda=0} ( \bd{k} ) =0$, we see from Eq.~(\ref{eq:nonana})
that  the classical self-energy $\Sigma_{\Lambda} ( \bd{k} )$
satisfies the initial condition
 \begin{equation} 
 {\Sigma}_{0} ( \bd{k} ) = T / b_0^{\prime},
 \end{equation}
with $b_0^{\prime} = S ( S+1)/3$, see Eq.~(\ref{eq:b0prime}).
To determine the initial values for the classical four-point vertices 
in Eq.~(\ref{eq:Gammahcomplete}) in a cutoff scheme with initially 
vanishing exchange interaction, we use the tree expansion~\cite{Kopietz10}
to relate these vertices to the corresponding four-spin correlation 
functions~\cite{Goll20,Kopietz10},
 \begin{subequations}
 \begin{eqnarray}
  G^{++--}_{\Lambda} ( \bd{k}_1 , \bd{k}_2 , \bd{k}_3 , \bd{k}_4 )
  & = &  - G_{\Lambda} ( \bd{k}_1 ) G_{\Lambda} ( \bd{k}_2 ) 
 G_{\Lambda} ( \bd{k}_3 ) G_{\Lambda} ( \bd{k}_4 ) 
 \nonumber
 \\
 & &  \hspace{-20mm} \times
 \Gamma_{\Lambda}^{--++} ( - \bd{k}_1 , - \bd{k}_2 , - \bd{k}_3 , - \bd{k}_4 ),
 \\
  G^{+-zz}_{\Lambda} ( \bd{k}_1 , \bd{k}_2 , \bd{k}_3 , \bd{k}_4 )
  & = &  - G_{\Lambda} ( \bd{k}_1 ) G_{\Lambda} ( \bd{k}_2 ) 
 G_{\Lambda} ( \bd{k}_3 ) G_{\Lambda} ( \bd{k}_4 ) 
 \nonumber
 \\
 & &  \hspace{-20mm} \times
 \Gamma_{\Lambda}^{-+zz} ( - \bd{k}_1 , - \bd{k}_2 , - \bd{k}_3 , - \bd{k}_4 ),
 \\
  G^{zzzz}_{\Lambda} ( \bd{k}_1 , \bd{k}_2 , \bd{k}_3 , \bd{k}_4 )
  & = &  - G_{\Lambda} ( \bd{k}_1 ) G_{\Lambda} ( \bd{k}_2 ) 
 G_{\Lambda} ( \bd{k}_3 ) G_{\Lambda} ( \bd{k}_4 ) 
 \nonumber
 \\
 & &  \hspace{-20mm} \times
 \Gamma_{\Lambda}^{zzzz} ( - \bd{k}_1 , - \bd{k}_2 , - \bd{k}_3 , - \bd{k}_4 ).
 \end{eqnarray}
 \end{subequations}
For vanishing exchange coupling 
 $
 G_0 ( \bd{k} ) = \beta b_0^{\prime} $
and
 \begin{subequations}
 \begin{eqnarray}
  G^{++--}_{0} ( \bd{k}_1 , \bd{k}_2 , \bd{k}_3 , \bd{k}_4 )
 & = &  \frac{2}{3} \beta^3 b_0^{\prime \prime \prime},
 \\
 G^{+-zz}_{0} ( \bd{k}_1 , \bd{k}_2 , \bd{k}_3 , \bd{k}_4 )
 & = &  \frac{1}{3} \beta^3 b_0^{\prime \prime \prime},
 \\
 G^{zzzz}_{0} ( \bd{k}_1 , \bd{k}_2 , \bd{k}_3 , \bd{k}_4 )
 & = &   \beta^3 b_0^{\prime \prime \prime},
 \end{eqnarray}
 \end{subequations}
where $b_0^{\prime \prime \prime}$ is the third order
coefficient in the Taylor expansion of the spin-$S$ 
Brillouin function $b ( y )$ given in Eq.~(\ref{eq:bydef}),
 \begin{equation}
 b ( y) = b_0^{\prime} y + \frac{1}{3!} b_0^{\prime \prime \prime} y^3 + 
 {\cal{O}} ( y^5 ).
 \end{equation}
Explicitly,
  \begin{subequations}
 \begin{eqnarray}
 b_0^{\prime} & = & \frac{ ( 2S+1)^2 -1}{12} = \frac{S (S+1)}{3},
 \\
 b_0^{\prime \prime \prime} & = & - \frac{ (  2 S+1 )^4 -1}{120}
 = - \frac{6}{5} b_0^{\prime} \left( b_0^{\prime} + \frac{1}{6} \right).
 \end{eqnarray}
 \end{subequations}
The initial values of the classical four-point vertices 
in Eq.~(\ref{eq:Gammahcomplete}) in a cutoff scheme where for $\Lambda =0$ the exchange
interaction is completely switched off 
are therefore
 \begin{subequations}
 \label{eq:initialGammaclassical}
 \begin{eqnarray}
  \Gamma_{0}^{--++} (  \bd{k}_1 ,  \bd{k}_2 ,  \bd{k}_3 ,  \bd{k}_4 )
 & = &  -  \frac{2T}{3} \frac{ b_0^{\prime \prime \prime}}{ (b_0^{\prime})^4 },
\label{eq:initialGammammpp}
 \\
   \Gamma_{0}^{-+zz} (  \bd{k}_1 ,  \bd{k}_2 ,  \bd{k}_3 ,  \bd{k}_4 )
 & = &  
-  \frac{T}{3} \frac{ b_0^{\prime \prime \prime}}{ (b_0^{\prime})^4 },
 \label{eq:initialGammampzz}
 \\
   \Gamma_{0}^{zzzz} (  \bd{k}_1 ,  \bd{k}_2 ,  \bd{k}_3 ,  \bd{k}_4 )
 & = &  -  T \frac{ b_0^{\prime \prime \prime}}{ (b_0^{\prime})^4 }.
 \label{eq:initialGammazzzz}
 \end{eqnarray}
 \end{subequations}

In order to calculate correlation functions at finite frequencies, 
we have to
include also the quantum vertices in the expansion of our 
generating functional $\Gamma_{\Lambda} [ \bd{m} , \bd{\eta} ]$,
where for notational simplicity we have renamed the quantum 
field $\bd{\eta}^q \rightarrow \bd{\eta}$.
Apart from pure quantum vertices involving only the $\bd{\eta}$-field, 
the vertex expansion contains also various types of mixed vertices, 
 \begin{widetext}
 \begin{eqnarray}
 \Gamma_{\Lambda} [ \bd{m} , \bd{\eta} ]
 & = &  \Gamma_{\Lambda} [  \bd{m} , 0 ] - \frac{1}{2} 
 \int_K \left[ \tilde{J}^{-1} ( \bd{k} )
 + \tilde{\Pi}_{\Lambda} ( K ) \right] \bd{\eta}_{-K} \cdot \bd{\eta}_K 
 \nonumber
 \\
 & + &   {\color{LimeGreen}{ [ \eta^- \eta^+ \eta^z ] + \frac{1}{3!} [ \eta^z \eta^z \eta^z ]}}
 +  {\color{red}{\frac{1}{(2!)^2} [ \eta^- \eta^- \eta^+ \eta^+ ]}}
 + {\color{Orange}{\frac{1}{2!} [ \eta^- \eta^+ \eta^z \eta^z ] }}
 + \frac{1}{4!} [  \eta^z \eta^z \eta^z \eta^z ]
 \nonumber
 \\
 & + & {\color{ForestGreen}{ 
   [  m^- \eta^+ \eta^z ]   + [ m^+ \eta^z \eta^- ]  + [ m^z \eta^- \eta^+ ]  }}  + \frac{1}{2!} [ m^z \eta^z \eta^z ]
 \nonumber
 \\
 & + &   {\color{blue}{[ m^- m^+ \eta^- \eta^+]}} +   
\frac{1}{(2!)^2} [ m^+ m^+ \eta^- \eta^- ] + \frac{1}{(2!)^2} [ m^- m^- \eta^+ \eta^+ ] 
+  \frac{1}{(2 ! )^2} [ m^z  m^z \eta^z \eta^z ]
 \nonumber
 \\
 & + & {\color{ProcessBlue}{  \frac{1}{2!} [ m^- m^+ \eta^z \eta^z ]}}  +  
 {\color{magenta}{ \frac{1}{2!} [ m^z m^z \eta^- \eta^+ ]     }} + [ m^- m^z \eta^+ \eta^z ]
 + [ m^+ m^z \eta^- \eta^z ]
 \nonumber
 \\
 & + & 
  \mbox{$[ m  \eta \eta  \eta ]$-vertices} \; + \;  \mbox{terms with $n > 4$ fields}.
 \label{eq:vertexhybrid}
 \end{eqnarray}
Here  $\int_K = \frac{1}{\beta N} \sum_{\bd{k} , \omega }$ and
$ K = ( \bd{k} , i \omega)$ is a collective label for wavevector and Matsubara frequency.
For later reference we have marked the vertices 
by various colors which match the colors in Fig.~\ref{fig:exactflow} and in the exact flow 
equations~(\ref{eq:Sigmaexactflow}) and (\ref{eq:Piexactflow})  given below.
We have also introduced the short notation
\begin{eqnarray}
 {\color{LimeGreen}{[ \eta^- \eta^+ \eta^z ]}} & = & \int_{K_1} 
 \int_{ K_2 } \int_{ K_3} \delta ( K_1 + K_2 + K_3 ) 
 \Gamma_{\Lambda}^{\eta^- \eta^+ \eta^z} ( K_1 , K_2 , K_3 )
\eta^-_{K_1}  \eta^+_{K_2} \eta^z_{K_3},
 \\
 {\color{red}{{[} \eta^- \eta^- \eta^+ \eta^+ {]}}} & = & \int_{K_1} 
 \int_{ K_2 } \int_{ K_3} \int_{K_3} \delta ( K_1 + K_2 + K_3 + K_4) 
 \Gamma_{\Lambda}^{\eta^- \eta^- \eta^+ \eta^+} ( K_1 , K_2 , K_3 , K_4)
\eta^-_{K_1} \eta^-_{K_2} \eta^+_{K_3} \eta^+_{K_4},
 \end{eqnarray}
where
$\delta (K ) = \beta N \delta_{\bd{k }, 0 } \delta_{\omega , 0 }$.
The other terms are defined similarly with
the convention that  in all expressions  involving classical fluctuations we should set $\bd{m}_{ K} = \beta \delta_{\omega , 0} \bd{m}_{\bd{k}}$, so that
the frequencies associated with the classical magnetization field $\bd{m}_K$ vanish.
Note that 
there are no vertices of the type $[ mm \eta ]$ and $ [mmm \eta]$ because the $m$-field
does not transfer any frequency.
%
The initial value of the vertices in Eq.~(\ref{eq:vertexhybrid})
in  a cutoff scheme where
the exchange interaction is initially switched off  are 
rather complicated and will be discussed in Sec.~\ref{sec:freespin}.

\subsection{Exact flow equations for the two-point vertices}

Substituting the vertex expansions (\ref{eq:Gammahcomplete}) and (\ref{eq:vertexhybrid})
into the generalized Wetterich equation (\ref{eq:Wetterichhybrid}) we obtain FRG flow equations
for the vertices. In particular, the classical self-energy $\Sigma_{\Lambda} ( \bd{k} )$ appearing
in the quadratic part of Eq.~(\ref{eq:Gammahcomplete}) satisfies
the exact flow equation
 \begin{eqnarray}
 \partial_{\Lambda} {\Sigma}_{\Lambda} ( \bd{k}  ) & = & 
 T \int_{  \bd{q}}
 \dot{G}_{\Lambda} ( \bd{q} ) \left[ 
 \Gamma_{\Lambda}^{--++} ( - \bd{k} , - \bd{q}, \bd{q} ,  \bd{k}) 
 + \frac{1}{2!} \Gamma_{\Lambda}^{-+zz} ( - \bd{k} ,\bd{k}, -  \bd{q} ,  \bd{q}) \right] 
 \nonumber
 \\
 & +& T \sum_{ \omega^{\prime} \neq 0} \int_{ \bd{q} }
 \dot{G}^{\eta}_{\Lambda} ( Q ) \left[ 
 {\color{blue}{ \Gamma_{\Lambda}^{m^-  m^+  \eta^- \eta^+} (  - \bd{k},  \bd{k} , -Q, Q)}} 
 +    {\color{ProcessBlue}{ \frac{1}{2!} \Gamma_{\Lambda}^{m^- m^+ \eta^z \eta^z} 
 ( -  \bd{k} , \bd{k}, - Q , Q) }} \right] 
 \nonumber
 \\
& - & T \sum_{ \omega^{\prime} \neq 0} \int_{ \bd{q}} \left[
   {G}^{\eta}_{\Lambda} ( Q )    {G}^{\eta}_{\Lambda} ( Q + \bd{k})      \right]^{\bullet} 
{ \color{ForestGreen}{ \Gamma^{m^- \eta^+ \eta^z}_{\Lambda} ( -  \bd{k} , -Q , Q + \bd{k}   )
                            \Gamma^{m^+ \eta^- \eta^z}_{\Lambda} ( \bd{k}, Q , - Q - \bd{k} ) }},
 \label{eq:Sigmaexactflow}
\end{eqnarray}
while the  interaction-irreducible subtracted dynamic susceptibility 
$\tilde{\Pi}_{\Lambda} ( K ) $ 
in the quadratic part of Eq.~(\ref{eq:vertexhybrid})
satisfies  
 \begin{eqnarray}
 -  \partial_{\Lambda} \tilde{\Pi}_{\Lambda} ( K ) & = & T \sum_{\omega^{\prime} \neq 0} \int_{ \bd{q}}
 \dot{G}^{\eta}_{\Lambda} ( Q ) \left[ 
 {\color{red}{
 \Gamma_{\Lambda}^{\eta^- \eta^- \eta^+ \eta^+} ( - K , -Q, Q , K) }}
 {\color{Orange}{+ \frac{1}{2!} \Gamma_{\Lambda}^{\eta^- \eta^+ \eta^z \eta^z} ( - K ,K, - Q , Q)  }}
 \right] 
 \nonumber
 \\
 & +& T \int_{  \bd{q}}
 \dot{G}_{\Lambda} ( \bd{q} ) \left[ 
 {\color{blue}{
 \Gamma_{\Lambda}^{m^-  m^+  \eta^- \eta^+} ( - \bd{q} , \bd{q} , - K ,  K) }}
 + {\color{magenta}{\frac{1}{2!} \Gamma_{\Lambda}^{m^z m^z \eta^- \eta^+} ( - \bd{q} , \bd{q} , - K ,K) 
 }} \right] 
 \nonumber
 \\
 & - & T \int_{  \bd{q}} 
 \Bigl\{
\left[
 {G}_{\Lambda} ( \bd{q} )    {G}^{\eta}_{\Lambda} (  \bd{q}  +K)      \right]^{\bullet} 
 {\color{ForestGreen}{
 \Gamma^{m^z \eta^- \eta^+ }_{\Lambda} ( - \bd{q}, - K , \bd{q} + K   )
                            \Gamma^{m^z \eta^- \eta^+ }_{\Lambda} ( \bd{q} , - \bd{q}  -K, K  ) }}
 + ( K \rightarrow - K ) \Bigr\}
 \nonumber
 \\
& - & T \sum_{\omega^{\prime} \neq 0} \int_{ \bd{q}} \left[
   {G}^{\eta}_{\Lambda} ( Q )    {G}^{\eta}_{\Lambda} (  Q +K)      \right]^{\bullet} 
  {\color{LimeGreen}{\Gamma^{\eta^- \eta^+ \eta^z}_{\Lambda} ( - K , K+Q , -Q )
                            \Gamma^{\eta^- \eta^+ \eta^z}_{\Lambda} ( - K -Q, K , Q ) }}
 \nonumber
 \\
 & - & \tilde{\Pi}^2_{\Lambda} ( K ) \partial_{\Lambda} \Sigma_{\Lambda} ( \bd{k} ).
 \label{eq:Piexactflow}
\end{eqnarray}
\end{widetext}
Here the external momentum-frequency label 
is denoted by $K = ( \bd{k} , i \omega )$, the loop 
momentum-frequency  is 
$Q = ( \bd{q} , i \omega^{\prime} )$, and 
the symbol  $ \bd{q} + K $ represents  $( \bd{q} + \bd{k} , i \omega )$ where
the frequency $\omega$ belongs to $K$.
The deformed classical propagator 
 \begin{equation}
 G_{\Lambda} ( \bd{k} )  = \frac{1}{ {J}_{\Lambda} ( \bd{k} )
 + {\Sigma}_{\Lambda} ( \bd{k} ) }
 \end{equation}
has already been defined in Eq.~(\ref{eq:Gclassreg}), and the corresponding single-scale propagator is
  \begin{eqnarray}
  \dot{G}_{\Lambda} ( \bd{k} ) & \equiv & -  G^{2}_{\Lambda} ( \bd{k} ) \partial_{\Lambda}  {J}_\Lambda ( \bd{k} )    
   = - G^2_{\Lambda} ( \bd{k} )  \partial_{\Lambda}  {R}_\Lambda ( \bd{k} ) .
 \hspace{9mm}
 \end{eqnarray}
The  quantum propagator and  its single-scale counterpart  are
 \begin{subequations}
 \begin{eqnarray}
 G^{\eta}_{\Lambda} ( K ) & = & - F_{\Lambda} ( K ) 
\equiv - \frac{ \tilde{J}_{\Lambda} (  \bd{k} ) }{ 1 + \tilde{J}_{\Lambda} (  \bd{k} )
 \tilde{\Pi}_{\Lambda} ( K ) },
 \label{eq:Fdef}
 \\
  \dot{G}^{\eta}_{\Lambda} ( K ) & = & - \dot{F}_{\Lambda} ( K )
 \equiv - \frac{  \partial_{\Lambda} J_\Lambda ( \bd{k} ) }{ [ 1 + 
 \tilde{J}_{\Lambda} ( \bd{k} ) \tilde{\Pi}_{\Lambda} ( K ) ]^2 }.
 \label{eq:etaprop}
 \hspace{7mm}
 \end{eqnarray}
 \end{subequations}
Graphical representations of the flow equations (\ref{eq:Sigmaexactflow}) and
(\ref{eq:Piexactflow}) are shown in Fig.~\ref{fig:exactflow}.
\begin{figure}[tb]
 \begin{center}
  \centering
\vspace{7mm}
 \includegraphics[width=0.47\textwidth]{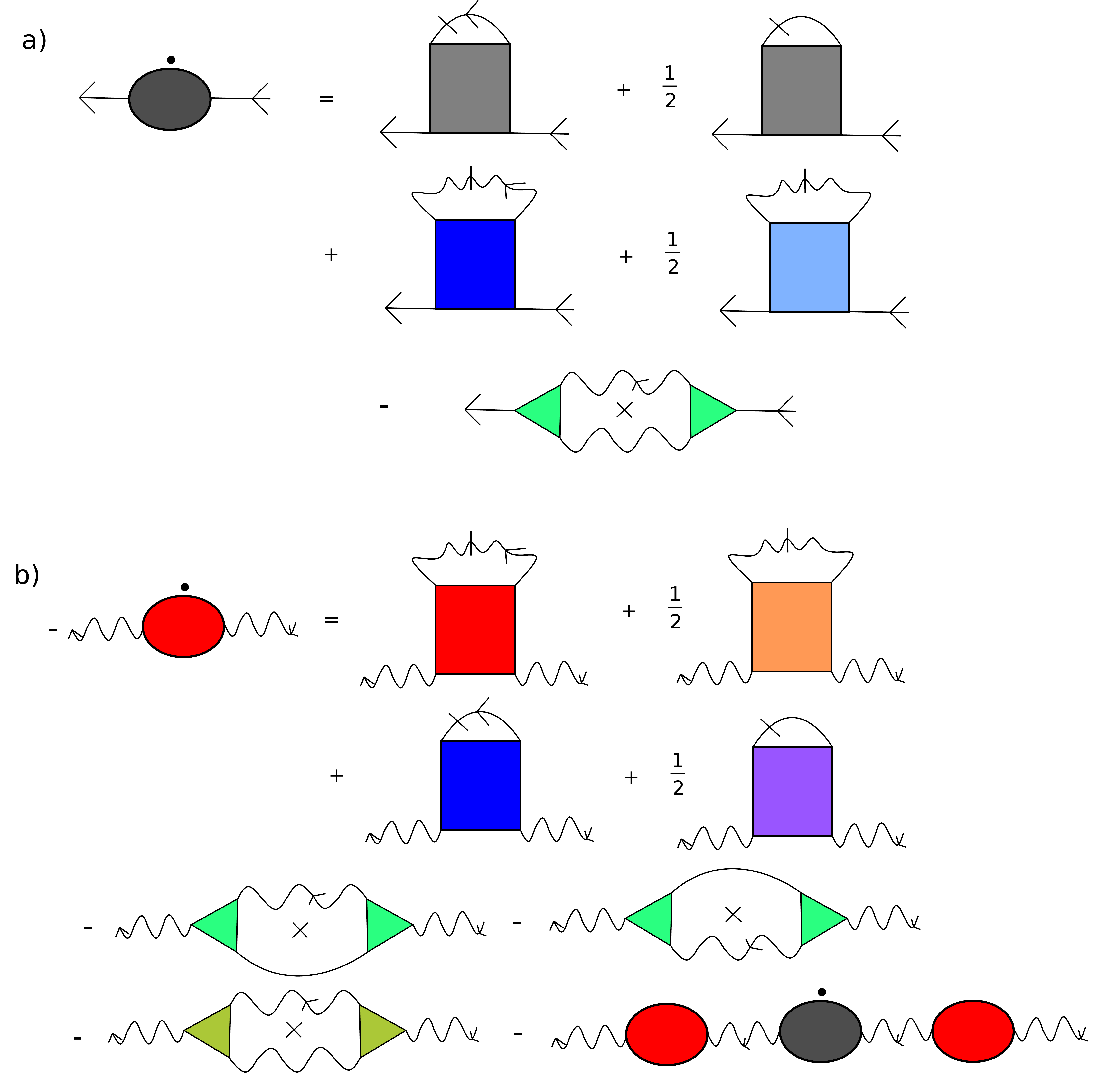}
   \end{center}
  \caption{%
The upper diagram (a) represents the exact flow equation~(\ref{eq:Sigmaexactflow}) 
for the classical self-energy $\Sigma_{\Lambda} ( \bd{k} )$,
while (b) represents the exact flow equation (\ref{eq:Piexactflow}) 
for  the subtracted irreducible dynamic susceptibility 
$\tilde{\Pi}_{\Lambda}  ( \bd{k} , i \omega )$.
We use the same color coding for the  vertices as in
the flow equations (\ref{eq:Sigmaexactflow}) and (\ref{eq:Piexactflow}).
Here the  dotted bubbles represent the scale derivatives
$\partial_{\Lambda} \Sigma_{\Lambda} ( \bd{k} )$ and
$\partial_{\Lambda} \tilde{\Pi}_{\Lambda}( \bd{k} , i \omega )$,
the solid arrows and lines represent the transverse and longitudinal
classical propagator $G_{\Lambda}^{-+} ( \bd{k} )$ and $ G^{zz}_{\Lambda} ( \bd{k} )$, 
while wavy arrows and lines represent the corresponding
quantum propagators $G^ {\eta, +-}_{\Lambda}  ( \bd{k} , i \omega )$ and $G^{\eta , zz}_{\Lambda} ( \bd{k} , i \omega )$.
Slashed lines represent the
 corresponding single-scale propagators,
and crosses inside loops mean that each of the propagators forming the loop 
should successively be replaced by the relevant single-scale propagator.
Note that spin-rotational invariance implies 
$G_{\Lambda}^{+-} ( \bd{k} ) = G^{zz}_{\Lambda} ( \bd{k} )
 = G_{\Lambda}( \bd{k} )$ and
  $G^ {\eta, +-} _{\Lambda} ( \bd{k} , i \omega ) = 
G^{\eta , zz}_{\Lambda} ( \bd{k} , i \omega ) = G^{\eta}_{\Lambda} ( \bd{k} , i \omega )$.
The above diagrams describe also the FRG flow in the presence of an external magnetic field
where transverse and longitudinal correlation functions should be distinguished.
}
\label{fig:exactflow}
\end{figure}

\section{Truncated flow equations
and integral equation for the irreducible dynamic susceptibility}
\label{sec:trunc}

We now specify our cutoff scheme. For our purpose, it is sufficient
to work with an interaction cutoff \cite{Krieg19} 
where the exchange interaction is initially switched off
at $\Lambda =0$ and  assumes the  physical value $J ( \bd{k} )$
at the final value $\Lambda =1$
of the deformation parameter.
Formally, this
scheme can be implemented via the regulator
 \begin{equation}
 R_{\Lambda} ( \bd{k} ) = ( \Lambda -1 ) {J} ( \bd{k} ),
 \; \; \; \Lambda \in [ 0, 1 ],
 \end{equation}
so that the deformed exchange interaction is
 \begin{equation}
 {J}_{\Lambda} ( \bd{k} ) = \Lambda {J} ( \bd{k} ) , \; \; \; \Lambda \in [ 0,1].
 \end{equation}

\subsection{Truncation with bare interaction vertices}
 \label{sec:freespin}

In the simplest truncation, we neglect the flow of  the three-point
and four-point vertices in Eqs.~(\ref{eq:Sigmaexactflow}) and
(\ref{eq:Piexactflow}). This amounts to neglecting the effect of the exchange
interaction on the higher-order spin correlations, which  
are then determined by the on-site SU(2)-algebra of a single non-interacting
spin. Although this truncation is too simple to give physically 
correct  results for the low-energy spin dynamics, it is instructive
to work out the explicit form the three-point and four-point  vertices
because it gives us a hint for more accurate truncations.

Using the initial values of the
classical four-point vertices given in 
Eq.~(\ref{eq:initialGammaclassical})  
we obtain for the relevant combination in Eq.~(\ref{eq:Sigmaexactflow}) at $\Lambda =0$, 
\begin{eqnarray}
 & &  \Gamma_{0}^{--++} ( - \bd{k} , - \bd{q}, \bd{q} ,  \bd{k}) 
 + \frac{1}{2!} \Gamma_{0}^{-+zz} ( - \bd{k} ,\bd{k}, -  \bd{q} ,  \bd{q}) 
 \nonumber
 \\
 & & = \frac{T}{ ( b_0^{\prime} )^3} \left( b_0^{\prime} + \frac{1}{6} \right).
 \label{eq:Gamma4init}
 \end{eqnarray}

Next, consider the initial conditions for the three-point vertices for vanishing exchange 
couplings. In this limit the pure quantum vertices in the last line of
Eq.~(\ref{eq:Piexactflow}) vanish, while the mixed three-legged vertices with one classical leg
are related to the corresponding Fourier transform of the
imaginary-time ordered three-spin correlation function  
via the tree expansion~\cite{Kopietz10,Goll19}
 \begin{subequations}
 \label{eq:tree3vertices}
 \begin{eqnarray}
 & & \tilde{J}_{\Lambda} ( \bd{k}_2 ) \tilde{J}_{\Lambda} ( \bd{k}_3 )
 G^{+-z}_{\Lambda}  ( - \bd{k}_1 , - K_2 , - K_3 ) =
 \nonumber
 \\
 &  &
  - G_{\Lambda} ( \bd{k}_1 ) 
 G^{\eta}_{\Lambda} ( K_2 ) G^{\eta}_{\Lambda} ( K_3 ) 
 {\color{ForestGreen}{\Gamma_{\Lambda}^{ m^- \eta^+ \eta^z } ( 
 \bd{k}_1 , K_2 , K_3 )
   }} , 
  \nonumber
 \\
 & &
 \\
 & & \tilde{J}_{\Lambda} ( \bd{k}_2 ) \tilde{J}_{\Lambda} ( \bd{k}_3 )
 G^{+-z}_{\Lambda}  ( - K_2 , - \bd{k}_1 , - K_3 ) =
 \nonumber
 \\
 &  &  
- G_{\Lambda} ( \bd{k}_1 ) 
 G^{\eta}_{\Lambda} ( K_2 ) G^{\eta}_{\Lambda} ( K_3 ) 
 {\color{ForestGreen}{\Gamma_{\Lambda}^{ m^+ \eta^- \eta^z } ( 
 \bd{k}_1 , K_2 , K_3 )
   }} , 
 \nonumber
 \\
 & &
\\
 & & \tilde{J}_{\Lambda} ( \bd{k}_2 ) \tilde{J}_{\Lambda} ( \bd{k}_3 )
 G^{+-z}_{\Lambda}  ( - K_2 , - K_3 , - \bd{k}_1 ) =
 \nonumber
 \\
 &  &  
-  G_{\Lambda} ( \bd{k}_1 ) 
 G^{\eta}_{\Lambda} ( K_2 ) G^{\eta}_{\Lambda} ( K_3 ) 
 {\color{ForestGreen}{\Gamma_{\Lambda}^{ m^z \eta^- \eta^+ } ( 
 \bd{k}_1 , K_2 , K_3 )
   }} .
  \nonumber
 \\
 & &
\end{eqnarray}
\end{subequations}
In the limit $\Lambda \rightarrow 0$ where $G^{\eta}_{\Lambda} ( K ) 
\rightarrow 0$ and $G_{\Lambda} ( \bd{k} ) \rightarrow \beta 
b_0^\prime $ these equations imply the initial conditions
 \begin{subequations}
 \begin{eqnarray}
 {\color{ForestGreen}{\Gamma_{0}^{ m^- \eta^+ \eta^z } ( 
     \bd{k}_1 , K_2 , K_3 ) }} & = & - \frac{1}{\beta b_0^{\prime}} G^{+-z}_0 ( 0 , - \omega_2 , \omega_2 ), \hspace{9mm}
 \\
 {\color{ForestGreen}{\Gamma_{0}^{ m^+ \eta^- \eta^z } ( 
     \bd{k}_1 , K_2 , K_3 ) }} & = & - \frac{1}{\beta b_0^{\prime}} G^{+-z}_0 ( - \omega_2 , 0,
 \omega_2 ),
 \\
 {\color{ForestGreen}{\Gamma_{0}^{ m^z \eta^- \eta^+ } ( 
     \bd{k}_1 , K_2 , K_3 ) }} & = & - \frac{1}{\beta b_0^{\prime}} G^{+-z}_0 ( - \omega_2 ,
 \omega_2, 0 ).
 \end{eqnarray}
\end{subequations}
Explicit expressions for the imaginary-time ordered connected spin correlation functions
in frequency space have first been derived by VLP \cite{Vaks68,Vaks68b}, see also
Refs.~[\onlinecite{Izyumov88,Tarasevych18,Goll19}].
In the zero-field limit of the mixed three-spin correlation  function is \cite{Tarasevych18}
 \begin{eqnarray}
  G_0^{+-z} ( \omega_1 , \omega_2 , \omega_3 )  & = & 
\beta  b_0^{\prime} ( 1 - \delta_{\omega_1,0} \delta_{\omega_2,0} \delta_{\omega_3,0} )
 \nonumber
 \\
 & \times  & 
 \left[ \frac{ \delta_{ \omega_1,0} }{ i \omega_2} 
 + \frac{ \delta_{\omega_2,0} }{ i \omega_3} 
 + \frac{ \delta_{  \omega_3,0} }{ i \omega_1} 
 \right]. \hspace{7mm}
 \end{eqnarray}
We conclude that
 \begin{eqnarray}
& &  {\color{ForestGreen}{\Gamma_{0}^{ m^- \eta^+ \eta^z } ( 
     \bd{k}_1 , K_2 , K_3 ) }} = -  {\color{ForestGreen}{\Gamma_{0}^{ m^+ \eta^- \eta^z } ( 
     \bd{k}_1 , K_2 , K_3 ) }} 
 \nonumber
 \\
 & = &  {\color{ForestGreen}{\Gamma_{0}^{ m^z \eta^- \eta^+ } ( 
     \bd{k}_1 , K_2 , K_3 ) }} = \frac{1}{i \omega_2},
 \label{eq:Gammapmzzero}
 \end{eqnarray}
so that for the relevant momentum-frequency labels in the flow 
equation~(\ref{eq:Sigmaexactflow}) we obtain
 \begin{eqnarray}
 &  & { \color{ForestGreen}{ \Gamma^{m^- \eta^+ \eta^z}_{0} ( -  \bd{k} , -Q , Q + \bd{k}   ) }}
 \nonumber
 \\
 & = & 
{\color{ForestGreen}{ \Gamma^{m^+ \eta^- \eta^z}_{0} ( \bd{k}, Q , - Q - \bd{k} ) }}
=  \frac{1}{  - i \omega^{\prime}  },
 \end{eqnarray}
and in the flow equation (\ref{eq:Piexactflow}) for the irreducible dynamic 
susceptibility the relevant initial vertices are
 \begin{eqnarray}
 & &
{\color{ForestGreen}{
 \Gamma^{m^z \eta^- \eta^+ }_{0} ( - \bd{q}, - K , \bd{q} + K   )}} 
 \nonumber
 \\
 & =  & {\color{ForestGreen}{
\Gamma^{m^z \eta^- \eta^+ }_{0} ( \bd{q} , - \bd{q}  -K, K  ) }} =
 \frac{1}{ - i \omega }.
 \end{eqnarray}

Let us now consider the quantum four-point 
vertices in the first line of Eq.~(\ref{eq:Piexactflow}) which are irreducible with respect to cutting a single interaction line.
These vertices are related to the connected four-spin correlation function via the
tree expansions~\cite{Goll19},
 \begin{widetext}
 \begin{eqnarray} 
 & &
 \left[ \prod_{i=1}^4 \tilde{J}_{\Lambda} ( \bd{k}_i ) \right]
 G^{++--}_{\Lambda} ( - K_1 , - K_2 , - K_3 , - K_4 ) = -
 \left[ \prod_{i=1}^4 G^{\eta}_{\Lambda} ( K_i ) \right]
 \Biggl\{
{\color{red}{ \Gamma_{\Lambda}^{\eta^- \eta^- \eta^+ \eta^+} ( K_1, K_2 , K_3  , K_4) }}
 \nonumber
 \\ 
 & & 
-
 \left[ {\color{ForestGreen}{\Gamma_{\Lambda}^{m^z \eta^- \eta^+} (  - \bd{k}_1 
 - \bd{k}_3 ,    K_1 , K_3  ) }}
\delta_{ \omega_1 + \omega_3,0}
G_{\Lambda} ( - \bd{k}_1 - \bd{k}_3 )  
 {\color{ForestGreen}{\Gamma_{\Lambda}^{m^z \eta^- \eta^+} ( 
   -  \bd{k}_2 - \bd{k}_4 ,   K_2 , K_4 )}} + ( K_3 \leftrightarrow K_4 ) \right]
  \nonumber
 \\
 & &  -
 \left[ {\color{LimeGreen}{\Gamma_{\Lambda}^{\eta^- \eta^+ \eta^z} ( K_1 , K_3 , - K_1 - K_3 ) }}
G^{\eta}_{\Lambda} ( - K_1 - K_3 )
 {\color{LimeGreen}{\Gamma_{\Lambda}^{\eta^- \eta^+ \eta^z} ( K_2 , K_4 , - K_2 - K_4 )}} + ( K_3 \leftrightarrow K_4 )
 \right] \Biggr\},
 \label{eq:treeGppmm}
 \end{eqnarray}
and
 \begin{eqnarray}
 & &
\left[ \prod_{i=1}^4 \tilde{J}_{\Lambda} ( \bd{k}_i ) \right]
 G^{+-zz}_{\Lambda} ( - K_1 , - K_2 , - K_3 , - K_4 ) = -
 \left[ \prod_{i=1}^4 G^{\eta}_{\Lambda} ( K_i ) \right]
 \Biggl\{
{\color{Orange}{ \Gamma_{\Lambda}^{\eta^- \eta^+ \eta^z \eta^z} ( K_1, K_2 , K_3  , K_4) }}
 \nonumber
 \\
 & &
-
 \left[ {\color{ForestGreen}{\Gamma_{\Lambda}^{m^- \eta^+ \eta^z} (  - \bd{k}_1 
 - \bd{k}_3 ,    K_1 , K_3  ) }}
\delta_{ \omega_1 + \omega_3,0}
G_{\Lambda} ( - \bd{k}_1 - \bd{k}_3 )
 {\color{ForestGreen}{\Gamma_{\Lambda}^{m^+ \eta^- \eta^z} ( 
   -  \bd{k}_2 - \bd{k}_4 ,   K_2 , K_4 )}} + ( K_3 \leftrightarrow K_4 ) \right]
  \nonumber
 \\
 & &  -
 \left[ {\color{LimeGreen}{\Gamma_{\Lambda}^{\eta^- \eta^+ \eta^z} ( K_1 , - K_1 - K_3, K_3 ) }}
G^{\eta}_{\Lambda} ( - K_1 - K_3 )
 {\color{LimeGreen}{\Gamma_{\Lambda}^{\eta^- \eta^+ \eta^z} ( -K_2 - K_4 ,  K_2 , K_4 )}} + ( K_3 \leftrightarrow K_4 )
 \right] 
 \nonumber
 \\
 & &   - \Gamma_{\Lambda}^{\eta^z \eta^z \eta^z} ( K_3 , K_4, - K_3 - K_4, )
G^{\eta}_{\Lambda} ( - K_3 - K_4 )
 {\color{LimeGreen}{\Gamma_{\Lambda}^{\eta^- \eta^+ \eta^z} ( K_1 , K_2 ,  -K_1 - K_2 )}}
\Biggr\}.
 \label{eq:treeGpmzz}
 \end{eqnarray}
Taking the limit  $\Lambda \rightarrow$ in 
Eqs.~(\ref{eq:treeGppmm}) and (\ref{eq:treeGpmzz}) 
we obtain the initial conditions
 \begin{eqnarray}
& &  {\color{red}{ \Gamma_{0}^{\eta^- \eta^- \eta^+ \eta^+} ( K_1, K_2 , K_3  , K_4) }}
 =  - G_0^{++--} ( - \omega_1 , - \omega_2 , - \omega_3 , - \omega_4 )
 \nonumber
 \\
 & &
+ 
  \left[ {\color{ForestGreen}{ \Gamma_{0}^{m^z \eta^- \eta^+} (  - \bd{k}_1 
 - \bd{k}_3 ,    K_1 , K_3  ) }}
\delta_{ \omega_1 + \omega_3,0} \beta b_0^{\prime}
 {\color{ForestGreen}{\Gamma_{0}^{m^z \eta^- \eta^+} ( 
   -  \bd{k}_2 - \bd{k}_4 ,   K_2 , K_4 )}} + ( K_3 \leftrightarrow K_4 ) \right]
 \nonumber
 \\
 & = & 
 - G_0^{++--} ( - \omega_1 , - \omega_2 , - \omega_3 , - \omega_4 )
 + \frac{1}{\beta b_0^{\prime}} \left( \delta_{ \omega_1 + \omega_3,0}
 + \delta_{ \omega_1 + \omega_4,0}  \right) 
 G_0^{+-z} ( - \omega_1 , \omega_1,0 )  G_0^{+-z} ( - \omega_2 , \omega_2 , 0 )
 \nonumber
 \\
 & = &  - G_0^{++--} ( - \omega_1 , - \omega_2 , - \omega_3 , - \omega_4 )
 - \frac{ \beta b_0^{\prime}}{\omega_1 \omega_2} 
 \left( \delta_{ \omega_1 + \omega_3,0}
 + \delta_{ \omega_1 + \omega_4,0}  \right)  ,
 \end{eqnarray}
and
 \begin{eqnarray}
 & & {\color{Orange}{ \Gamma_{0}^{\eta^- \eta^+ \eta^z \eta^z} ( K_1, K_2 , K_3  , K_4) }} =  - G_0^{+-zz} ( - \omega_1 , - \omega_2 , - \omega_3 , - \omega_4 )
 \nonumber
 \\
 & & + 
 \left[ {\color{ForestGreen}{\Gamma_{0}^{m^- \eta^+ \eta^z} (  - \bd{k}_1 
 - \bd{k}_3 ,    K_1 , K_3  ) }}
\delta_{ \omega_1 + \omega_3,0} \beta b_0^{\prime}
 {\color{ForestGreen}{\Gamma_{0}^{m^+ \eta^- \eta^z} ( 
   -  \bd{k}_2 - \bd{k}_4 ,   K_2 , K_4 )}} + ( K_3 \leftrightarrow K_4 ) \right]
 \nonumber
 \\
 &  = &  - G_0^{+-zz} ( - \omega_1 , - \omega_2 , - \omega_3 , - \omega_4 )
 + \frac{1}{\beta b_0^{\prime}} \left( \delta_{ \omega_1 + \omega_3,0}
 + \delta_{ \omega_1 + \omega_4,0}  \right) 
 G_0^{+-z} ( 0, - \omega_1 , \omega_1 )  G_0^{+-z} ( - \omega_2 , 0, \omega_2 )
 \nonumber
 \\
 & = &  - G_0^{+-zz} ( - \omega_1 , - \omega_2 , - \omega_3 , - \omega_4 )
 + \frac{ \beta b_0^{\prime}}{\omega_1 \omega_2} 
 \left( \delta_{ \omega_1 + \omega_3,0}
 + \delta_{ \omega_1 + \omega_4,0}  \right)  .
 \end{eqnarray}
 \end{widetext}
For the frequency combinations needed in the first line of Eq.~(\ref{eq:Piexactflow}) 
we obtain
\begin{subequations}
  \begin{eqnarray}
 & &{\color{red}{ G^{++--}_{0} (  \omega ,  \omega^{\prime} , - \omega^{\prime}  , - \omega  )}} 
 \nonumber
 \\
 & = &
  \frac{2}{3} \beta^3 b_0^{\prime \prime \prime} \; \; \; \mbox{if $\omega = \omega^{\prime} =0$},
 \\
 & = &    \frac{\beta b_0^{\prime}}{( \omega^{\prime})^2 }   \; \; \; \;  \; \mbox{if $\omega = 0$ and $
\omega^{\prime}  \neq 0$},
 \\
& = &    \frac{\beta b_0^{\prime}}{ \omega^2 }   \hspace{7mm} \mbox{if $\omega \neq 0$ and $
\omega^{\prime}  = 0$},
 \\
& =  &  - 2 \frac{\beta b_0^{\prime}}{ \omega^2 }   \hspace{3mm} \mbox{if $\omega = \omega^{\prime}
\neq 0$},
\\
& =  &    \frac{\beta b_0^{\prime}}{ \omega^2 }   \hspace{7mm} \mbox{if $\omega = - \omega^{\prime}
\neq 0$},
\\
 & = & 0  \hspace{12mm} \mbox{else},
 \end{eqnarray}
 \end{subequations}
and
\begin{subequations}
  \begin{eqnarray}
 & &  {\color{Orange}{{G}^{+-zz}_{0} (  \omega , - \omega,  \omega^{\prime} , - \omega^{\prime}    )}} 
 \nonumber
 \\
 & = & 
  \frac{1}{3} \beta^3 b_0^{\prime \prime \prime} \; \; \; \; \; \mbox{if $\omega = \omega^{\prime} =0$},
 \\
 & = &  2  \frac{\beta b_0^{\prime}}{( \omega^{\prime})^2 }   \; \; \; \; \;  \mbox{if $\omega = 0$ and $
\omega^{\prime}  \neq 0$},
 \\
 & = & 
 2  \frac{\beta b_0^{\prime}}{ \omega^2 }   \; \; \; \; \; \;  \; \mbox{if $\omega \neq 0$ and $
\omega^{\prime}  = 0$},
 \\
 & = &  - \frac{ \beta b_0^{\prime}}{ \omega^2} \hspace{6mm} \mbox{if $| \omega |  =
 | \omega^{\prime} |   \neq  0$},
 \\
 & = & 0 \hspace{13mm} \mbox{else}.
 \end{eqnarray}
 \end{subequations}
Keeping in mind that both Matsubara frequencies in the first 
line of Eq.~(\ref{eq:Piexactflow}) are non-zero, the initial condition of the
relevant linear combination is
 \begin{eqnarray}
 & & {\color{red}{
 \Gamma_{0}^{\eta^- \eta^- \eta^+ \eta^+} ( - K , -Q, Q , K) }}
 {\color{Orange}{+ \frac{1}{2!} \Gamma_{0}^{\eta^- \eta^+ \eta^z \eta^z} ( - K ,K, - Q , Q)  }}
 \nonumber
 \\
 &  &  =  \frac{ \beta b_0^{\prime}}{\omega^2}
 \left[   \frac{5}{2}  \delta_{ \omega , \omega^{\prime} }  
  - \frac{1}{2!}  \delta_{ \omega , - \omega^{\prime}  } \right]
 \nonumber
 \\
 & & \hspace{3mm}
 - {\color{ForestGreen}{ \frac{\beta b_0^{\prime}}{ \omega \omega^{\prime}}
 ( \delta_{\omega , \omega^{\prime}} +1 ) 
-  \frac{\beta b_0^{\prime}}{  \omega^2} \frac{1}{2!}
 ( \delta_{\omega , \omega^{\prime}} + \delta_{ \omega , - \omega^{\prime}} ) 
}}
 \nonumber
 \\
 &  & = \frac{\beta b_0^{\prime}}{ \omega^2} ( \delta_{\omega , \omega^{\prime}}
 - \delta_{\omega , - \omega^{\prime}} ) - \frac{\beta b_0^{\prime}}{\omega 
 \omega^{\prime}}
 \nonumber
 \\
 & & = \frac{\beta b_0^{\prime}}{ \omega \omega^{\prime}} 
 \left(  \delta_{\omega , \omega^{\prime}} +  \delta_{\omega , - \omega^{\prime}} -1 \right).
 \label{eq:quantumvertexinit}
\end{eqnarray}
Noting that this is an odd function of $\omega^{\prime}$
while in the paramagnetic phase the single-scale propagator
$\dot{G}_{\Lambda}^{\eta} ( \bd{q} , i \omega^{\prime} )$
in the first line of the flow equation (\ref{eq:Piexactflow}) as an even function
of $\omega^{\prime}$, we conclude that the contribution from the quantum  
vertices in the
first line of  Eq.~(\ref{eq:Piexactflow}) vanishes if we approximate the vertices
by their initial values.

Next, consider the four-point vertices with two classical and two quantum fields
 in Eqs.~(\ref{eq:Sigmaexactflow}) and
(\ref{eq:Piexactflow}).
The relevant tree expansions are
 \begin{widetext}
 \begin{subequations}
 \label{eq:tree123}
 \begin{eqnarray}
 & & \tilde{J}_\Lambda ( \bd{k}_3 ) \tilde{J}_{\Lambda} ( \bd{k}_4 ) G^{++--}_{\Lambda}
 ( - \bd{k}_1 , -K_3 , - \bd{k}_2 , - K_4 )  =  - G_{\Lambda} ( \bd{k}_1 )
 G_{\Lambda} ( \bd{k}_2 )
 G^{\eta}_{\Lambda} ( K_3 )  G^{\eta}_{\Lambda} ( K_4 )
 \Biggl\{
  {\color{blue}{ \Gamma_{\Lambda}^{m^-  m^+  \eta^- \eta^+} (  \bd{k}_1, \bd{k}_2 , 
 K_3 ,  K_4)}} 
 \nonumber
 \\
& & \hspace{20mm}
-  {\color{ForestGreen}{\Gamma_{\Lambda}^{m^- \eta^+ \eta^z} 
(   \bd{k}_1  ,    K_4 , - \bd{k}_1 - K_4  ) }}
G^{\eta}_{\Lambda} ( - \bd{k}_1 - K_4 )
 {\color{ForestGreen}{\Gamma_{\Lambda}^{m^+ \eta^- \eta^z} ( 
  \bd{k}_2 ,   K_3 , -  {\bd{k}}_2 - K_3)}} 
   \Biggr\},
 \label{eq:tree1}
 \end{eqnarray}
 \begin{eqnarray} 
 & & \tilde{J}_\Lambda ( \bd{k}_3 ) \tilde{J}_{\Lambda} ( \bd{k}_4 ) G^{+-zz}_{\Lambda}
 ( - \bd{k}_1 , - \bd{k}_2 , -K_3, - K_4 )  =  - G_{\Lambda} ( \bd{k}_1 )
 G_{\Lambda} ( \bd{k}_2 )
 G^{\eta}_{\Lambda} ( K_3 )  G^{\eta}_{\Lambda} ( K_4 )
 \Biggl\{
  {\color{ProcessBlue}{ \Gamma_{\Lambda}^{m^-  m^+  \eta^z \eta^z} (  \bd{k}_1, \bd{k}_2 , 
 K_3 ,  K_4)}} 
 \nonumber
 \\
& & \hspace{20mm}
-  \left[ {\color{ForestGreen}{\Gamma_{\Lambda}^{m^- \eta^+ \eta^z} 
(   \bd{k}_1  ,   - {\bd{k}}_1 - K_3 , K_3 ) }}
G^{\eta}_{\Lambda} ( -  {\bd{k}}_1 - K_3 )
 {\color{ForestGreen}{\Gamma_{\Lambda}^{m^+ \eta^- \eta^z} ( 
  \bd{k}_2 ,   -  {\bd {k}}_2 - K_4 , K_4)}} + ( K_3 \leftrightarrow K_4 ) \right]
   \Biggr\}, \hspace{7mm}
 \label{eq:tree2}
 \end{eqnarray}
 \begin{eqnarray}
 & & \tilde{J}_\Lambda ( \bd{k}_3 ) \tilde{J}_{\Lambda} ( \bd{k}_4 ) G^{+-zz}_{\Lambda}
 ( -K_3, -K_4 , - \bd{k}_1,   - \bd{k}_2  )  =  - G_{\Lambda} ( \bd{k}_1 )
 G_{\Lambda} ( \bd{k}_2 )
 G^{\eta}_{\Lambda} ( K_3 )  G^{\eta}_{\Lambda} ( K_4 )
 \Biggl\{
  {\color{magenta}{ \Gamma_{\Lambda}^{m^z  m^z  \eta^- \eta^+} (  \bd{k}_1, \bd{k}_2 , 
 K_3 ,  K_4)}} 
 \nonumber
 \\
& & \hspace{20mm}
-  \left[ {\color{ForestGreen}{\Gamma_{\Lambda}^{m^z \eta^- \eta^+} 
(   \bd{k}_1  ,    K_3 , - \bd{k}_1 - K_3  ) }}
G^{\eta}_{\Lambda} ( - \bd{k}_1 - K_3 )
 {\color{ForestGreen}{\Gamma_{\Lambda}^{m^z \eta^- \eta^+} ( 
  \bd{k}_2  , -  {\bd{k}}_2 - K_4, K_4)}}  + ( \bd{k}_1 \leftrightarrow \bd{k}_2 ) \right]
   \Biggr\}.
 \label{eq:tree3}
 \end{eqnarray}
 \end{subequations}
Using the initial conditions (\ref{eq:Gammapmzzero}) for the three-point vertices and the fact that for vanishing exchange coupling
$G^{\eta}_0 ( K ) = - \tilde{J}_0 ( \bd{k} ) = - 1 / ( \beta b_0^{\prime})$
we obtain from Eq.~(\ref{eq:tree123})
for  $\Lambda \rightarrow  0$,
 \begin{subequations}
 \begin{eqnarray}
   & & {\color{blue}{ \Gamma_{0}^{m^-  m^+  \eta^- \eta^+} (  \bd{k}_1, \bd{k}_2 , 
 K_3 ,  K_4)}}   
 \nonumber
 \\
 &  & 
=   - \frac{1}{( \beta b_0^{\prime} )^2} G_0^{++--} ( 0, - \omega_3 , 0 ,  \omega_3 ) 
- \frac{1}{ \beta b_0^{\prime} } 
 {\color{ForestGreen}{\Gamma_{0}^{m^- \eta^+ \eta^z} 
(   \bd{k}_1  ,    K_4 , - \bd{k}_1 - K_4  ) }}
 {\color{ForestGreen}{\Gamma_{0}^{m^+ \eta^- \eta^z} ( 
  \bd{k}_2 ,   K_3 , -  {\bd{k}}_2 - K_3)}} 
 \nonumber
 \\
 & & 
 = - \frac{1}{\beta b_0^{\prime} \omega_3^2} +  \frac{1}{\beta b_0^{\prime} \omega_3^2} =0,
 \\
 & &  {\color{ProcessBlue}{ \Gamma_{0}^{m^-  m^+  \eta^z \eta^z} (  \bd{k}_1, \bd{k}_2 , 
 K_3 ,  K_4)}} 
 \nonumber
 \\
 & & =  - \frac{1}{( \beta b_0^{\prime} )^2} G_0^{+-zz} ( 0, 0  , -\omega_3 ,   \omega_3 )
 -  \frac{1}{\beta b_0^{\prime} }
\left[ {\color{ForestGreen}{\Gamma_{0}^{m^- \eta^+ \eta^z} 
(   \bd{k}_1  ,   - {\bd{k}}_1 - K_3 , K_3 ) }}
 {\color{ForestGreen}{\Gamma_{0}^{m^+ \eta^- \eta^z} ( 
  \bd{k}_2 ,   -  {\bd {k}}_2 - K_4 , K_4)}} + ( K_3 \leftrightarrow K_4 ) \right] 
 \nonumber 
\\
& &  = - \frac{2}{\beta b_0^{\prime} \omega_3^2} +  \frac{2}{\beta b_0^{\prime} \omega_3^2} =0,
 \\
 & & {\color{magenta}{ \Gamma_{0}^{m^z  m^z  \eta^- \eta^+} (  \bd{k}_1, \bd{k}_2 , 
 K_3 ,  K_4)}} 
\nonumber
 \\
 & & =  - \frac{1}{( \beta b_0^{\prime} )^2} G_0^{+-zz} (  -\omega_3 ,   \omega_3 ,0,0)
  - \frac{1}{\beta b_0^{\prime} }
 \left[ {\color{ForestGreen}{\Gamma_{0}^{m^z \eta^- \eta^+} 
(   \bd{k}_1  ,    K_3 , - \bd{k}_1 - K_3  ) }}
 {\color{ForestGreen}{\Gamma_{0}^{m^z \eta^- \eta^+} ( 
  \bd{k}_2  , -  {\bd{k}}_2 - K_4, K_4)}}  + ( \bd{k}_1 \leftrightarrow \bd{k}_2 ) \right]
 \nonumber 
\\
& &  = - \frac{2}{\beta b_0^{\prime} \omega_3^2} +  \frac{2}{\beta b_0^{\prime} \omega_3^2} =0.
 \end{eqnarray}
 \label{eq: Gamma4mixedinitial}
\end{subequations}
 \end{widetext}

In summary, if we approximate the three-point and four-point vertices in the
exact flow equations (\ref{eq:Sigmaexactflow}) and (\ref{eq:Piexactflow}) by
their initial values for vanishing exchange couplings, we obtain the following
truncated system of flow equations,
 \begin{widetext}
  \begin{eqnarray}
 \partial_{\Lambda} {\Sigma}_{\Lambda} ( \bd{k}  ) & = & \frac{T^2}{ ( b_0^{\prime} )^3}
 \left( b_0^{\prime} + \frac{1}{6} \right) \int_{\bd{q}} 
 \dot{G}_{\Lambda} ( \bd{q} ) +
 T \sum_{ \omega \neq 0} \frac{1}{\omega^2}
 \int_{ \bd{q}} 
\bigl[  {F}_{\Lambda } ( \bd{q} , i \omega )   {F}_{\Lambda } ( \bd{q} + \bd{k} , i \omega ) \bigr]^{\bullet},
 \label{eq:Sigmabareflow}
 \\
 \partial_{\Lambda} \tilde{\Pi}_{\Lambda} ( \bd{k} , i \omega )
  & = & \frac{T}{\omega^2} \int_{\bd{q}} 
 \Bigl\{  
  \bigl[ F_{\Lambda} ( \bd{q} , i \omega )  G_{\Lambda} ( \bd{q} + \bd{k} ) \bigr]^{\bullet}
+ ( \bd{k} \rightarrow - \bd{k} )
  \Bigr\}   +  \tilde{\Pi}^2_{\Lambda} ( \bd{k} , i \omega  ) \partial_{\Lambda} \Sigma_{\Lambda} ( \bd{k} ),
 \label{eq:Pibareflow}
 \end{eqnarray}
\end{widetext}
where the effective dynamical interaction
$F ( \bd{q} , i \omega ) = - G^\eta ( \bd{q} , i \omega )$ 
is the negative of the $\eta$-propagator defined in  Eq.~(\ref{eq:Fdef}), and we have introduced the notation
 \begin{eqnarray}
 & & \bigl[ F_{\Lambda} ( \bd{q} , i \omega )  G_{\Lambda} ( \bd{q} + \bd{k} ) \bigr]^{\bullet}    
 \nonumber
 \\
 & = & 
 \dot{F}_{\Lambda} ( \bd{q} , i \omega )  G_{\Lambda} ( \bd{q} + \bd{k} ) +
  {F}_{\Lambda} ( \bd{q} , i \omega )  \dot{G}_{\Lambda} ( \bd{q} + \bd{k} ).
 \hspace{7mm}
 \end{eqnarray}
Unfortunately, the truncated flow equation (\ref{eq:Pibareflow}) violates the Ward identity
$\tilde{\Pi}_{\Lambda} ( \bd{k} =0, i \omega \neq 0 ) =0$ due to the conservation of the total spin,
see Eq.~(\ref{eq:WardPi}) below.
Moreover, the last term $\tilde{\Pi}^2_{\Lambda} ( \bd{k} , i \omega  ) \partial_{\Lambda} \Sigma_{\Lambda} ( \bd{k} )$ leads to a violation of the
continuity condition (\ref{eq:continuity}), which is obvious by writing the
corresponding contribution to the flow equation as
$ \partial_{\Lambda} \tilde{\Pi}^{-1}_{\Lambda} (  \bd{k} , i \omega ) = - \partial_{\Lambda}
 \Sigma_{\Lambda} ( \bd{k} ) + \ldots $.
Given the fact that the constraints imposed by Ward identities are expected to be
essential for a correct description of the spin dynamics,
we conclude that the  truncation in this subsection with bare three-point and four-point vertices is 
not sufficient to obtain reliable results for the spin dynamics.

\subsection{Vertex corrections}
 \label{sec:vertex}

In principle, we could  now write down flow equations for the three-point and four-point vertices in Eqs.~(\ref{eq:Sigmaexactflow}) and (\ref{eq:Piexactflow}) which depend in turn 
on various types of higher-order vertices. We thus obtain an infinite hierarchy
of the flow equations for the vertices generated 
by $\Gamma_{\Lambda} [ \bd{m}^c , \bd{h}^q ]$.
The construction of
sensible approximation  schemes for this infinite hierarchy  is one of the main 
technical challenges of our SFRG approach.
A powerful  strategy to construct truncation strategies
for FRG flow equations
is based on the use of Ward identities providing exact relations
between vertices of different order.
This strategy 
has been adopted previously in different contexts in 
Refs.~[\onlinecite{Schuetz05,Bartosch09,Kopietz10}] and we will use it again 
in this work to express the four-point vertices in the FRG flow equations
for the irreducible spin susceptibility in terms of two-point vertices.

\subsubsection{Equations of motion}

Ward identities for imaginary-time ordered spin correlation functions of different order
can be derived using the Heisenberg equations of motion of the spin operators and
the resulting equations of motion for the correlation functions, as described in 
Ref.~[\onlinecite{Goll19}]. After transforming the equations of motion to 
momentum-frequency space, we find that
the two-spin correlation function $G ( K ) = G ( \bd{k} , i \omega )$ 
is related to the mixed three-spin correlation function 
$G^{+-z} ( Q+K, -Q , - K )$
via the integral equation
 \begin{equation}
 i \omega G ( K ) =
 \int_Q [ J ( \bd{q} ) - J ( \bd{q} + \bd{k} ) ] G^{+-z} ( Q+K, -Q , - K ),
 \label{eq:eom3}
 \end{equation}
where we assume that the  spin-rotational invariance   
is not spontaneously broken.
Setting $\bd{k} =0$ for finite frequency $\omega \neq 0 $
we obtain
 \begin{equation}
 G (  \bd{k} =0 , i\omega  \neq 0 ) = \frac{ \tilde{\Pi} ( 0 , i \omega ) }{ 1 + \tilde{J} ( 0 )
 \tilde{\Pi} ( 0 , i \omega ) } =0,
 \end{equation}
and hence
 \begin{equation}
  \tilde{\Pi} ( \bd{k} =0, i \omega \neq 0 ) =0.
 \label{eq:WardPi}
 \end{equation}
Similarly, we can derive the following
equation of motion for the mixed three-spin correlation function,
 \begin{eqnarray}
 & & i \omega G^{+-z} ( Q + K  , - Q , -K )  =  G ( Q ) - G ( Q + K  )
 \nonumber
 \\
 &   &   +  [ J ( \bd{q} ) - J ( \bd{q} + \bd{k}  )  ] G ( Q ) G ( Q+ K ) \nonumber 
 \\
 &  &  - \int_{Q^{\prime}} [ J ( \bd{q}^{\prime} ) - J ( \bd{q}^{\prime}  + \bd{k} )  ] 
 \nonumber
 \\
 & & \hspace{5mm} \times 
 G^{++--} ( Q + K  , - Q^{\prime} -K  , Q^{\prime} , - Q ),
 \label{eq:eom4}
 \end{eqnarray}
which depends on the connected four-spin correlation function
$ G^{++--} ( Q + K  , - Q^{\prime} -K  , Q^{\prime} , - Q )$.

\subsubsection{Fixing four-point vertices via
Ward identity and continuity condition}

By approximating the three-point and four-point vertices in Eq.~(\ref{eq:Piexactflow})
by their non-interacting limits we have neglected the
contribution from the finite-frequency (quantum)  four-point vertex
 \begin{eqnarray}
  {\color{red}{\Gamma_{\Lambda}^{\eta \eta \eta \eta} ( -K , K , -Q , Q )}}
 & = &  {\color{red}{
 \Gamma_{\Lambda}^{\eta^- \eta^- \eta^+ \eta^+} ( - K , -Q, Q , K) }}
 \nonumber
 \\
 &  & \hspace{-15mm} +
 {\color{Orange}{ \frac{1}{2!} \Gamma_{\Lambda}^{\eta^- \eta^+ \eta^z \eta^z} ( - K ,K, - Q , Q)  }},
 \label{eq:Gamma4quantumapprox}
 \hspace{7mm}
 \end{eqnarray}
as well as the contribution from the mixed classical-quantum four-point vertex
 \begin{eqnarray}
 {\color{blue}{
  \Gamma_{\Lambda}^{mm  \eta \eta} ( - \bd{q} , \bd{q} , - K ,  K) }}
 & = & 
 {\color{blue}{
 \Gamma_{\Lambda}^{m^-  m^+  \eta^- \eta^+} ( - \bd{q} , \bd{q} , - K ,  K) }}
 \nonumber
 \\
 & & \hspace{-15mm}
 + {\color{magenta}{\frac{1}{2!} \Gamma_{\Lambda}^{m^z m^z \eta^- \eta^+} ( - \bd{q} , \bd{q} , - K ,K) .
 }}
  \label{eq:Gamma4approx}
 \end{eqnarray}
Although for $\Lambda =0$
these vertices do not contribute to the flow of the irreducible susceptibility,
for finite $\Lambda$ this is not true any more, which is the reason 
for the violation of the Ward identity (\ref{eq:WardPi}) within a truncation where all 
higher-order  vertices are approximated by their initial values.
To restore the Ward identity, we should therefore take the flow of at least one of 
the  above vertices  into account. 
For simplicity let  us still approximate the quantum four-point vertex
${\color{red}{\Gamma_{\Lambda}^{\eta \eta \eta \eta} ( -K , K , -Q , Q )}}$ 
by its initial value given in Eq.~(\ref{eq:quantumvertexinit}),
 \begin{eqnarray}
 {\color{red}{\Gamma_{\Lambda}^{\eta \eta \eta \eta} ( -K , K , -Q , Q )}}
 & \approx &  {\color{red}{\Gamma_{0}^{\eta \eta \eta \eta} ( -K , K , -Q , Q )}}
 \nonumber
 \\
 & = &  \frac{\beta b_0^{\prime}}{ \omega \omega^{\prime}} 
 \left(  \delta_{\omega , \omega^{\prime}} +  \delta_{\omega , - \omega^{\prime}} -1 \right),
 \hspace{10mm}
 \label{eq:quantumapprox}
  \end{eqnarray}
so that this vertex does not contribute to the flow of $\tilde{\Pi}_{\Lambda} ( K )$.
This leaves us with the
mixed classical-quantum vertex
${\color{blue}{\Gamma_{\Lambda}^{mm  \eta \eta} ( - \bd{q} , \bd{q} , - K ,  K)}}$
to  restore the  Ward identity (\ref{eq:WardPi}).
To simplify the algebra, let us also neglect the
dependence of this vertex
on the momentum $\bd{q}$ of the classical field,
\begin{equation}
 {\color{blue}{\Gamma_{\Lambda}^{mm  \eta \eta} ( - \bd{q} , \bd{q} , - K ,  K) 
 \approx \Gamma_{\Lambda}^{mm  \eta \eta} ( 0  , 0 , - K ,  K) .}}
  \label{eq:Gamma4approx2}
 \end{equation}
With these approximations, the exact flow equation (\ref{eq:Piexactflow})
for the subtracted irreducible susceptibility reduces to
 \begin{eqnarray}
 & & \partial_{\Lambda} \tilde{\Pi}_{\Lambda} ( \bd{k} , i \omega ) 
    =    
\tilde{\Pi}^2_{\Lambda} ( \bd{k} , i \omega  ) \partial_{\Lambda} \Sigma_{\Lambda} ( \bd{k} ) 
 \nonumber
 \\
 & & - T \int_{\bd{q}}
    \dot{G}_{\Lambda} ( \bd{q}  )    {\color{blue}{\Gamma^{mm \eta \eta }_{\Lambda}
 (  0  , 0 , - K , K  ) }}  
 \nonumber
 \\
 &  & +
  \frac{T}{\omega^2} \int_{\bd{q}}
 \Bigl\{ 
  \bigl[ F_{\Lambda} ( \bd{q} , i \omega )  G_{\Lambda} ( \bd{q} + \bd{k} ) \bigr]^{\bullet}
 + ( \bd{k} \rightarrow - \bd{k} ) 
  \Bigr\}, \hspace{7mm}
 \label{eq:PiWIflow}
 \end{eqnarray}
which replaces Eq.~(\ref{eq:Pibareflow}).
Instead of writing down an additional flow equation for
 ${\color{blue}{\Gamma^{mm \eta \eta }_{\Lambda}
 (  0  , 0 , - K , K  ) }}  $,
we now fix this vertex by demanding that the solution of the
flow equation (\ref{eq:PiWIflow}) satisfies the  Ward identity (\ref{eq:WardPi}) 
as well as the continuity condition (\ref{eq:continuity})
for all values of the deformation parameter $\Lambda$, i.e.,
 \begin{eqnarray}
 \tilde{\Pi}_{\Lambda} ( 0  , i \omega \neq 0 ) & = & 0, \label{eq:ward2}
 \\
 \tilde{\Pi}^{-1}_{\Lambda} ( \bd{k} \neq 0 , 0 ) & = & 0.
 \label{eq:continuity2}
 \end{eqnarray}  
The simplest way to satisfy these constraints 
which is compatibe with the initial conditions at $\Lambda =0$
is to choose the
scale-dependent mixed four-point vertex as follows,
 \begin{equation}
{\color{blue}{\Gamma^{mm \eta \eta }_{\Lambda}
 (  0 , 0 , - K , K  )}}  =  \frac{1}{\omega^2}
 \left[  W_{\Lambda} ( i \omega ) +    C_{\Lambda} ( \bd{k} , i \omega )   \right],
 \label{eq:continuityconstraint}
 \end{equation}
where the Ward identity (\ref{eq:ward2}) is enforced by the contribution
 \begin{equation}
 W_{\Lambda} (  i \omega ) \equiv
 \frac{  2  \int_{\bd{q}}  \bigl[ F_{\Lambda} ( \bd{q} , i \omega )  G_{\Lambda} ( \bd{q} ) \bigr]^{\bullet}   }{
  \int_{\bd{q}} \dot{G}_{\Lambda} ( \bd{q} ) },
 \label{eq:Wconstraint}
 \end{equation}
while the continuity condition (\ref{eq:continuity2}) is enforced by
 \begin{equation}
 C_{\Lambda} ( \bd{k} , i \omega ) \equiv
 \frac{ \omega^2 \tilde{\Pi}^2_{\Lambda} ( \bd{k} , i \omega  ) \partial_{\Lambda} \Sigma_{\Lambda} ( \bd{k} ) }{ 
  T \int_{\bd{q}} \dot{G}_{\Lambda} ( \bd{q}  ) }.
 \label{eq:Cconstraint}
 \end{equation}
Note that $C_{\Lambda} ( \bd{k} , i \omega )$
cancels the term $\tilde{\Pi}^2_{\Lambda} ( \bd{k} , i \omega  ) \partial_{\Lambda} \Sigma_{\Lambda} ( \bd{k} )$
on the right-hand side of Eq.~(\ref{eq:PiWIflow}) which would otherwise violate
the continuity condition (\ref{eq:continuity2}).
It is important to note that our choice (\ref{eq:continuityconstraint})
of the  mixed four-point vertex is consistent with
the initial condition~(\ref{eq: Gamma4mixedinitial})  
at $\Lambda =0$ where the deformed exchange coupling $J_{\Lambda =0} ( \bd{k} )$ and 
hence also the vertex ${\color{blue}{\Gamma^{mm \eta \eta }_{\Lambda=0}
 (  0 , 0 , - K , K  )}}$ vanish.
This follows from the fact that for small $J_{\Lambda}$
the expressions in the numerator  of Eqs.~(\ref{eq:Wconstraint}) and (\ref{eq:Cconstraint})
vanish as $J_{\Lambda}^2$ while
the integral $\int_{\bd{q}} \dot{G}_{\Lambda} ( \bd{q} )$
in the denominator
vanishes as $J_{\Lambda}$, implying $ W_{\Lambda =0} ( i \omega ) =
 C_{\Lambda=0} ( \bd{k} , i \omega )=0$.
Substituting Eqs.~(\ref{eq:continuityconstraint}),~(\ref{eq:Wconstraint}), 
and (\ref{eq:Cconstraint}) into
the flow equation~(\ref{eq:PiWIflow}) we obtain the following  flow equation for the
dynamic irreducible susceptibility,
  \begin{eqnarray}
 \partial_{\Lambda} \tilde{\Pi}_{\Lambda} ( \bd{k} , i \omega )
  & =  & \frac{T}{\omega^2} \int_{\bd{q}}
  \Bigl[ F_{\Lambda} ( \bd{q} , i \omega ) [ G_{\Lambda} ( \bd{q} + \bd{k} )
 \nonumber
 \\
 & & \hspace{10mm}
 + G_{\Lambda} ( \bd{q} - \bd{k} ) - 2 G_{\Lambda} ( \bd{q} ) ] \Bigr]^{\bullet} .
 \hspace{7mm}
 \label{eq:Piflow2}
 \end{eqnarray}
The vanishing of the integrand on the right-hand side for $\bd{k} =0$ guarantees
that the solution of Eq.~(\ref{eq:Piflow2}) satisfies the 
Ward identity (\ref{eq:ward2}). The fact that the solution
of Eq.~(\ref{eq:Piflow2}) satisfies also the continuity 
condition (\ref{eq:continuity2}) is guaranteed by the prefactor of $1/ \omega^2$ which implies that
the inverse of $\tilde{\Pi}_{\Lambda} ( \bd{k} \neq 0 , i \omega )$ vanishes for $\omega \rightarrow 0$.

\subsubsection{Renormalized three-point vertex}

It turns out that the flow equation (\ref{eq:Piflow2}) still
does not include all vertex corrections which are necessary to calculate the
dynamic spin susceptibility for finite momentum $\bd{k}$.
To see this, consider the equation of motion (\ref{eq:eom4})
for the mixed three-spin correlation function. 
The approximations (\ref{eq:quantumapprox}) and (\ref{eq:Gamma4approx2}) are consistent with neglecting the
momentum dependence of the 
four-spin correlation function $G^{++--} ( Q+K, - Q^{\prime} - K , Q^{\prime} , Q )$
in the last line of Eq.~(\ref{eq:eom4}). By shifting
the  loop momentum  $\bd{q}^{\prime} \rightarrow \bd{q}^{\prime} + \bd{k}$
it is then easy to see that this term does not contribute to the equation of motion,
which therefore reduces to
 \begin{eqnarray}
 & & i \omega G^{+-z} ( Q + K  , - Q , -K )  =  G ( Q ) - G ( Q + K  )
 \nonumber
 \\
 &   &  \hspace{19mm} +  [ J ( \bd{q} ) - J ( \bd{q} + \bd{k}  )  ] G ( Q ) G ( Q+ K )
 \nonumber
 \\
 & = &  G ( Q ) \left[ 1 +    \frac{{J} ( \bd{q}  ) -  {J} ( \bd{q} + \bd{k} ) }{2}   G ( Q+K) \right]
 \nonumber
 \\
 &- &  G ( Q + K  ) \left[ 1 +    \frac{ {J} ( \bd{q} + \bd{k}  )  - {J} ( \bd{q} ) }{2}  G ( Q ) \right].
 \label{eq:eompmz}
 \end{eqnarray}
Taking the  limit $J \rightarrow 0$ and assuming $\omega \neq 0$ this reduces to
 \begin{equation}
 G_0^{+-z} ( Q + K  , - Q , -K )  =  \frac{  \beta b_0^{\prime}  }{i \omega}  
 \left[   \delta_{\omega^{\prime} , 0}    -  
 \delta_{\omega^{\prime} + \omega , 0} \right],
 \label{eq:Gpmz0}
 \end{equation}
where again $K = ( \bd{k} , i \omega )$ and $Q = ( \bd{q} , i \omega^{\prime} )$.
Eq.~(\ref{eq:Gpmz0}) is equivalent with the zeroth-order approximation (\ref{eq:Gammapmzzero})
for the mixed three-point vertices which we have used  to derive Eqs.~(\ref{eq:Pibareflow}) and
(\ref{eq:Piflow2}). To construct an approximation 
consistent with the equation of motion (\ref{eq:eompmz})  for finite $J$, we retain the terms in the square braces 
in the last two lines of Eq.~(\ref{eq:eompmz}) neglecting
the frequency dependence of the propagators.
Then Eq.~(\ref{eq:Gpmz0}) should be replaced by
  \begin{eqnarray}
 G_\Lambda^{+-z} ( Q + K  , - Q , -K )  & = &
  \frac{  \beta b_0^{\prime}  }{i \omega}  
 \Bigl[   \delta_{\omega^{\prime} , 0} Y_{\Lambda} ( \bd{q} , \bd{q} + \bd{k} )    
 \nonumber
 \\
 & - &    \delta_{\omega^{\prime} + \omega , 0} Y_{\Lambda} ( \bd{q} + \bd{k} , \bd{q})    \Bigr],
 \hspace{7mm}
 \label{eq:Gpmzvert}
 \end{eqnarray}
with scale-dependent vertex correction factor
 \begin{equation}
 Y_{\Lambda} ( \bd{q} + \bd{k} , \bd{q}) \equiv
  1 +    \frac{ {J}_{\Lambda} ( \bd{q} + \bd{k}  )  - {J}_{\Lambda} ( \bd{q} ) }{2}  G_{\Lambda} ( \bd{q} ).
 \end{equation}
Using the tree expansion (\ref{eq:tree3}) to calculate the corresponding irreducible three-point vertices
and defining
 \begin{equation}
 Z_{\Lambda} ( \bd{q} , \bd{k} ) \equiv Y^2_{\Lambda} ( \bd{q} + \bd{k} , \bd{q} ),
 \end{equation}
we  obtain   instead of Eq.~(\ref{eq:Piflow2}) for the flow of the
irreducible dynamic susceptibility, 
 \begin{widetext}
  \begin{eqnarray}
 \partial_{\Lambda} \tilde{\Pi}_{\Lambda} ( \bd{k} , i \omega )
  & =  & \frac{T}{\omega^2} \int_{\bd{q}}
  \Bigl\{  \bigl[ F_{\Lambda} ( \bd{q} , i \omega )  G_{\Lambda} ( \bd{q} + \bd{k} )  \bigr]^{\bullet}
  Z_{\Lambda} ( \bd{q} , \bd{k} )
 + \bigl[ F_{\Lambda} ( \bd{q} , i \omega )  G_{\Lambda} ( \bd{q} - \bd{k} )  \bigr]^{\bullet}
  Z_{\Lambda} ( \bd{q} , - \bd{k} ) 
 - 2 \bigl[ F_{\Lambda} ( \bd{q} , i \omega ) 
 G_{\Lambda} ( \bd{q} )  \bigr]^{\bullet}  \Bigl\}.
 \nonumber
 \\
 & &
 \label{eq:Piflow3}
 \end{eqnarray}
Moreover, taking into account 
the flow of the purely classical four-point vertex 
 \begin{eqnarray}
 \Gamma^{(4)}_{\Lambda} ( - \bd{k} , \bd{k} , - \bd{q} , 
 \bd{q} ) & = & \Gamma_{\Lambda}^{--++} ( - \bd{k} , - \bd{q}, \bd{q} ,  \bd{k}) 
 + \frac{1}{2!} \Gamma_{\Lambda}^{-+zz} ( - \bd{k} ,\bd{k}, -  \bd{q} ,  \bd{q}),
 \end{eqnarray}
as well as all vertex corrections discussed above
we obtain instead of Eq.~(\ref{eq:Sigmabareflow})
for the  flow equation of the static self-energy,
  \begin{eqnarray}
 \partial_{\Lambda} {\Sigma}_{\Lambda} ( \bd{k}  ) & = & T
 \int_{\bd{q}} 
 \dot{G}_{\Lambda} ( \bd{q} )  \Gamma^{(4)}_{\Lambda} ( - \bd{k} , \bd{k} , - \bd{q} ,  \bd{q} )
 +  T \sum_{ \omega \neq 0} 
 \int_{ \bd{q}}   \frac{ \dot{F}_{\Lambda } ( \bd{q} , i \omega ) }{\omega^2}
 \biggl[ - W_{\Lambda} ( i \omega ) - C_{\Lambda} ( \bd{q} , i \omega )
 \nonumber
 \\
 & &
 \hspace{45mm}  +   {F}_{\Lambda } ( \bd{q} + \bd{k} , i \omega ) Z_{\Lambda} ( \bd{k} , \bd{q} )
 + {F}_{\Lambda } ( \bd{q} -  \bd{k} , i \omega ) Z_{\Lambda} ( - \bd{k} , \bd{q} )
 \biggr].
 \label{eq:Sigmaflow3}
 \end{eqnarray}
\end{widetext}
Here the energies $W_{\Lambda} ( i \omega )$ and $C_{\Lambda} ( \bd{q} , i \omega )$
are defined in Eqs.~(\ref{eq:Wconstraint}) and (\ref{eq:Cconstraint}); these terms
are due to the mixed classical-quantum four-point vertices in  the second line
of the exact flow equation (\ref{eq:Sigmaexactflow})
which we approximate again by Eq.~(\ref{eq:continuityconstraint}).

We conclude this subsection with three remarks:

\begin{enumerate}

\item
To obtain a closed system of flow equation, we should add a flow equation for the
classical four-point vertex  
$\Gamma^{(4)}_{\Lambda} ( - \bd{k} , \bd{k} , - \bd{q} ,  \bd{q} )$ 
which contributes to the flow of the static self-energy in Eq.~(\ref{eq:Sigmaflow3}).
In  the simplest approximation we can replace this vertex  by  its initial value
given in Eq.~(\ref{eq:Gamma4init}).

\item
Keeping in mind that $Z_{\Lambda} ( \bd{q} , 0 )=1$, we note that the solution of
the flow equation (\ref{eq:Piflow3}) satisfies the Ward identity
$\tilde{\Pi}_{\Lambda} ( 0, i \omega \neq 0 ) =0$ for all values of the deformation parameter $\Lambda$.

\item Within our truncation the  flow equation (\ref{eq:Piflow3}) 
for the irreducible dynamic susceptibility does not involve any frequency summations.
The Matsubara frequency $i \omega $ therefore plays the role of an external parameter
so that the analytic continuation to real frequencies can be trivially performed.
Obviously, within our truncation only  elastic scattering processes are taken into account for the calculation
of $\tilde{\Pi} ( \bd{k} , i \omega )$.
On the other hand,  our flow equation (\ref{eq:Sigmaflow3}) for the static self-energy
$\Sigma ( \bd{k} )$ involves a frequency summation, so that it takes also inelastic scattering 
processes into account.

\end{enumerate}

\subsection{Integral equation for the irreducible dynamic susceptibility}

Although Eqs.~(\ref{eq:Piflow3}) and (\ref{eq:Sigmaflow3}) can be used to
calculate the static self-energy $\Sigma ( \bd{k} )$ and thus detect possible magnetic instabilities,
in this work we will focus  on  
the finite-frequency spin dynamics in the paramagnetic phase at high temperatures.
To this end, it is sufficient to simplify the above system of flow equations
by ignoring the flow equation (\ref{eq:Sigmaflow3}) for the static self-energy, assuming
that the static  two-spin correlation function
$G ( \bd{k} ) = 
 G_{\Lambda = 1} ( \bd{k} )$ can  be determined by some other method.
In fact, at  
high temperatures, we can simply calculate  $G ( \bd{k} )$
via an expansion in powers of $J/T$, as will be discussed in Sec.~\ref{sec:high}.  
The vertex correction factor
 \begin{eqnarray}
Z ( \bd{q} , \bd{k} ) & = & Y^2 ( \bd{q} + \bd{k} , \bd{q} )
 = \biggl[ 1 +    \frac{ {J} ( \bd{q} + \bd{k}  )  
 - {J} ( \bd{q} ) }{2} G ( \bd{q} )   \biggr]^2
 \nonumber
 \\
 & &
 \label{eq:Zvertdef}
 \end{eqnarray}
is  then independent of the deformation parameter $\Lambda$ so that our flow 
equation (\ref{eq:Piflow3}) for the dynamic susceptibility reduces to
  \begin{eqnarray}
 \partial_{\Lambda} \tilde{\Pi}_{\Lambda} ( \bd{k} , i \omega )
  & =  & \frac{T}{\omega^2} \int_{\bd{q}}
   \dot{F}_{\Lambda} ( \bd{q} , i \omega ) \bigl[ G ( \bd{q} + \bd{k} ) Z ( \bd{q} , \bd{k} )
 \nonumber
 \\
 & & \hspace{3mm}
 + G ( \bd{q} - \bd{k} ) Z ( \bd{q} , - \bd{k} ) 
- 2 G ( \bd{q} ) \bigr]  .
 \hspace{7mm}
 \label{eq:Piflow4}
 \end{eqnarray}
To convert this integro-differential equation into an integral equation we 
use the Katanin substitution \cite{Katanin04}, which amounts to 
replacing  the
single-scale propagator $\dot{F}_{\Lambda} ( \bd{q} , i \omega )$ by a total scale-derivative
$\partial_{\Lambda} F_{\Lambda} ( \bd{q} , i \omega )$.
The right-hand side of Eq.~(\ref{eq:Piflow4}) is then a total 
$\Lambda$-derivative so that
by integrating both sides over $\Lambda$ we obtain an integral equation for the irreducible dynamic 
susceptibility  $\tilde{\Pi} ( \bd{k} , i \omega ) =
 \tilde{\Pi}_{\Lambda =1 } ( \bd{k} , i \omega )$. 
The lower limit $\Lambda =0$ does not contribute because
for finite $\omega$ the function $\tilde{\Pi}_{\Lambda =0} ( \bd{k} , i \omega ) $ vanishes. 
Then Eq.~(\ref{eq:Piflow4})  reduces to the following integral equation for the
subtracted irreducible dynamic susceptibility,
\begin{eqnarray}
\tilde{\Pi} ( \bd{k} , i\omega ) 
 & = &  \frac{1}{ \omega^2} \int_{ \bd{q}}
    \frac{  \tilde{V} ( \bd{k} , \bd{q} ) }{   G ( \bd{q} ) + \tilde{\Pi} ( \bd{q} , i \omega )} 
  ,
 \label{eq:Pidynint}
 \end{eqnarray}
where the dimensionless kernel $ \tilde{V} ( \bd{k} , \bd{q} )$ is defined by
 \begin{eqnarray}
 \tilde{V} ( \bd{k} , \bd{q} ) & \equiv &   T \bigl[   {G} ( \bd{q} + \bd{k} ) Z ( \bd{q} , \bd{k} )
  +   {G} ( \bd{q} -  \bd{k}  ) Z ( \bd{q} , - \bd{k} ) 
 \nonumber
 \\
 & & \hspace{3mm}  -   2 {G} ( \bd{q} ) \bigr].
 \label{eq:kerneldef}
 \end{eqnarray}
It is convenient to parametrize the subtracted irreducible dynamic susceptibility 
in terms of an  energy $\Delta ( \bd{k} , i \omega )$ as follows
 \begin{equation}
 \tilde{\Pi} ( \bd{k} , i \omega ) = G ( \bd{k} ) \frac{ \Delta ( \bd{k} , i \omega )}{ | \omega | }.
 \label{eq:Deltadef}
 \end{equation}
Substituting this definition  into 
Eq.~(\ref{eq:Gtildepi}) relating $\tilde{\Pi} ( \bd{k} , i \omega )$  to
the dynamic spin-spin correlation function $G ( \bd{k} , i \omega )$ 
and using the fact that by construction $\tilde{J} ( \bd{k} ) = G^{-1} ( \bd{k} )$,
we obtain
 \begin{equation}
 G ( \bd{k} , i \omega ) = G ( \bd{k} ) \frac{ \Delta ( \bd{k} , i  \omega ) }{ 
  \Delta ( \bd{k} , i \omega )  + | \omega |}.
 \label{eq:GMatsubara}
 \end{equation}
After analytic continuation to real frequencies 
$ i \omega \rightarrow \omega + i 0^+$ this reduces to Eq.~(\ref{eq:Gall}).
We call  $\Delta ( \bd{k} ,  \omega )$ the dissipation energy, because
a purely real value of this energy
implies a pole of the retarded
spin-spin correlation function $G ( \bd{k} , \omega )$ on the imaginary axis in
the complex frequency plane. The energy  $\Delta ( \bd{k} ,  \omega )$ can then be 
identified with  the energy scale associated with  the dissipative decay of spin fluctuations 
with wavevector $\bd{k}$.
Substituting the definition (\ref{eq:Deltadef})
into the integral equation ~(\ref{eq:Pidynint}) we find that within our truncation
of the FRG flow equations
the dissipation energy  $\Delta ( \bd{k} , i \omega )$ satisfies the integral equation
 \begin{equation}
 \Delta ( \bd{k} , i \omega ) = \int_{\bd{q}} \frac{ V ( \bd{k} , \bd{q} ) }{
  \Delta ( \bd{q} , i \omega )  +   | \omega |  },
 \label{eq:Deltaint}
 \end{equation}
where the kernel
 \begin{eqnarray}
 V ( \bd{k} , \bd{q} ) & = &  G^{-1} ( \bd{k} ) G^{-1} ( \bd{q} )  \tilde{V} ( \bd{k} , \bd{q} )
  =  T  G^{-1} ( \bd{k} ) G^{-1} ( \bd{q} ) 
 \nonumber
 \\
  & & \hspace{-17mm} \times
\bigl[   {G} ( \bd{q} + \bd{k} ) Z ( \bd{q} , \bd{k} )
 +   {G} ( \bd{q} -  \bd{k}  ) Z ( \bd{q} , - \bd{k} ) 
   -   2 {G} ( \bd{q} ) \bigr]
 \nonumber
 \\
 & &
 \label{eq:Wdef}
 \end{eqnarray}
has units of energy squared.
Here the vertex renormalization  factor $Z ( \bd{q} , \bd{k} )$ is
defined in Eq.~(\ref{eq:Zvertdef}).
Assuming that for small wavevectors 
the dissipation  energy $\Delta ( \bd{k} , i \omega )$
can be expanded as
 \begin{equation}
 \Delta ( \bd{k} , i \omega )    =
{\cal{D}} ( i \omega )  k^2 + {\cal{O}} (  k^4 ),
 \end{equation}
we conclude that if the dynamics is indeed  diffusive, then the spin-diffusion coefficient
is given by
 \begin{equation}
   {\cal{D}}  =  {\cal{D}} ( 0 ) = \lim_{ k \rightarrow 0} \frac{ \Delta ( \bd{k} , 0 ) }{k^2}.
 \label{eq:Ddifdef}
 \end{equation}

The non-linear integral equation (\ref{eq:Deltaint})  can be solved for the
 dissipation energy 
$\Delta ( \bd{k} , i \omega )$ if the static spin-spin correlation function 
$G ( \bd{k})$ has been  determined by some other method.
We could now go back to the system of flow equations (\ref{eq:Piflow3})
and (\ref{eq:Sigmaflow3}) to determine both
the dynamic susceptibility $\tilde{\Pi}_{\Lambda} ( \bd{k} , i \omega )$ and
the static self-energy $\Sigma_{\Lambda} ( \bd{k} )$.
However, the explicit solution of this system of equations
requires extensive numerical calculations which are 
beyond the scope of this work. In the rest of this work  we will focus 
on the dissipative dynamics at high temperatures where
the static spin-spin correlations can be obtained  by means of an expansion in powers of
$J / T$ which can then be used to determine 
the kernel $V ( \bd{k} , \bd{q}) $ in
the integral equation (\ref{eq:Deltaint}). 

To conclude this section, let us point out  
that within the framework of the so-called mode-coupling theory \cite{Kawasaki66} (see Refs.~[\onlinecite{Goetze99,Das04}] for
reviews) 
a similar parametrization of the retarded spin-spin correlation function is
used. Typically, 
in mode-coupling theory
one starts from a generalized Langevin equation for the Kubo relaxation function \cite{Mori65}, 
where the memory kernel, involving higher order correlations, is 
closely related to the dissipation energy $\Delta(\bm{k},\omega)$ used by us.  After applying several approximations to the kernel, one arrives at a closed integro-differential equation for the relaxation function, which has been 
extensively studied in the literature \cite{Kawasaki66,Blume70,Hubbard71,Goetze99,Das04}. 
However,
in contrast to our integral equation (\ref{eq:Deltaint}),
the integro-differential equation for the relaxation function 
obtained in mode-coupling theory is non-local in frequency-space.

\section{Dissipative spin dynamics at infinite temperature}
\label{sec:high}

The problem of spin diffusion in quantum Heisenberg magnets
at infinite temperature has been discussed by many authors. Older works
focused on three-dimensional systems 
\cite{DeGennes58,Mori62,Bennett65,Resibois66,Redfield68,TahirKheli69,Blume70,Morita72,Morita75,Kopietz93},
while recently the high-temperature spin dynamics in one-dimensional Heisenberg magnets 
has attracted considerable attention \cite{Ljubotina17,Gopalakrishnan19a,Gopalakrishnan19b,Nardis19,Nardis20,Bulchandani20,Dupont20,Bulchandani21}. 
Even at $ T =   \infty$ the problem of calculating the 
dynamic spin-spin correlation function of Heisenberg magnets 
is non-trivial and requires non-perturbative resummation  and extrapolation schemes.
In fact, up until now,  a resummation scheme 
based on the diagrammatic approach to quantum spin systems developed by 
Vaks, Larkin and Pikin \cite{Vaks68,Vaks68b,Izyumov88}
which generates a  diffusive pole in the
spin-spin correlation function  has not been found.
We now show that by solving  the integral equation (\ref{eq:Deltaint}) 
we obtain such a non-perturbative
resummation. Although in this work we focus on the limit
of  infinite temperature, we have preliminary evidence \cite{Tarasevych21}  
that our approach gives sensible results
in the entire paramagnetic regime, including the temperature range in the vicinity
of the critical point.

Once we have calculated the dissipation energy $\Delta ( \bd{k} , i \omega )$
by solving the integral equation (\ref{eq:Deltaint}), we can obtain the retarded spin-spin correlation function by analytic continuation to real frequencies,
$ i \omega \rightarrow \omega + i 0$, which amounts to 
the replacement $ | \omega | \rightarrow - i \omega$.
From Eq.~(\ref{eq:GMatsubara}) we then obtain the
retarded spin-spin correlation function in the form (\ref{eq:Gall}),
where the retarded dissipation energy $\Delta ( \bd{k} , \omega )$ 
is in general a complex function which we decompose into real and imaginary part,
 \begin{equation}
 \Delta ( \bd{k} , \omega ) = \Delta_R ( \bd{k} , \omega ) +
 i  \Delta_I ( \bd{k} , \omega ) .
 \end{equation}
The dynamic structure factor can then be obtained with the help of the
fluctuation-dissipation theorem,
 \begin{eqnarray}
  S( \bd{k} , \omega ) &  = &   \left[ 
1 + \frac{1}{ e^{  \beta \omega } -1} \right]
   \frac{1}{\pi} {\rm Im} G ( \bd{k} , \omega )
 \nonumber
 \\
 & = &  \frac{ \omega G ( \bd{k} ) }{ 1- e^{ - \beta \omega}  }
 \frac{1}{\pi} \frac{  \Delta_R ( \bd{k} , \omega ) }{ 
  \Delta_R^2 ( \bd{k} , \omega ) + [ \omega - \Delta_I ( \bd{k} , \omega ) ]^2 }.
 \label{eq:dynstruc}
 \hspace{7mm}
 \end{eqnarray}
In the limit of infinite temperature this reduces to
 \begin{equation}
  S( \bd{k} , \omega ) = \frac{b_0^{\prime}}{\pi} 
\frac{  \Delta_R ( \bd{k} , \omega ) }{ 
  \Delta_R^2 ( \bd{k} , \omega ) + [ \omega - \Delta_I ( \bd{k} , \omega ) ]^2 }.
 \label{eq:SinfiniteT}
 \end{equation}

\subsection{General strategy}

At temperatures $T \gg J$ it is sufficient to approximate the
static self-energy by its truncated expansion in powers of $1/T$ up to order $J^2/T$,
\begin{equation}
 \Sigma ( \bd{k} ) = \frac{T}{ b_0^{\prime}} + \frac{ \Sigma_2 ( \bd{k} ) }{T} 
 + {\cal{O}} \left( \frac{ J^3}{T^2} \right),
 \end{equation}
where
 \begin{eqnarray}
 \Sigma_2 ( \bd{k} ) & = &  \frac{1}{12} \int_{\bd{q}}
    J ( \bd{q} ) J ( \bd{q} + \bd{k} )
+   \Bigl( b_0^{\prime} + \frac{1}{6} \Bigr)  \int_{\bd{q}} 
 J^2 ( \bd{q} ).
 \label{eq:I2def}
 \hspace{9mm}
 \end{eqnarray}
The kernel $V ( \bd{k} , \bd{q} )$ defined in Eq.~(\ref{eq:Wdef}) then reduces to
  \begin{eqnarray}
   V ( \bd{k} , \bd{q} )  &  =  &  
\frac{b_0^{\prime}}{4}  \left[ \left[    J ( \bd{q} ) -   J ( \bd{q} + \bd{k} )  \right]^2
 +  \left[    J ( \bd{q} ) -   J ( \bd{q} - \bd{k} )  \right]^2 \right]
 \nonumber
 \\
 &  +  &  
2 \Sigma_2 ( \bd{q} ) - \Sigma_2 ( \bd{q} + \bd{k} ) - \Sigma_2 ( \bd{q} - \bd{k} ) 
   + {\cal{O}} \left( \frac{J^3}{T}  \right).
 \nonumber
 \\
 & &
 \label{eq:Khigh}
 \end{eqnarray}
Note that at high temperatures $V ( \bd{k} , \bd{q} )$
is a non-trivial function of order $J^2$
satisfying  $V ( 0 , \bd{q} ) =0$.
To explicitly solve the integral equation (\ref{eq:Deltaint}) for the dissipation energy 
$\Delta ( \bd{k} , i \omega )$,
we note that for exchange couplings $J_{ij}$ with finite range  
the kernel $V ( \bd{k} , \bd{q} )$ can be expanded as
\begin{equation}
  V ( \bd{k} , \bd{q} ) = \sum_{\alpha =1}^n e^{ i \bd{k} \cdot \bd{R}_{\alpha} } 
 V_{\alpha} ( \bd{q} ),
 \label{eq:Wexpansion}
 \end{equation}
where $\bd{R}_1, \bd{R}_2, \ldots, \bd{R}_n$ is a finite set of vectors of the underlying Bravais lattice which depends on the precise form of $J ( \bd{k} )$ and on the geometry and dimensionality of the lattice.
The solution of Eq.~(\ref{eq:Deltaint}) is then of the form
 \begin{equation}
 \Delta ( \bd{k} , i \omega ) = 
\sum_{\alpha =1}^n  e^{ i \bd{k} \cdot \bd{R}_{\alpha} }  \Delta_{\alpha} ( i \omega ) ,
 \label{eq:fexpansion}
 \end{equation}
where the $n$ coefficients $\Delta_1( i \omega ) , \ldots , \Delta_n ( i \omega )$ 
satisfy the following system of non-linear equations,
 \begin{equation}
 \Delta_{\alpha}  ( i \omega )=  
 \int_{\bd{q}} \frac{ V_{\alpha} ( \bd{q} )}{ | \omega | + \sum_{\alpha^{\prime} =1}^n 
 e^{ i \bd{q} \cdot \bd{R}_{\alpha^{\prime}}}
 \Delta_{\alpha^{\prime}}  ( i \omega )}, \; \; \;  \alpha = 1, \ldots , n.
 \label{eq:falpha}
 \end{equation}

To obtain an explicit solution of these equations, let us assume here
for simplicity
that the spins are located on a $d$-dimensional hypercubic lattice
with spacing $a$ and that the exchange couplings $J_{ij}$ connect only pairs
of nearest neighbors.  In Sec.~\ref{sec:bcc}  and in Appendix~B
we will discuss more 
general models including next-nearest-neighbor exchange.
Denoting by $J$
the strength of the nearest-neighbor coupling,
the Fourier transform of the exchange couplings 
on a $d$-dimensional hypercubic lattice is
 \begin{equation}
 J ( \bd{k} ) = J \sum_{\bd{\delta}} e^{ i \bd{k} \cdot \bd{\delta} } = 2 d J {\gamma}_{\bd{k} } , 
 \end{equation}
where the sum is over the $2d$  vectors $\bd{\delta}$ 
with  length $| \bd{\delta} | =a$ connecting a given site to its
nearest neighbors. For later convenience we have introduced the normalized 
nearest-neighbor hypercubic form factor
 \begin{equation}
 {\gamma}_{\bd{k} } = \frac{1}{2d} \sum_{\bd{\delta}} e^{ i \bd{k} \cdot \bd{\delta} } =
 \frac{1}{d}  \sum_{\mu =1}^d \cos ( k_{\mu} a ).
 \label{eq:gammakdef}
 \end{equation}
Using
 \begin{equation}
 \int_{\bd{q}} \gamma_{\bd{q}}  \gamma_{\bd{q} + \bd{k} } 
 = \frac{ \gamma_{\bd{k}}}{2d},
 \end{equation} 
the integral $\Sigma_2 ( \bd{k} )$ defined in Eq.~(\ref{eq:I2def}) 
is easily evaluated,
 \begin{equation}
  \Sigma_2 ( \bd{k} ) =   2 d J^2 \left[ \frac{ \gamma_{\bd{k}}}{12} +  b_0^{\prime} + \frac{1}{6} 
  \right] .
 \end{equation}
We conclude that for nearest-neighbor exchange on a hypercubic lattice
 \begin{eqnarray}
& &  2 \Sigma_2 ( \bd{q} ) - \Sigma_2 ( \bd{q} + \bd{k} ) -  \Sigma_2 ( \bd{q} - \bd{k} ) 
 \nonumber
 \\
 & = & \frac{d J^2}{6} \left[ 2 \gamma_{\bd{q}} - \gamma_{ \bd{q} + \bd{k}}  - \gamma_{ \bd{q} - \bd{k} } \right].
 \end{eqnarray}
At this point it is convenient 
to measure all energies in units of $| J | \sqrt{b_0^{\prime} }$, defining the dimensionless quantities
 \begin{equation}
 \tilde  \Delta ( \bd{k},i\omega ) \equiv \frac{\Delta(\bm{k},i\omega) }{|J|\sqrt{b'_0}}, \; \; \;
 \tilde{\omega} \equiv \frac{\omega}{ |J|\sqrt{b'_0}} .
 \label{eq:tildeDelta}
 \end{equation}

\subsection{Spin-diffusion coefficient in $d \geq 2$}
 \label{sec:cubic}

The cubic symmetry and the condition $ \Delta( \bd{k} =0 , i \omega) =0$
imply that for nearest-neighbor coupling
the expansion~(\ref{eq:fexpansion}) 
can be expressed in terms of only three independent form factors. 
In dimensionless form the expansion is therefore
 \begin{eqnarray}
  \tilde \Delta ( \bd{k},i\omega ) & = & ( 1 - \gamma_{\bd{k}} )   \tilde{\Delta}_1(i\omega) + 
 ( 1 - \gamma_{2\bd{k}} ) \tilde{\Delta}_{2}^{\parallel}(i\omega)
 \nonumber \\ &+ &   ( 1 - \gamma^{\bot}_{\bd{k}} ) \tilde{\Delta}_{2}^{\bot}(i\omega),
 \label{eq:fabc}
 \end{eqnarray}
where we have introduced the  
off-diagonal next-nearest-neighbor form factor
 \begin{equation}
  \gamma^{\bot}_{\bd{k}} = \frac{2}{d (d-1 )} \sum_{ 1 \leq \mu 
< \mu^{\prime}  \leq d}
 \cos ( k_{\mu} a ) \cos ( k_{\mu^{\prime}} a ).
 \label{eq:gammabot}
 \end{equation}
In $d$ dimensions we find from Eq.~(\ref{eq:falpha}) that the
three amplitudes in Eq.~(\ref{eq:fabc}) satisfy the following system of
equations,
 \begin{subequations}
 \label{eq:selfcondelta}
 \begin{eqnarray}
  \tilde{\Delta}_{1}(i {\omega}) 
& =  &  2d \int_{\bm{q}}\frac{1}{|\tilde{\omega}| + \tilde \Delta(\bm{q},i {\omega})}    
 + \frac{d}{3b'_0}\int_{\bm{q}}\frac{\gamma_{\bm{q}}}{|\tilde{\omega}|+ \tilde \Delta(\bm{q},i {\omega})}     
 \nonumber
 \\
 &  & -  2  \tilde{\Delta}^{\parallel}_{2}(i \omega) -  2 \tilde{\Delta}^{\bot}_{2}(i {\omega} ) ,
   \label{eq: eqsc}
 \\
  \tilde{\Delta}^{\parallel}_{2}(i {\omega}) & = &  -d \int_{\bm{q}}\frac{\gamma_{2\bm{q}}}{|\tilde{\omega}| + \tilde \Delta(\bm{q},i {\omega})},
  \label{eq: pareqsc}
  \\
 \tilde{\Delta}^{\bot}_{2}(i  {\omega}) & = &  -2d(d - 1) \int_{\bm{q}}\frac{\gamma^\perp_{\bm{q}}}{|\tilde{\omega}| + \tilde \Delta(\bm{q},i {\omega})}.
 \label{eq: boteqsc}
 \end{eqnarray}
\end{subequations}
It turns out that for $d > 2$ the  self-consistent solution of these equations have a finite limit
for $\omega \rightarrow 0$, implying that the static
dissipation energy $\Delta(\bm{k},i\omega =0)$ is finite. \cite{footnoteCor}
According to Eq.~(\ref{eq:Ddifdef}) the
spin-diffusion coefficient ${\cal{D}}$
can then be obtained from the quadratic term in the expansion of
$\Delta ( \bd{k} , 0 )$ for small $\bd{k}$. From Eqs.~(\ref{eq:tildeDelta})
and (\ref{eq:fabc}) we obtain
\begin{equation}
 {\cal{D}} = \frac{|J|\sqrt{b'_0}a^2}{2 d} \left[ \tilde{\Delta}_{1}(0) + 4 \tilde{\Delta}_{2}^{\parallel}(0) + 2 
 \tilde{\Delta}_{2}^{\bot}(0) \right].  
 \label{eq:Dcubic}
 \end{equation}
In the limit of high dimensions the solution of Eqs.~(\ref{eq:selfcondelta}) simplifies
because to leading order in $1/d$ the first term   on the right-hand side  of Eq.~(\ref{eq: eqsc})
without form factor dominates. In this limit we obtain
$\tilde \Delta_1(0) = \sqrt{2d}$ and $ \ \tilde{\Delta}^{\parallel}_2(0) = \tilde{\Delta}^{\bot}_2(0) = 0$,
implying $\mathcal{D} = |J|a^2 \sqrt{b'_0/2d}$.
In the physically relevant case of three dimensions we have to solve
Eqs.~(\ref{eq:selfcondelta}) numerically to obtain the value of ${\cal{D}}$.
In Table \ref{tab:dres} we present our numerical results for ${\cal{D}}$ in three dimensions
for different spin quantum numbers  $S$.
 \begin{table}
 \begin{center}
 \begin{tabular}{|c || c | c | c | c | c |} 
 \hline
 $S$ & $\frac{1}{2}$ & $1$ &  $\frac{3}{2}$ &  $2$ & $\infty$ \\ [0.5ex]  \hline
 $\frac{{\cal{D}}}{|J|a^2\sqrt{4b'_0}}$ & $0.217$  & $0.189$  & $0.179$ & $0.175$ & $0.167$ \\ 
 \hline
\end{tabular}
\end{center}
\caption{Spin-diffusion coefficient 
${\cal{D}}$ of the nearest-neighbor spin-$S$ Heisenberg model on a cubic lattice
with lattice spacing $a$ at infinite temperature obtained from Eq.~(\ref{eq:Dcubic}).
Note that for $S=1/2$ the  normalization factor 
$ \sqrt{ 4 b_0^{\prime}} = \sqrt{4 S(S + 1)/3}$ is unity. 
The quantum term proportional to $d / ( 3 b_0^{\prime} )$
in Eq.~(\ref{eq: eqsc}) gives rise to some additional spin-dependence which vanishes
for $S \rightarrow \infty$.
For large  $S$ we obtain ${\cal{D}} = 0.193 \, S |J| a^2$ to leading order.}
 \label{tab:dres}
 \end{table}
In the special case of  $S = \frac{1}{2}$ our result $\mathcal{D} \approx 0.217 |J|a^2$ is roughly 
$30 \%$ smaller than theoretical results obtained by extrapolating the  short-time expansion of 
suitable correlation functions
to long times \cite{DeGennes58,Mori62,Bennett65,Resibois66,Redfield68,TahirKheli69,Morita72,Morita75,Kopietz93,Boehm94}. 
Surprisingly, controlled  numerical results for the spin-diffusion coefficient of the
three-dimensional Heisenberg model at infinite temperature are not available.
Note, however, that there is  experimental evidence \cite{Labrujere82} that
extrapolations based on the short-time expansion tend to overestimate the numerical value of ${\cal{D}}$.
In the following subsection we will use our method to calculate the high-temperature value
of ${\cal{D}}$ 
for a Heisenberg model on a body-centered cubic lattice relevant to the
experiment of Ref.~\cite{Labrujere82}.

\subsection{Spin diffusion on a body-centered cubic lattice including next-nearest-neighbor exchange}
%
\label{sec:bcc}
\begin{figure}[tb]
 \begin{center}
  \centering
\vspace{7mm}
\includegraphics[width=0.3\textwidth]{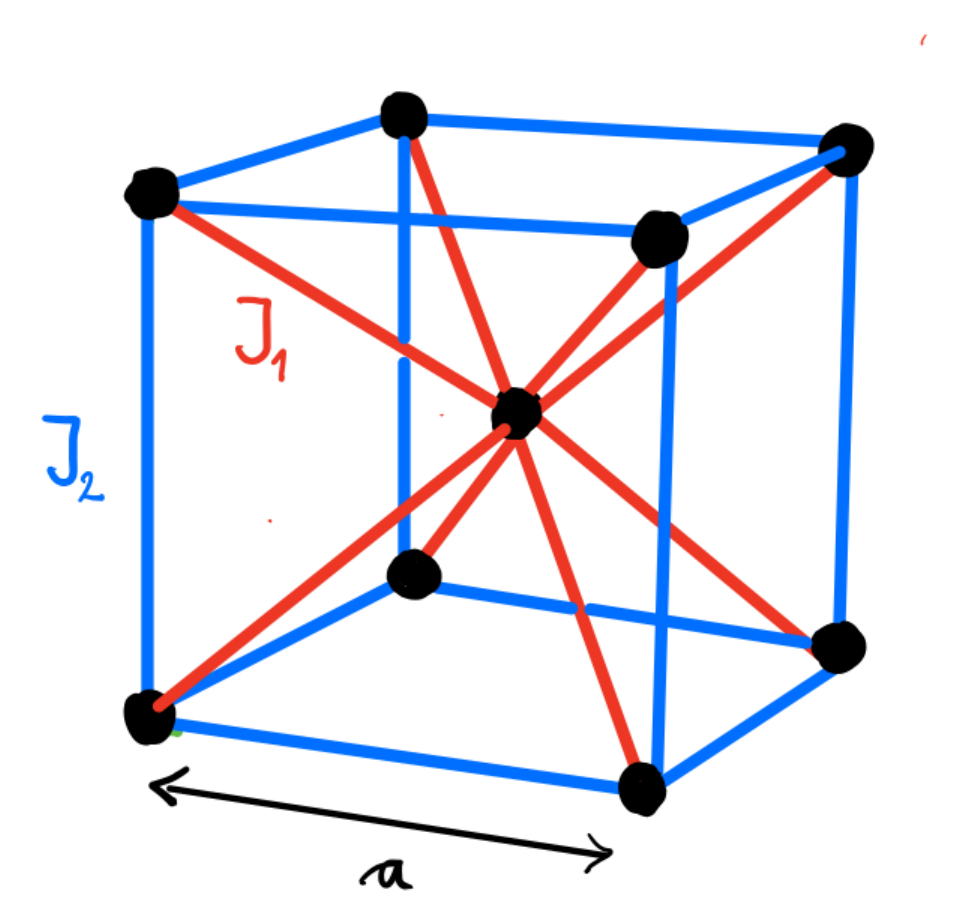}
   \end{center}
  \caption{%
Conventional unit cell of the body-centered cubic lattice 
with lattice spacing $a$.
The black dots represent
the magnetic copper ions in Rb$_2$CuBr$_4\cdot 2$H$_2$O
which are coupled by nearest-neighbor exchange $J_1$ (red lines) 
and next-nearest-neighbor exchange
$J_2$ (blue lines). The non-magnetic atoms  
of Rb$_2$CuBr$_4\cdot 2$H$_2$O
are omitted.
}
\label{fig:bccgeometry}
\end{figure}

The measurement of the spin-diffusion coefficient ${\cal{D}}$ in the ferromagnetic insulator
Rb$_2$CuBr$_4\cdot 2$H$_2$O by
Labrujere {\it{et al.}}~[\onlinecite{Labrujere82}]
seems to be the only published experimental determination of
${\cal{D}}$ in a three-dimensional Heisenberg magnet 
at high temperatures. The
 magnetic properties of the copper ions in this 
material can be described by 
a ferromagnetic spin $S =1/2$ Heisenberg model on a body-centered cubic (bcc) lattice with
nearest-neighbor exchange \cite{Labrujere82,footnoteJ}  $| J_1 | /2  \approx  0.49$ K and next-nearest-neighbor exchange
$| J_2 | /2  \approx  0.29$ K, as illustrated  in 
Fig.~\ref{fig:bccgeometry}. 
The spin-diffusion coefficient was measured at two different temperatures $T = 20$ K and $T= 77$ K, which are 
two orders magnitude larger than the energy scales $| J_1 | S^2 $ and
$|J_2 | S^2$ associated with the exchange couplings.
Since the bcc lattice is a Bravais lattice, we can use  the formalism developed in this work
to calculate the
spin-diffusion coefficient. 
To explicitly solve the integral equation (\ref{eq:Deltaint}),
we need the Fourier transform of the exchange couplings for the geometry shown in
Fig.~\ref{fig:bccgeometry},
 \begin{subequations}
 \begin{eqnarray}
 J(\bm{k}) & = & J_1(\bm{k}) + J_2(\bm{k}),
 \\
J_1(\bm{k})  & = & 8J_1\gamma^{\rm bcc}_{\bm{k}},
 \\
J_2(\bm{k}) & = & 6 J_2 \gamma_{\bm{k}}.
 \end{eqnarray}
 \end{subequations}
Here the normalized bcc form factor is
\begin{equation}
 \gamma^{\rm bcc}_{\bm{k}} = \cos\Big(\frac{k_x a}{2}\Big)\cos\Big(\frac{k_y a}{2}\Big) \cos\Big(\frac{k_z a}{2}\Big),  
\label{eq: bccJ_1}
\end{equation}
and the normalized cubic form factor $\gamma_{\bd{k}}$ can be obtained 
by setting $d=3$ 
in Eq.~(\ref{eq:gammakdef}), i.e.,
\begin{equation}
 {\gamma}_{\bd{k} } = 
 \frac{1}{3} \left[ \cos ( k_{x} a ) + \cos ( k_y a ) + \cos ( k_z a ) \right].
 \label{eq:gammak3def}
 \end{equation}

As in the calculation of the dissipation energy $\Delta ( \bd{k} , 0 )$
for the cubic lattice described in Sec.~\ref{sec:cubic}, we decompose
$\Delta(\bm{k},i\omega)$ into a finite number of form factors 
and solve the resulting non-linear equations for the amplitudes 
at $\omega = 0$ numerically.
For a bcc lattice with nearest-neighbor and next-nearest-neighbor exchange
six independent form factors are necessary  to obtain a closed system of equations.
Technical details of the calculation are given in Appendix~A.
In the simplified case of only nearest-neighbor exchange we obtain
 $\mathcal{D}_{ {\rm bcc}}^{ J_2 =0} \approx 0.18|J_1|a^2$ for $S=1/2$, which is roughly a factor of
$5/6$ smaller than our result on a cubic lattice for the same value of 
$J_1$.
Our result for the ratio ${\cal{D}}_{\rm bcc} / {\cal{D}}_{ \rm cubic}$ agrees with the corresponding ratio 
obtained by Morita \cite{Morita72, Morita75} using a
different method.  According to Ref.~[{\onlinecite{Labrujere82}],
in the experimentally 
studied material Rb$_2$CuBr$_4\cdot 2$H$_2$O the ratio of exchange couplings is
$J_2/ J_1 \approx 0.6$; with this value  we obtain on a bcc lattice
 \begin{equation}
\mathcal{D}_{ {\rm bcc}}^{ J_2 / J_1 = 0.6   } \approx 0.23|J_1|a^2  = 
0.46| {J}^{\prime}_{1}|a^2,
 \label{eq:Dbccres}
 \end{equation}
where we have  set
$|{J}^{\prime}_1| = |J_1| /2$ to facilitate the comparison \cite{footnoteJ}
with Ref.~[\onlinecite{Labrujere82}], where the experimental result
 \begin{equation}
 {\cal{D}}_{\rm exp} \approx (0.31 \pm 0.03)|{J}^{\prime}_1|a^2
 \label{eq:Dexp}
 \end{equation}
 is  presented in terms  of $J_1^{\prime}$. 
Our theoretical  prediction (\ref{eq:Dbccres}) for the high-temperature
spin-diffusion coefficient
in Rb$_2$CuBr$_4\cdot 2$H$_2$O is about
$30 \%$ larger than
the corresponding experimental result in Eq.~(\ref{eq:Dexp}).
With the exception of the method developed by Bennett and Martin \cite{Bennett65}
(which gives a prefactor $0.40$ instead of our $0.46$ in Eq.~(\ref{eq:Dbccres}))
other theoretical approaches \cite{Mori62,Resibois66,TahirKheli69}  
predict even larger values for ${\cal{D}}_{\rm bcc}$.  
We conclude that at high temperatures
the measured value of the spin-diffusion coefficient
in the ferromagnetic insulator Rb$_2$CuBr$_4\cdot 2$H$_2$O is significantly smaller than  all available theoretical predictions.

A possible explanation for this discrepancy
is that
at high temperatures the relevant value of the 
next-nearest-neighbor coupling $J_2 $ is not given by
$J_2 /J_1 = 0.6$ but has a value somewhere
in the range $ -0.4 \lesssim J_2 / J_1  \lesssim 0$. 
As shown in  Fig.~\ref{fig:diffbcc}, in this range  ${\cal{D}}_{ \rm bcc}$
exhibits a broad minimum as a function of $J_2/J_1$ which is reasonably
close to the experimental value.
\begin{figure}[tb]
 \begin{center}
  \centering
\vspace{7mm}
\includegraphics[width=0.45\textwidth]{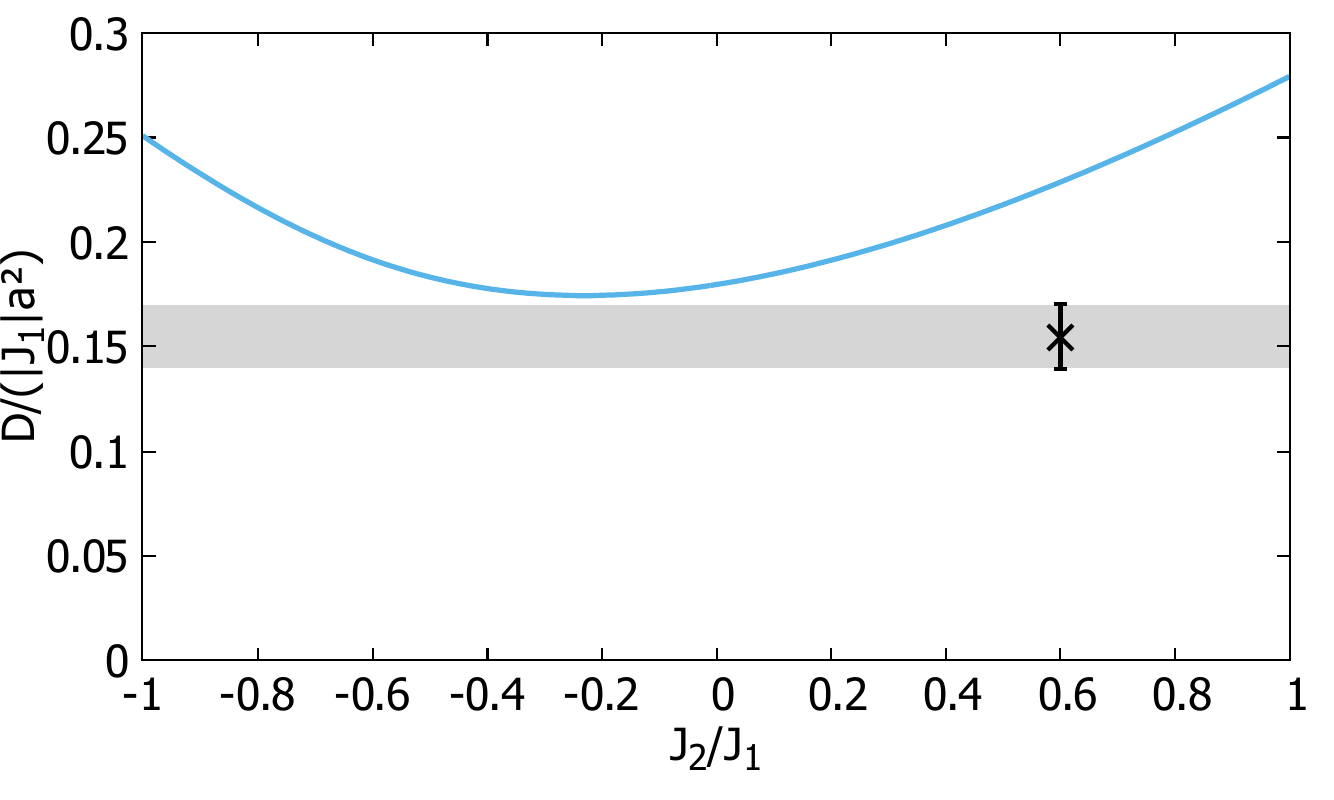}
   \end{center}
  \caption{%
Spin-diffusion coefficient $\mathcal{D}_{\rm bcc}$ for a spin $1/2$ Heisenberg magnet
with nearest-neighbor exchange $J_1$ and next-nearest-neighbor exchange $J_2$ 
on a bcc lattice at infinite temperature as a function of the ratio $J_2 /J_1$. The blue  curve is our result  obtained from the
solution of the integral equation (\ref{eq:Deltaint}) for the dissipation energy $\Delta ( \bd{k} , 0 )$.
The black cross at $J_2/J_1 = 0.6$ with error bar marks
the experimental result of Labrujere {\it{ et al.}}~\cite{Labrujere82}
obtained in the high-temperature regime of the magnetic insulator
Rb$_2$CuBr$_4\cdot 2$H$_2$O; the shaded area represents the experimental uncertainty assuming
that the true value of $J_2/J_1$ is not known.
}
\label{fig:diffbcc}
\end{figure}
Although this agreement might be accidental,
a possible reason  
for the deviation of $J_2 / J_1$ from the value
$0.6$
used in Ref.~[\onlinecite{Labrujere82}] could be 
a significant  temperature-dependence of $J_2$ 
in the high-temperature regime probed in the experiment.
This hypothesis is supported by the fact that
in the related compound K$_2$CuCl$_4\cdot 2$H$_2$O
a strong temperature-dependence of the nearest-neighbor exchange interaction has been observed~\cite{Kennedy70}, which decreases by a factor of five when raising $T$ from 77 K to 300 K. As a possible reason the authors of \cite{Kennedy70} identified a low-lying optical phonon.

\subsection{Anomalous spin diffusion in reduced dimensions}

We now come back to the nearest-neighbor spin-$S$ Heisenberg model 
on a hypercubic lattice and consider the case  $d \leq  2$.
Then it is not allowed to approximate
$\tilde{\Delta} ( \bd{q} , i \omega ) \approx \tilde{\Delta} ( \bd{q} , 0 )$
in Eq.~(\ref{eq:selfcondelta}) because the frequency-dependence of
$\tilde{\Delta} ( \bd{q} , i \omega )$ is essential to cut the infrared divergence 
of the integrals.
For small frequencies $  | \omega |  \ll   | J | \sqrt{b_0^{\prime}} $
the leading behavior of the relevant integrals can be obtained by expanding
the integrands to leading order in $\bd{q}$,
 \begin{equation}
 \tilde{\Delta} ( \bd{q} , i \omega ) = \tilde{\cal{D}} ( i \omega )   {q}^2 + \ldots,
 \end{equation}
where
 \begin{equation}
 \tilde{\cal{D}} ( i \omega ) \equiv \frac{ {\cal{D}} ( i \omega ) }{| J | \sqrt{b_0^{\prime}}}.
 \end{equation}
The leading singular part of the integrals in Eq.~(\ref{eq:selfcondelta}) can then be obtained by approximating,
\begin{equation}
\int_{\bm{q}}\frac{f(\bm{q})}{|\tilde{\omega}| + \tilde \Delta(\bm{q},i {\omega})} \approx f(\bm{0})\int_{\bm{q}}\frac{1}{|\tilde{\omega}| + \tilde{\cal{D}} (i{\omega})
 q^2},
  \label{eq:cutintegrallowd}
\end{equation}
where $f ( \bd{q} )$ is any of the enumerators in Eq.~(\ref{eq:selfcondelta}).
Note that from  Eq.~(\ref{eq:fabc}) we find that the 
coefficient of order $k^2$ in the expansion of
$\tilde{\Delta} ( \bd{k} , i \omega )$ satisfies
\begin{equation}
 \tilde{\cal{D}} ( i{\omega})  = \frac{a^2}{6b'_0}\int_{\bm{q}}\frac{\gamma_{\bm{q}}}{|\tilde{\omega}| + \tilde \Delta(\bm{q},i {\omega})} +  
 a^2 \int_{\bm{q}}\frac{1 - \gamma_{2\bm{q}}}{|\tilde{\omega}| + \tilde \Delta(\bm{q}.
 i {\omega})}.
\end{equation}
From this expression we conclude that the singular
part of the spin-diffusion coefficient 
is completely determined  by the self-energy contribution 
 $2 \Sigma_2(\bm{q}) - \Sigma_2(\bm{q + k})   -   \Sigma_2(\bm{q - k}) \propto  1/ b'_0$ 
to  the high-temperature expansion (\ref{eq:Khigh}) of the kernel
$V ( \bd{k} , \bd{q} )$ of the integral equation (\ref{eq:Deltaint}).
In dimensions $d \leq 2$ the 
leading singular part of the generalized diffusion coefficient 
can therefore be obtained from the solution of
 \begin{equation}
 \tilde{\cal{D}} ( i{\omega})  = \frac{a^2}{6b'_0}\int_{\bm{q}}\frac{1}{|\tilde{\omega}| +   \tilde{\cal{D}} ( i \omega ) q^2 }.
 \end{equation}
In terms of dimensionful quantities this can also be written as
 \begin{equation}
 {\cal{D}} ( i{\omega})  = \frac{J^2 a^2}{6}\int_{\bm{q}}\frac{1}{|{\omega}| +   {\cal{D}} ( i \omega ) q^2 }.
 \label{eq:Dsing}
 \end{equation}

Consider first the case of one dimension, where the solution of Eq.~(\ref{eq:Dsing})
yields for the singular part of the generalized diffusion coefficient
\begin{equation}
{\cal{D}}(i\omega) = \left(\frac{|J|}{144|\omega|}\right)^{\frac{1}{3}}|J|a^2,
 \; \; \; \; d=1.
\end{equation}
To obtain the retarded spin-spin correlation function and the dynamic structure factor, 
we should analytically continue
${\cal{D}} ( i \omega )$ to real frequencies, $ i \omega \rightarrow \omega + i 0$,
which amounts to replacing $  |  \omega | \rightarrow - i \omega $. 
The correct branch of the multi-valued function  $(- i \omega )^{-1/3}$
is determined by the condition that the real part of
${\cal{D}} ( \omega )$ must be positive to guarantee the positiveness of
the dynamic structure factor in Eq.~(\ref{eq:SinfiniteT}).
This implies a complex anomalous diffusion coefficient,
 \begin{equation}
 {\cal{D}}(\omega  ) = \left(\frac{|J|}{144 |\omega | }\right)^{\frac{1}{3}} 
 | J | a^2
 \left( \frac{\sqrt{3}}{2} + \frac{i}{2} {\rm sgn} \omega \right),
 \label{eq:D1res}
 \end{equation}
where the real part
${\rm Re} {\cal{D}} ( \omega ) = \sqrt{3}
 {\rm Im} {\cal{D}} ( \omega ) {\rm sgn} \omega$ 
has the same order of magnitude as the imaginary part.
The corresponding dynamic structure factor $S ( {k}_x , \omega )$ defined via Eq.~(\ref{eq:SinfiniteT}) has, as a function of $k_x$, a broad maximum at $k_x = k_{\ast}$ determined by the condition
 \begin{equation}
| {\cal{D}} ( \omega)|k_{\ast}^2 = \omega,
 \end{equation}
 implying
 \begin{equation}
 k_{\ast} \propto \omega^{2/3}.
 \label{eq:z1d}
 \end{equation}
In Fig.~\ref{fig:dynstruc1d}  we show the momentum dependence
of the dynamic structure factor $S ( k_x , \omega )$ 
for small momenta and three different frequencies.
The dynamic exponent $z = 3/2$ implied by Eq.~(\ref{eq:z1d}) and the superdiffusive
singularity
${\cal{D}} ( \omega ) \propto | \omega |^{-1/3}$ are in agreement with
recent calculations for integrable
isotropic Heisenberg chains with nearest-neighbor coupling
\cite{Ljubotina17,Gopalakrishnan19a,Gopalakrishnan19b,Nardis19,Nardis20,Bulchandani20,Dupont20,Nardis21,Bulchandani21}. 
On the other hand, for  non-integrable chains with larger spin $S > 1/2$  
the situation is less clear \cite{Bulchandani21}: some authors obtained normal diffusion \cite{Dupont20}, recognizing broken integrability as its cause, while others found that  superdiffusion 
persists even for non-integrable chains \cite{Nardis19}. 
The fact that for non-integrable chains  our approach yields the same
superdiffusive high-temperature spin dynamics as for integrable chains might be related to the fact our integral equation (\ref{eq:Deltaint})
takes only elastic scattering into account,  as pointed out at the end of 
Sec.~\ref{sec:vertex}.
While for integrable chains this approximation seems to be justified, 
in the case of non-integrable chains it might break down at very low energies.
\begin{figure}[tb]
 \begin{center}
  \centering
\vspace{7mm}
  \includegraphics[width=0.47\textwidth]{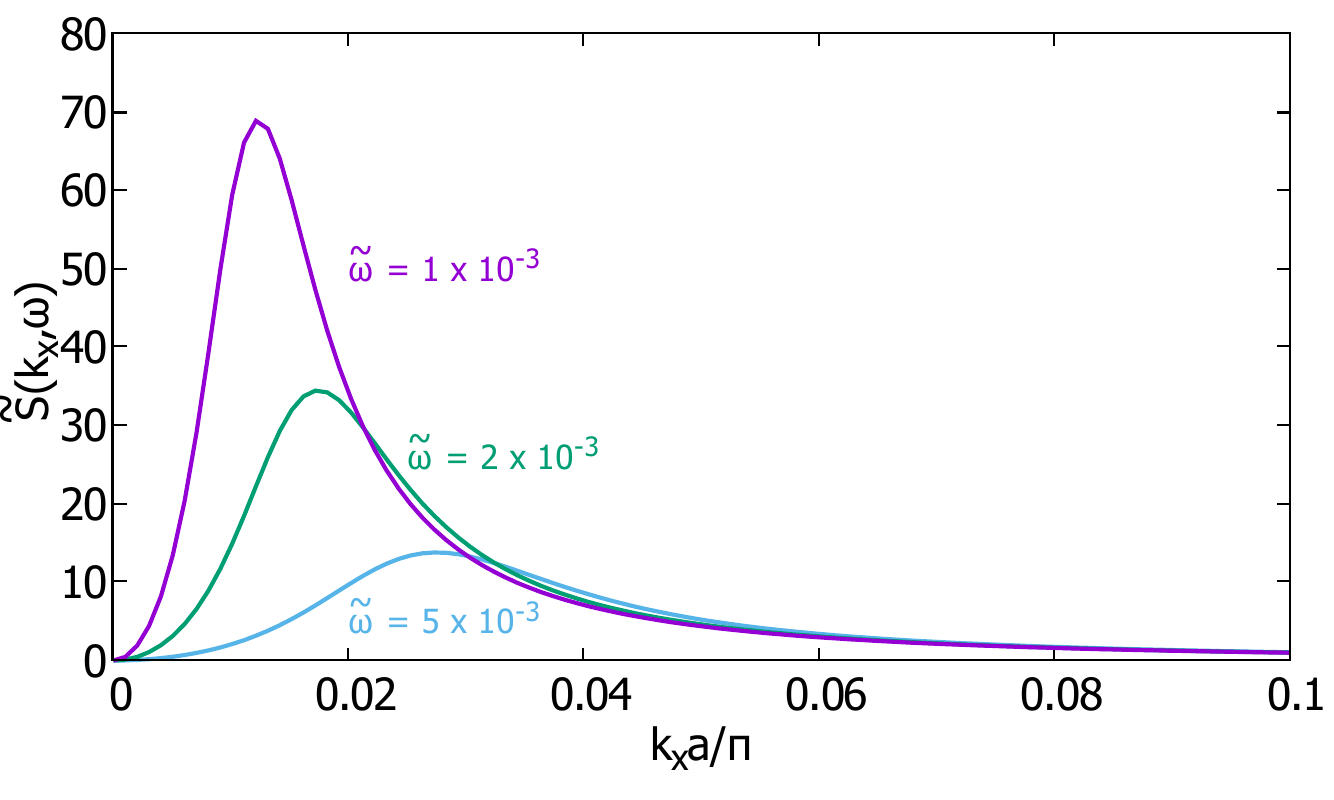}
   \end{center}
  \caption{%
Momentum dependence of the dimensionless 
dynamic structure factor $\tilde{S}( k_x , \omega ) = S ( k_x , \omega ) |J|\sqrt{b'_0}$ 
for small momenta $ | k_x | \ll \pi /a$  and frequencies 
$\tilde{\omega} = \omega / ( | J| \sqrt{b_0^{\prime} } ) = 10^{-3} $ (violet curve), 
$2  \times 10^{-3}$ (green curve) and $5 \times 10^{-3}$  (blue curve)
of a spin $1/2$ Heisenberg chain
with nearest-neighbor exchange $J$ at infinite temperature.}
\label{fig:dynstruc1d}
\end{figure}

Let us now consider the marginal case of $d=2$ where
the integral in Eq.~(\ref{eq:Dsing}) has a logarithmic singularity  which is cut by the
frequency $ | \omega |$. Retaining only the leading logarithm we obtain
\begin{equation}
{\cal{D}}(i\omega) = \sqrt{\ln\left(\frac{{\cal{D}}(i{\omega})}{a^2| \omega|}\right) } 
 \frac{ | J|a^2}{\sqrt{24 \pi} }, \; \; \; \; d=2.
 \label{eq:diffusion2log}
\end{equation}
The solution of this implicit equation can be expressed in terms of the
so-called Lambert $W$-function  (product logarithm) \cite{Corless96} which 
satisfies $W (x) = \ln [ x / W (x) ]$. 
Here we are interested only in the leading logarithm,
which can be obtained by a simple iteration of  the self-consistency 
equation~(\ref{eq:diffusion2log}).
After analytic continuation to real frequencies we obtain for
$ | \omega | \ll J$,
\begin{equation}
 {\cal{D}} ( \omega )   = \frac{ | J | a^2}{\sqrt{24 \pi}}
 \left[ \sqrt{ \ln \left( \frac{ | J | }{ \sqrt{24 \pi} | \omega | } \right)}
 + i \frac{\pi}{4} {\rm sgn} \omega \right].
 \label{ew:diffusion2res}
 \end{equation}
Note that for $\omega \rightarrow 0$ the real part of ${\cal{D}} ( \omega )$
is logarithmically  larger than the imaginary part, whereas in $d=1$ the real- and 
imaginary part of ${\cal{D}} ( \omega )$ 
in Eq.~(\ref{eq:D1res}) have the same order of magnitude.

\section{Dissipation energy and 
dynamic structure factor for all wavevectors}
 \label{sec:dissip}
So far we have focused on the leading term in the expansion
$\Delta ( \bd{k} , i \omega ) = {\cal{D}} ( i \omega ) k^2 + {\cal{O}} ( k^4)$ 
of the dissipation energy for small wavevectors
which determines the frequency-dependent spin-diffusion coefficient ${\cal{D}} ( i \omega )$. 
However,
the solution of the integral equation (\ref{eq:Deltaint}) gives
the dissipation energy 
$\Delta ( \bd{k} , i \omega )$ and hence the dynamic structure factor for arbitrary wavevectors.  
The momentum dependence of $\Delta ( \bd{k} , i \omega )$ is of particular interest
for Heisenberg magnets with exchange
interactions beyond nearest neighbors because in this case 
$\Delta ( \bd{k} , i \omega )$ and the corresponding
dynamic structure factor $S ( \bd{k} , \omega )$ defined via Eq.~(\ref{eq:dynstruc})
can have characteristic features in  the
first  Brillouin zone which can be used derive constraints on competing exchange interactions.
As far as we know, this effect has not been noticed before.
In order to illustrate this effect, we have solved 
the integral equation (\ref{eq:Deltaint}) for $\Delta ( \bd{k} , i \omega )$
in the low-frequency limit $ | \omega | \ll | J_1 |$
for a Heisenberg model with nearest-neighbor exchange $J_1$ and next-nearest neighbor exchange $J_2$ on cubic lattices in dimensions $d=1,2,3$.
Technical details of the calculation are given in Appendix~B. 
For convenience we measure energies in units of  $|J_1 | \sqrt{b_0^{\prime}}$, defining
 \begin{subequations} 
\begin{eqnarray}
 \tilde \Delta(\bd{k}, {\omega})  & \equiv & \frac{
 \Delta(\bd{k}, \omega) }{ |J_1|\sqrt{b'_0}},
 \\ 
 \tilde \omega & \equiv &  \frac{\omega }{   |J_1|\sqrt{b'_0} },
 \\
  \tilde S(\bm{k},\omega) & \equiv &  S(\bm{k},\omega)|J_1|\sqrt{b'_0}.
 \label{eq:tildeSdyn}
 \end{eqnarray}
 \end{subequations}
For a discussion of the
dynamic structure factor $S(\bd{k},\omega)$ 
as a function of the wavevector $\bd{k}$ in the first Brillouin zone, 
we note that for small frequencies 
and for $k a = \mathcal{O}(1)$
we may approximate
\begin{equation}
\tilde S(\bm{k},\omega) \approx \frac{b'_0}{\pi}\frac{\tilde \Delta_R(\bd{k},\omega)}{|\tilde \Delta(\bd{k},\omega)|^2}, 
\label{eq:dynstruclargemomenta}
\end{equation}
which allows us to deduce the qualitative behavior of 
$S(\bm{k},\omega)$ from $\Delta(\bm{k},\omega)$ and vice versa.
In particular, we see that minima of $\Delta ( \bd{k} , \omega )$ correspond to
maxima of $S ( \bd{k} , \omega )$, while  maxima 
of $\Delta ( \bd{k} , \omega )$ correspond to
minima of $S ( \bd{k} , \omega )$.

\subsection{One dimension}

Let us first consider the case $d=1$, where
according to Eq.~(\ref{eq: 1dnnn})
the dimensionless dissipation energy can be written as
 \begin{eqnarray}
 \tilde{\Delta}(k_x ,i {\omega}) = \sum_{j = 1}^4 [ 1 - \cos(j k_x a)]
 \tilde{\Delta}_j(i {\omega}).
\label{eq:1dDeltares}
\end{eqnarray}
The dimensionless amplitudes $\tilde{\Delta}_j ( i \omega )$ at $T = \infty$ can be obtained analytically in the low-frequency limit by applying the approximation   (\ref{eq:cutintegrallowd}) to the integrals in the self-consistency  equations (\ref{eq:1damplitudes}).
In Fig.~\ref{fig:dissenergy1d} we show the momentum-dependent part 
 \begin{equation}
\tilde \Delta(k_x) \equiv \frac{2}{\sqrt{3}}|\tilde \omega|^{\frac{1}{3}}
{\rm Re} \tilde{\Delta} (k_x,   \omega + i 0 )
 \label{eq:tildedelta1}
 \end{equation}
of the dimensionless 
dissipation energy of a $J_1$-$J_2$ chain with spin $1/2$ 
as a function of $J_2/J_1$.
\begin{figure}[tb]
 \begin{center}
  \centering
\vspace{7mm}
  \includegraphics[width=0.47\textwidth]{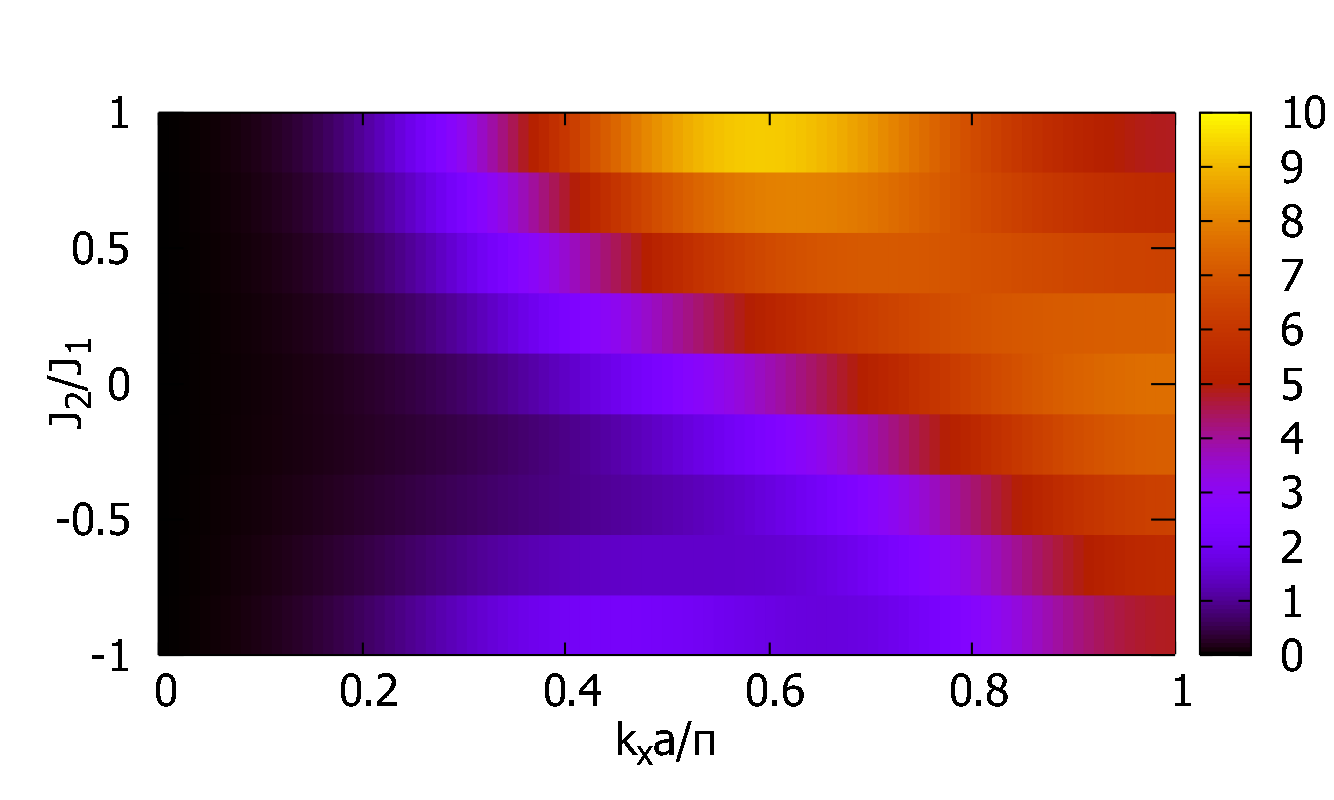}
   \end{center}
  \caption{%
Contour plot of the
momentum-dependent part $\tilde \Delta(k_x)$ of the dimensionless 
dissipation energy defined in  Eq.~(\ref{eq:tildedelta1})
for a $J_1$-$J_2$ chain with spin $1/2$ as a function of the coupling ratio $\mu = J_2 / J_1$ in the interval
$-1 \leq \mu  \leq 1$.}
\label{fig:dissenergy1d}
\end{figure}
In a range of negative coupling ratios starting at  $J_2/J_1 = \mu_2  \approx -0.67$ 
and extending beyond $J_2/J_1 = - 1$, the function $\tilde \Delta(k_x)$
exhibits a two-peak structure, with one maximum located at $k_x a = \pi$, a second 
maximum  at $k_x a \lesssim {\pi}/{2}$, and a minimum somewhere  in the interval
$ [ {\pi}/{2},\pi ]$. 
If the coupling ratio $\mu = J_2 / J_1$ is smaller than a 
certain value $\mu_- < -1$  (not shown in Fig.~\ref{fig:dissenergy1d}),  the second maximum at $k_x a \lesssim {\pi}/{2}$ becomes the global maximum.
On the other hand, for positive coupling ratio $J_2 / J_1 > 0$ 
such a structure cannot be observed. For values of $J_2/J_1$ larger than the threshold
$\mu_+ \approx 0.28$ the peak at $k_x a =
\pi$ evolves into the global maximum in the interval 
 $[ \frac{\pi}{2},\pi ]$ and a local minimum at $k_x a = \pi$. 
This non-trivial momentum dependence gives rise to a 
two-peak structure in the dynamic structure factor, which 
according to Eqs.~(\ref{eq:tildeSdyn}), (\ref{eq:dynstruclargemomenta})
and (\ref{eq:tildedelta1}) can for small frequencies $| \omega | \ll | J_1 |$
and large wavevectors $ | k_x a | = {\cal{O}} ( 1 )$
be written as
 \begin{equation}
 \tilde S(k_x,\omega) = \frac{\sqrt{3}b'_0|\tilde \omega|^{\frac{1}{3}}}{2\pi \tilde \Delta(k_x)}.
 \label{eq:S1Delta1}
 \end{equation}
The momentum dependence of the dynamic structure factor in this regime is therefore
given by the inverse of the function $\tilde{\Delta} ( k_x )$
defined in Eq.~(\ref{eq:tildedelta1}),
which we plot in Fig.~\ref{fig:dissenergy_inv_1d} for three different values of $J_2/J_1$, chosen as $-1,0,1$ in order to display all qualitative features.
\begin{figure}[tb]
 \begin{center}
  \centering
\vspace{7mm}
\includegraphics[width=0.47\textwidth]{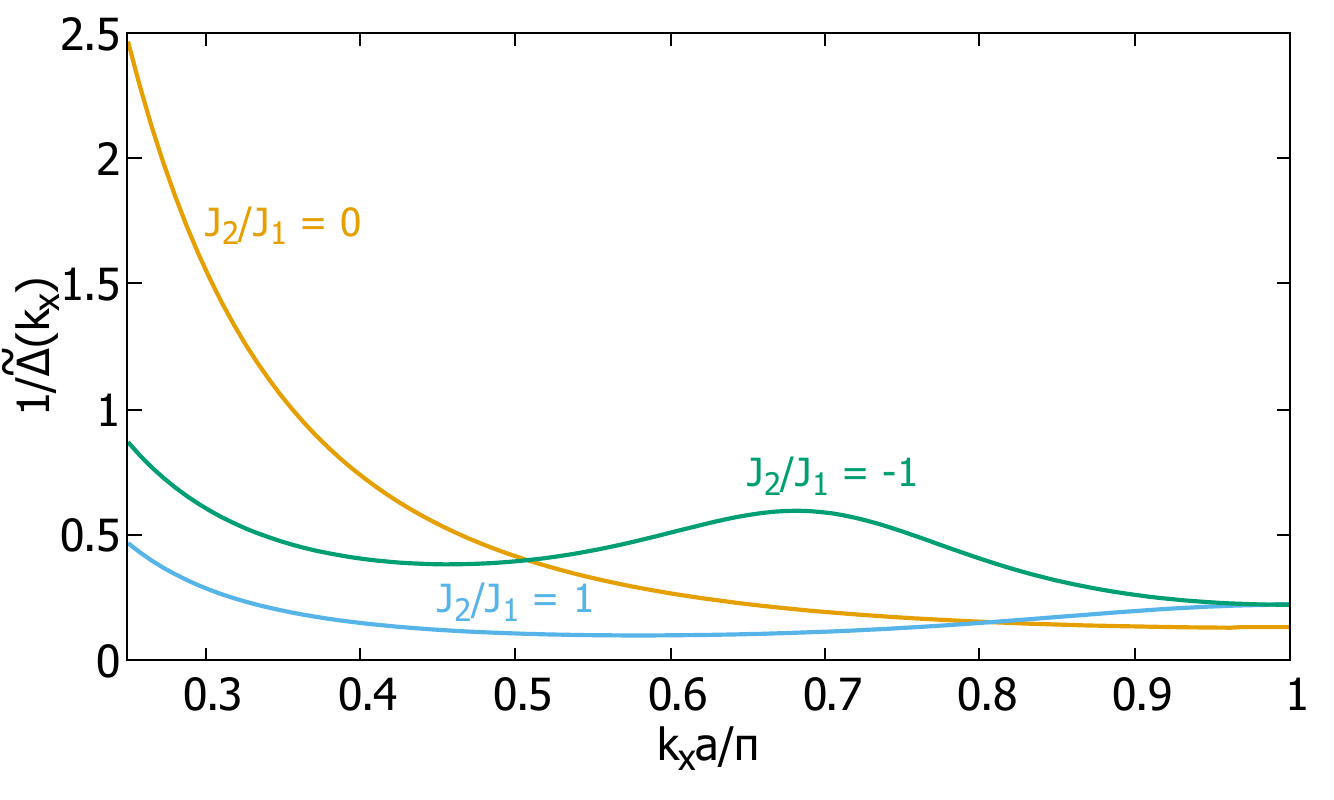}
   \end{center}
  \caption{%
Inverse of the momentum-dependent part $\tilde \Delta(k_x)$ of the dimensionless
dissipation energy defined in Eq.~(\ref{eq:tildedelta1})
for a spin $1/2$ chain 
for large momenta $k_x a > \pi /4$ and $J_2/J_1 = -1, 0, 1$ (green, orange, blue). 
According to Eq.~(\ref{eq:S1Delta1}), for $| \omega | \ll | J_1 | $  this quantity is proportional to the
dynamic structure factor.}
\label{fig:dissenergy_inv_1d}
\end{figure}
Note that in Fig.~\ref{fig:dissenergy_inv_1d}  we draw a different momentum range than in 
Fig.~\ref{fig:dynstruc1d}, so that the
dominant peak for small wavevectors is not visible. 
The lineshape in Fig.~\ref{fig:dissenergy_inv_1d}
exhibits a second peak at short wavelengths, which  
moves from $k_x a = \pi$ for $J_2 / J_1 = 1$ to 
a value in the interval $k_x a \in [ \frac{\pi}{2} , \pi ]$
for $J_2/J_1 = -1$. In the latter case the peak is surrounded by two local minima which is a direct consequence  
of   the two maxima of  $\tilde \Delta(k_x)$ which emerge for $ J_2/J_1 < \mu_2 \approx  -0.67$.

\subsection{Two dimensions}

Next, consider the case of two dimensions, where
the dissipation energy $\Delta ( \bd{k} , i \omega )$ exhibits a logarithmic
dependence on the frequency $\omega$, which in the long-wavelength limit
can be expressed in terms of  the anomalous diffusion coefficient
${\cal{D}} ( i \omega )$ defined in Eq.~(\ref{eq:diffusion2log}). The asymptotic limit $\omega \rightarrow 0$ of $\Delta ( \bd{k} , i \omega )$  can be calculated analytically from the self-consistency equations (\ref{eq:2damplitudes}) for the amplitudes of its Fourier expansion (\ref{eq:2dnnn}), using  again the approximation (\ref{eq:cutintegrallowd}).
Since the  logarithmic frequency-dependence  survives also at short wavelengths, it is
convenient to scale out the frequency-dependence
by defining the momentum dependent dimensionless dissipation energy
 \begin{equation}
 \tilde \Delta(\bm{k})  \equiv  \frac{ {\rm Re} \tilde{\Delta} ( \bd{k} , \omega + i 0 ) }{ \sqrt{ \ln\Big(\frac{\sqrt{J^2_1 +2J_2^2}}{\sqrt{24\pi}|\omega|}\Big)}}
 .
 \label{eq:Deltak2}
 \end{equation}
Our results for $\tilde \Delta(\bm{k})$ in the first quadrant of the
Brillouin zone  for different values of $J_2 / J_1$ are shown
in Fig.~\ref{fig:dissenergy2d}. 
For sufficiently large negative values of $J_2 / J_1$ starting at $J_2/ J_1  = \mu_2  \approx -0.52$
and extending again beyond $J_2/J_1 = -1$,
the function   $\tilde \Delta(\bm{k})$ then exhibits two peaks  at $ \bd{k} a = \big(0, \pi\big)$ and 
 $ \bd{k} a =\big( \pi , \pi \big)$.
Similar to the case  of one dimension,
for negative $J_2 /J_1$
the global maximum is located at the corner $\bm{k}a  = \big( \pi , \pi \big)$ 
of the Brillouin zone
for much larger values of $|J_2|$ than for positive $J_2 / J_1$.
On the other hand, for $J_2/J_1 > 0$ the function $\tilde \Delta(\bm{k})$
is more sensitive to the presence of $J_2$; at $J_2/J_1 = \mu_+ \approx 0.52$ 
the wavevector where $\tilde \Delta(\bm{k})$ exhibits a 
maximum shifts
from $ \bd{k} a = \big( \pi  , \pi \big)$ to $ \bd{k} a = \big(0, \pi \big)$.
For $\omega \rightarrow 0$ and large wavevectors $|\bm{k}a| = \mathcal{O}(1)$ the dynamic structure factor can be obtained from
\begin{equation}
\tilde S(\bm{k},\omega) = \sqrt{\ln \left( \frac{\sqrt{J^2_1 +2J_2^2}}{\sqrt{24\pi}|\omega|} \right)}\frac{b'_0}{\pi \tilde \Delta(\bm{k})}.
 \label{eq:S2Delta2}
\end{equation}
As in one dimension,
the momentum dependence of $\tilde S(\bm{k},\omega)$ is proportional to the inverse of 
$\tilde \Delta(\bm{k})$ which is plotted in Fig.~\ref{fig:dissenergy_inv_2d}  
along the path  $ \bd{k} a = \big(0,\pi) \rightarrow  \big( \pi ,\pi)$  
for $J_2/J_1 = -1, 0, 1$. One sees that for $J_2/J_1 = 1$ 
the short-wavelength peak is located at $ \bd{k} a = \big( \pi , \pi)$, while for  $J_2 / J_1 = -1$
the dynamic structure factor exhibits a 
 maximum on the path connecting the two local minima at $ \bd{k} a = \big( 0 ,\pi)$ and $ \big( \pi, \pi)$. 
%
%
\begin{figure}[tb]
 \begin{center}
  \centering
\vspace{7mm}
\includegraphics[width=0.45\textwidth]{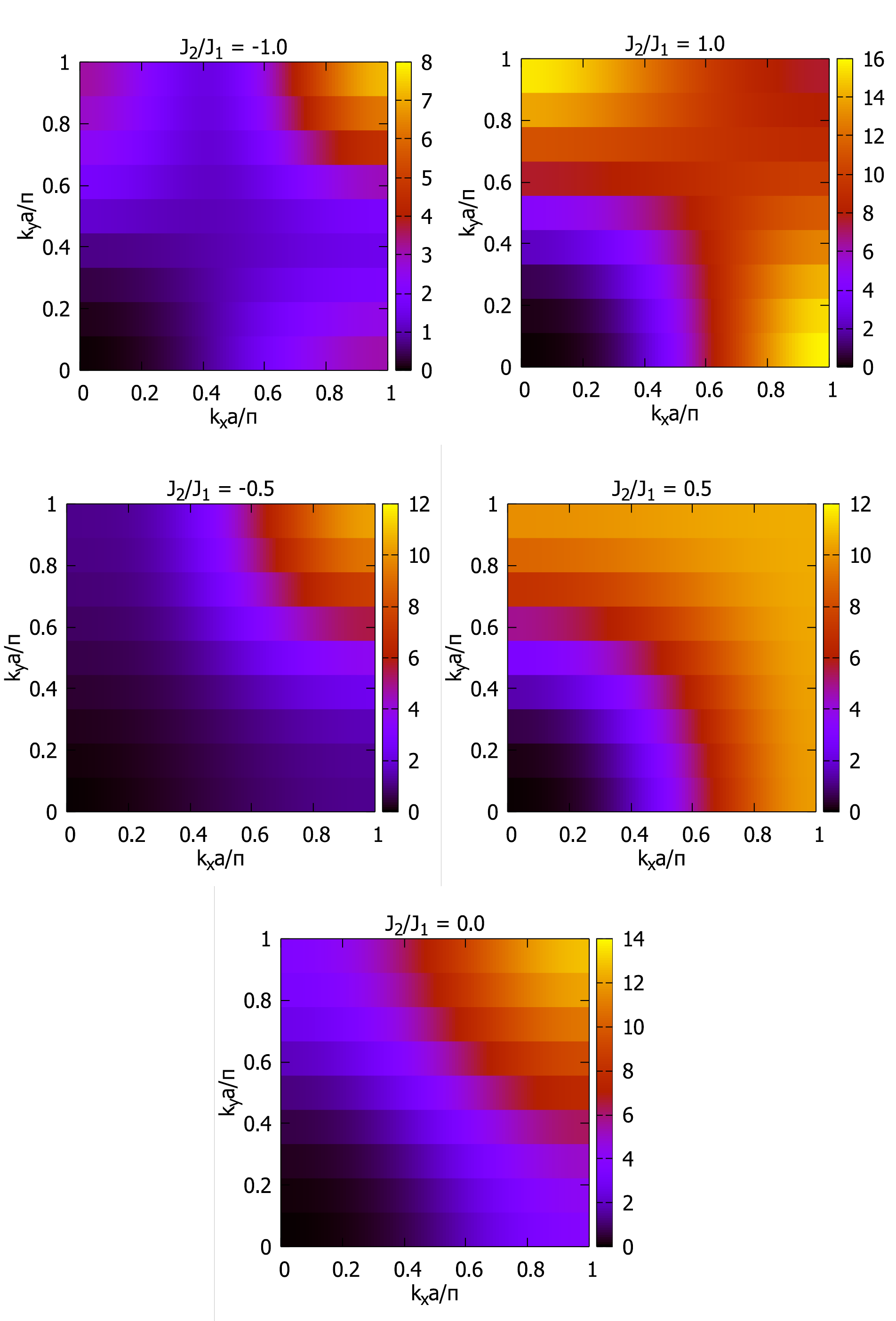}
   \end{center}
  \caption{%
Momentum dependent part of the rescaled dissipation energy
$\tilde{\Delta} ( \bd{k} )$ at infinite temperature 
of a spin-$1/2$ $J_1$-$J_2$ Heisenberg model 
on a square lattice, see  Eq.~(\ref{eq:Deltak2}).
The contour plots are for $J_2/J_1 = -1, -0.5, 0, 0.5, 1$ (counterclockwise, starting from top left). 
}
\label{fig:dissenergy2d}
\end{figure}
\begin{figure}[tb]
 \begin{center}
  \centering
\vspace{7mm}
\includegraphics[width=0.47\textwidth]{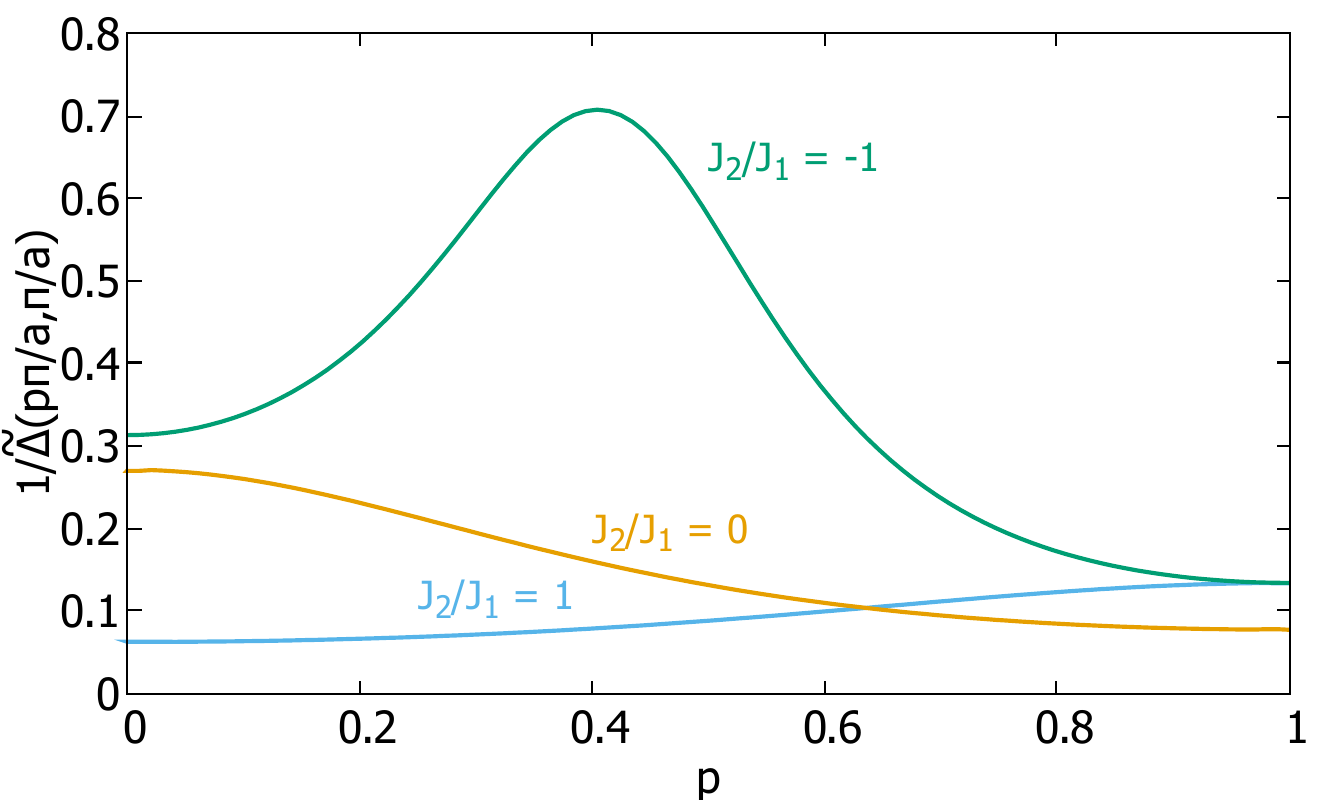}
   \end{center}
  \caption{%
Inverse of momentum dependent part $\tilde \Delta\big( \bm{k} \big)$ 
of the dimensionless dissipation energy for a spin-$1/2$ square lattice $J_1$-$J_2$ Heisenberg model
at infinite temperature. 
The three curves represent the momentum dependence along the path 
 $\bm{k}(p) = \frac{\pi}{a}\big(p,1\big)$  for $J_2/J_1 = -1, 0, 1$ (green, orange, blue). 
According to Eq.~(\ref{eq:S2Delta2}) the curves are proportional to the low-frequency limit of the dynamic
structure factor.
}
\label{fig:dissenergy_inv_2d}
\end{figure}

\subsection{Three dimensions}
In $d = 3$ the dimensionless dissipation energy $\tilde{\Delta} ( \bd{k} , \omega)$
has  a finite limit $\tilde \Delta(\bm{k},0)$ for $\omega \rightarrow 0$, which can be 
obtained  by numerically solving  the system (\ref{eq:3damplitudes}) of equations 
for the amplitudes introduced in Eq.~(\ref{eq:fabcNNN}). Our results for 
$\tilde \Delta(\bm{k},0)$
are shown  in Fig. \ref{fig:dissenergy3dpiplane}
 as a function of $k_x, k_y \geq 0$ 
in the plane $k_z  = \pi /a$ for different values of $J_2/J_1$.
The main qualitative features of the momentum dependence are similar to the
behavior in
reduced dimensions discussed above.
\begin{figure}[tb]
 \begin{center}
  \centering
\vspace{7mm}
\includegraphics[width=0.37\textwidth]{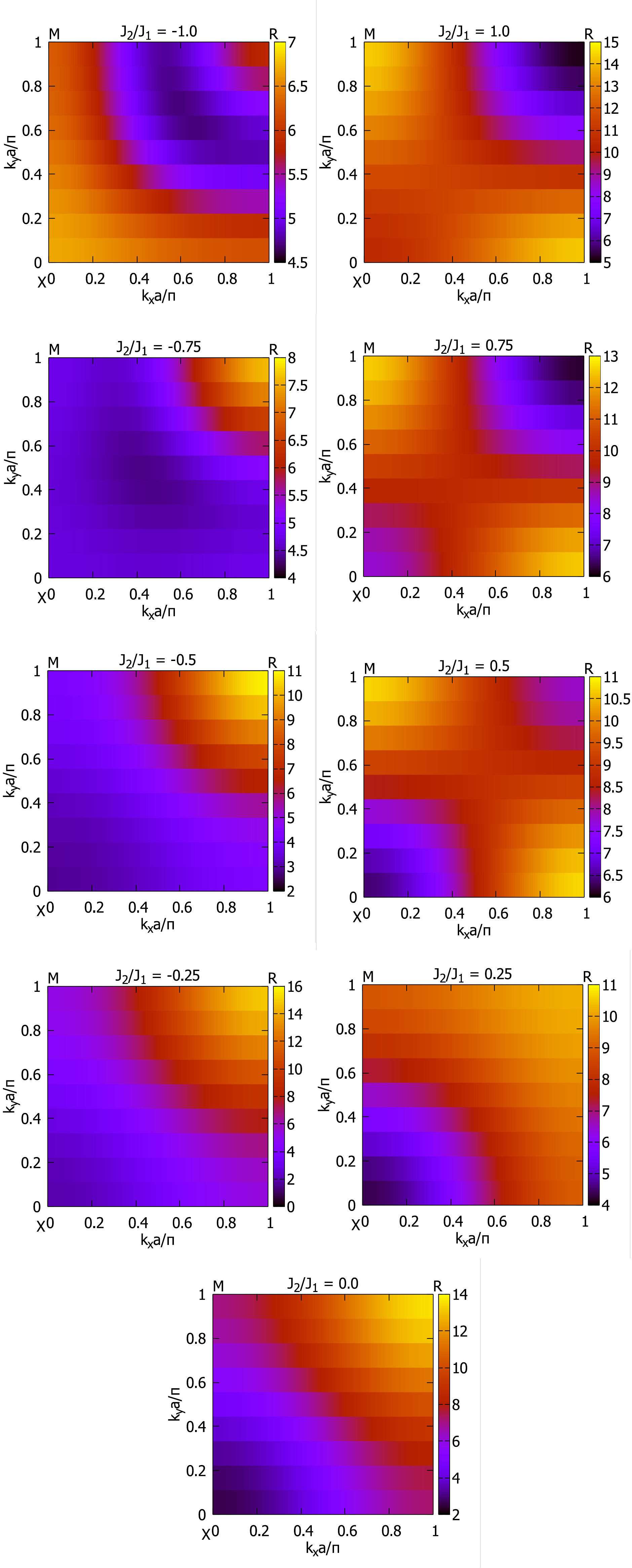}
   \end{center}
  \caption{%
Dimensionless dissipation energy $\tilde \Delta( \bm{k} , \omega =0)$ 
in the plane $k_z = \pi /a$
of the spin-$1/2$ $J_1$-$J_2$ Heisenberg model on a simple cubic lattice at infinite temperature.
The contour plots are  for $J_2/J_1 = -1,...,1$ in steps of $0.25$ 
(counterclockwise, starting from top left). 
}
\label{fig:dissenergy3dpiplane}
\end{figure}
\begin{figure}[tb]
 \begin{center}
  \centering
\vspace{7mm}
\includegraphics[width =0.47\textwidth]{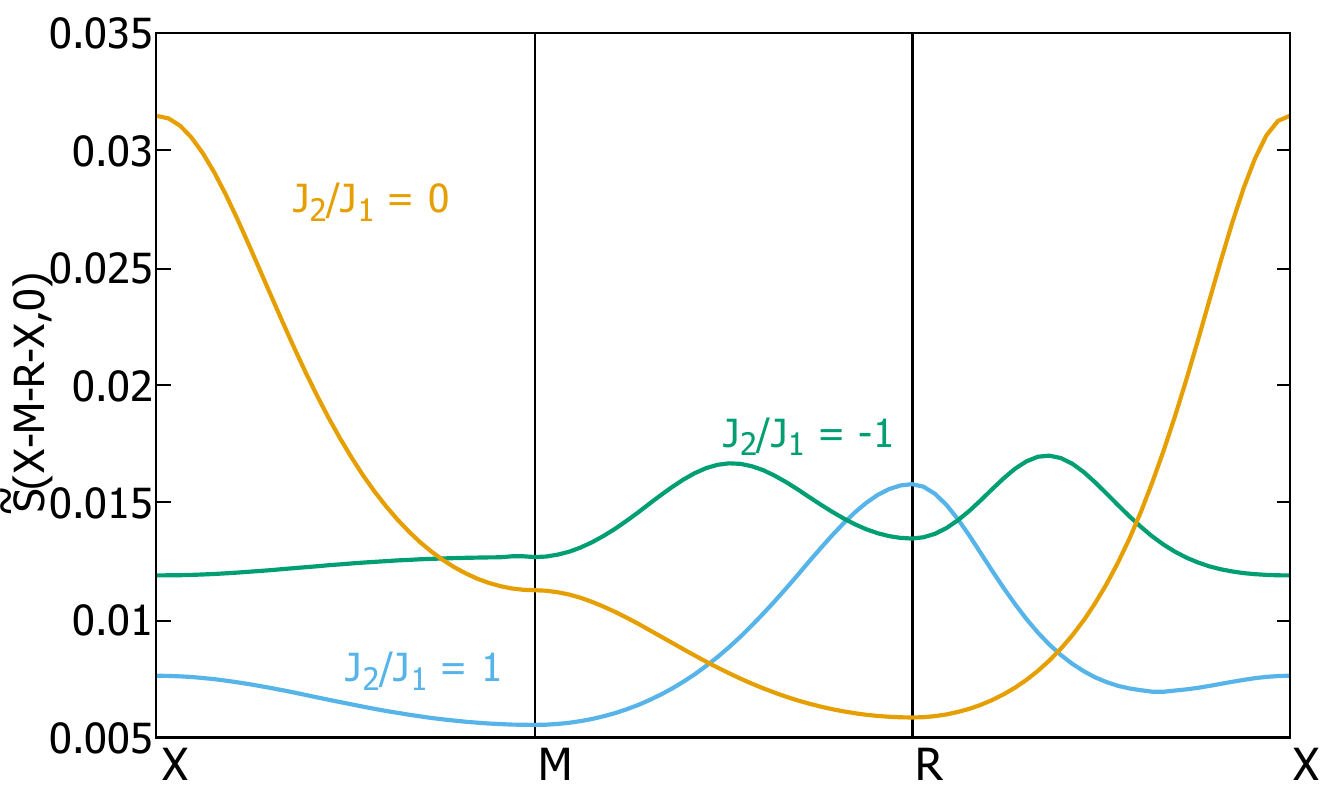}
   \end{center}
  \caption{%
Dimensionless dynamic structure factor $\tilde S ( \bd{k} ,0) = S ( \bd{k} , 0 ) | J_1 |
 \sqrt{b_0^{\prime}}$ 
given in Eq.~(\ref{eq:S3Delta3}) for a three-dimensional spin-$1/2$
Heisenberg magnet in a simple cubic lattice with nearest-neighbor exchange $J_1$ and
next-nearest-neighbor exchange $J_2$ at infinite temperature.  The plot is 
along the closed path in the first Brillouin zone $\bm{X} - \bm{M} - \bm{R} - \bm{X}$ 
described in the text
for $J_2 / J_1 = -1,0,1$ (green, orange, blue), see also  Fig. \ref{fig:dissenergy3dpiplane}.
}
\label{fig:dynstruc3dpathXMRX}
\end{figure}
For $J_2/J_1 < 0$ the maximum of $\tilde{\Delta} ( \bd{k} , 0 )$
at the corner
$\bm{R} = \frac{\pi}{a}\big(1,1,1\big)$ of the Brillouin zone is more stable
than for $J_2 / J_1 > 0$.
Furthermore, a two-peak structure emerges at 
$\bm{R}$ and $\bm{X} = \frac{\pi}{a}\big(0,0,1\big)$ 
with $\bm{M} = \frac{\pi}{a}\big(0,1,1\big)$ remaining a saddle point. The degeneracy 
point where the peaks at $\bm{X}$ and $\bm{R}$ 
have equal height 
is $J_2/J_1 = \mu_- \approx -0.97$,  in contrast to low dimensions where  $ \mu_- < - 1$. 
For positive  $J_2/J_1$ a simple crossover from $\bm{R}$ to $\bm{M}$ takes place 
at $J_2/J_1 = \mu_+ \approx 0.32$. 
Using Eq.~(\ref{eq:dynstruclargemomenta}) the low-frequency limit  $\tilde S (\bd{k} ,0)$
of the dynamic structure factor becomes
\begin{equation}
\tilde S (\bd{k} ,0) = \frac{b'_0}{\pi \tilde \Delta(\bm{k},0)},
 \label{eq:S3Delta3}
\end{equation}
which is shown in Fig.~\ref{fig:dynstruc3dpathXMRX}  for  $J_2/J_1 = -1, 0, 1$  
along the closed path $\bm{X} - \bm{M} - \bm{R} - \bm{X}$.  
For $J_2 / J_1 =1$ we obtain a second peak at $\bm{R}$. 
On the other hand, for $J_2 / J_1 = -1$  the dynamic structure factor exhibits local minima at
$\bm{R}$ and  $\bm{X}$ while assuming intermediate maxima on the paths $\bm{X} - \bm{R}$ and $\bm{M} - \bm{R}$.

\subsection{Common features in all dimensions}

To conclude this section, let us summarize the
robust features of the
dissipation energy 
 $\Delta(\bd{k},i\omega)$ and the resulting dynamic structure factor $S(\bm{k},\omega)$ 
at infinite temperature which are independent of the dimensionality of the system.
For negative $J_2 / J_1$ these quantities are less sensitive to the
next-nearest neighbor coupling $J_2$ than for positive $J_2 / J_1$.
In particular, for  $J_2 / J_1 < 0 $ 
 the corner of the Brillouin zone $\bd{R}  = ( \pi /a , \ldots , \pi/a )$
remains a maximum of 
$\Delta ( \bd{k} , \omega )$ -- and hence a minimum of $S ( \bd{k} , \omega )$ --
in a larger range of $| J_2 / J_1 |$ than for  $J_2 / J_1 > 0$.
For sufficiently large negative $J_2/J_1 < \mu_2(d)< 0 $ the
function $\Delta(\bd{k},i\omega)$ develops a second local maximum at a wavevector $\bd{Q}$ distinct from $\bd{R}$.
The corresponding dynamic structure factor $S ( \bd{k} , \omega )$ then exhibits  a local maximum
somewhere on a path connecting ${\bd{R}}$ to $\bd{Q}$.
This structure also  persists for $J_2/J_1 < - 1$. 
In the case of $J_2/J_1 > 0$ the position of the global maximum of $\Delta(\bd{k},i\omega)$ changes at  $J_2/J_1 = \mu_+(d)$, which leads for $J_2 / J_1 > \mu_+$ to a  short-wavelength peak 
of  $S(\bm{k},\omega)$ at $\bd{R}  = ( \pi /a , \ldots , \pi/a )$.
We conclude that for positive $J_2 /J_1 > \mu_+ ( d )$
the dynamic structure factor exhibits in all dimensions
a second peak  at the corner $\bd{R}$ of the first Brillouin zone.
This  peak  is absent for  negative $J_2 / J_1$, where in the regime $J_2 / J_1 < \mu_2 ( d ) < 0$
the dynamic structure factor exhibits local maxima along lines connecting local minima.

\section{Summary and conclusions}

 \label{sec:summary}

In this work we have studied the spin dynamics of 
quantum Heisenberg models with arbitrary spin-rotationally invariant exchange couplings 
by means of a 
new variant of the functional renormalization  group approach to quantum spin systems 
proposed  in 
Ref.~[\onlinecite{Krieg19}] and further developed in Refs.~[\onlinecite{Tarasevych18,Goll19,Goll20}].
In our quest to establish the SFRG as a useful tool
for calculating the spin dynamics of Heisenberg magnets  
without long-range magnetic order
we have encountered a number of
challenging technical problems which required 
non-trivial modifications of the established FRG formalism \cite{Berges02,Pawlowski07,Kopietz10,Metzner12,Dupuis21}:

\begin{enumerate}

\item
First of all,  we have avoided  the problem of the
non-existence of the Legendre transform of the generating functional of the
connected  correlation functions of an isolated spin
by introducing a hybrid functional $\Gamma_\Lambda [\bd{m}^c, \bd{\eta}^q]$ 
[see Eq.~(\ref{eq:Gammahybrid})]
where the
static (classical) fluctuations associated with the magnetization field
$\bd{m}^c$ are treated  differently from the dynamic (quantum) fluctuations associated with the
exchange field $\bd{\eta}^q$. 
Our construction is motivated by the fact that
in the classical  sector the Legendre transform of the generating functional of static 
spin correlation functions
is well-defined even for vanishing exchange couplings.
Moreover, we know from previous calculations \cite{Krieg19} that
a Legendre transform  to classical propagator-irreducible vertices yields better results for thermodynamic quantities than
a formulation in terms of interaction-irreducible 
vertices~\cite{Vaks68,Vaks68b,Izyumov88}. 

\item
Another technical subtlety of our approach is that
at finite frequencies we define the notion of
irreducibility with respect to the flowing inverse 
static propagator $\tilde J_\Lambda(\bm{q}) = G^{-1}_\Lambda(\bm{q})$ 
instead of the deformed bare  exchange coupling. 
This results in a convenient parametrization of 
$G(\bm{k},i\omega)$ which  is crucial for
implementing the  restoration of ergodicity for any finite value of the exchange couplings. 

\item
To obtain a closed system of FRG flow equations for the static self-energy and the
irreducible dynamic susceptibility 
which is compatible with the Ward identities due to
spin-rotational invariance and the ergodicity for finite exchange couplings, 
we had 
to take the flow of the three-spin and four-spin vertices
into account. 
We have done this with the help of 
the Ward identity $G_\Lambda( \bd{k} =0 ,i\omega \neq  0) = 0$
and the continuity condition   $G_\Lambda(\bd{k} \neq 0, i\omega \rightarrow 0) = G_\Lambda(\bm{k},0)$ due to  ergodicity.

\item
By assuming that the static 
spin correlations can be determined by some other method 
(such as a controlled high-temperature expansion)
we have been able to
transform the flow equation (\ref{eq:Piflow3})
for the irreducible dynamic susceptibility 
into a closed integral equation (\ref{eq:Deltaint}) 
for the dissipation energy $\Delta ( \bd{k} , i \omega )$
which determines the dynamic spin-spin correlation function 
via Eq.~(\ref{eq:Gall}).

\end{enumerate}

Although we have preliminary evidence \cite{Tarasevych21} that
our integral equation (\ref{eq:Deltaint})  
can be used to calculate
the low-frequency spin dynamics in the entire paramagnetic regime,
in this work we have focused on the
high temperature regime $T \gg | J |$
where the
static spin-spin correlation function $G ( \bd{k} )$ can be 
obtained via a controlled expansion in powers of $J / T$.
We use the resulting $G ( \bd{k})$ 
as an input 
for our integral equation (\ref{eq:Deltaint}) for the dissipation energy
$\Delta ( \bd{k} , i \omega )$.
We emphasize that our approach does not make any a priori assumptions 
regarding the existence of normal spin diffusion, nor does it rely on an extrapolation of a high-frequency (short-time) expansion.

We have used our approach to calculate 
the spin-diffusion coefficient ${\cal{D}}$
in three-dimensional Heisenberg magnets  with
nearest-neighbor and next-nearest-neighbor exchange
on simple cubic and body centered cubic lattices.
Our numerical results for ${\cal{D}}$ are
by a factor of up to two smaller than older
predictions based on the extrapolation of the
short-time expansion~\cite{DeGennes58,Mori62,Bennett65,Redfield68,Morita72,Morita75,Kopietz93,Boehm94}, although the experimental result for ${\cal{D}}$
reported in Ref.~[\onlinecite{Labrujere82}] is still somewhat smaller than our prediction.
Furthermore, contrary to these older approaches~\cite{DeGennes58,Mori62,Bennett65,Redfield68,Morita72,Morita75,Kopietz93,Boehm94},  
our method predicts anomalous diffusion in reduced dimensions $d \leq 2$.
In particular, in $d=1$ 
our result  ${\cal{D}} ( \omega )
 \propto | \omega |^{-1/3}$   for the frequency-dependence of 
the generalized diffusion coefficient
agrees with recent investigations of spin chains~[\onlinecite{Ljubotina17,Gopalakrishnan19a,Gopalakrishnan19b,Nardis19,Nardis20,Bulchandani20,Dupont20,Nardis21,Bulchandani21}],
at least in cases where convergence 
of different numerical and analytic approaches has been 
achieved. Finally,  we have also used
our approach to calculate the 
full $\bm{k}$-dependence of the dynamic structure factor $S ( \bd{k} , \omega)$ at high temperatures, 
which allows us to relate the  short-distance behavior
of $S ( \bd{k} , \omega)$
to the nature of competing exchange interactions.

The methods developed in this work can be extended in many directions.
Although here we have focused on the solution of the integral equation (\ref{eq:Deltaint})
for the dissipation energy $\Delta ( \bd{k} , i \omega )$
at high temperatures,
we have preliminary evidence \cite{Tarasevych21} that
Eq.~(\ref{eq:Deltaint}) gives sensible results in the entire paramagnetic regime.
In particular, by solving this  integral equation 
for temperatures slightly above the critical temperature 
we can  investigate the critical spin dynamics of Heisenberg magnets.
Our method can also be used 
as an unbiased approach to frustrated quantum spin systems 
where even the calculation of thermodynamics like the phase diagram poses a serious challenge. In this context FRG approaches employing
representations of the spin operators  in terms of 
Abrikosov pseudo-fermions~\cite{Reuther10,Reuther11,Reuther11a,Buessen16,Thoenniss20,Kiese20}
have been successfully used to calculate 
static ground state properties of quantum spin systems.
On the other hand, dynamic properties such as the dynamic structure
factor have  so far not been calculated  
within the   pseudofermion FRG; in fact, at this point it is not clear 
whether the corresponding technical problems will be solved in the near future.
Moreover, at finite temperatures the pseudofermion FRG becomes
inaccurate because it introduces unphysical Hilbert space sectors.
Although this problem can be elegantly  avoided 
using an SO(3)-symmetric representation of the 
spin operators in terms of Majorana fermions \cite{Niggemann20},
this pseudo-Majorana FRG exhibits an unphysical divergence in the limit of vanishing temperature.
In contrast, our SFRG approach allows us to calculate the
spin-spin correlation function $G ( \bd{k} , \omega )$  for vanishing and 
finite frequencies at all temperatures where the spin-rotational invariance is not
spontaneously broken. In fact, by numerically solving the
flow equations~(\ref{eq:Piflow3}) and (\ref{eq:Sigmaflow3})
we can in principle obtain both the static 
spin self-energy  $\Sigma ( \bd{k} )$ and the dynamic dissipation energy
$\Delta ( \bd{k} , i \omega )$.  
Although the direct 
numerical solution of the flow equations~(\ref{eq:Piflow3}) and (\ref{eq:Sigmaflow3})
is beyond the scope of this work, we believe that the numerical solution of these equations
will be very rewarding because it
will allow us to obtain
the dynamic structure factor $S ( \bd{k} , \omega )$ of frustrated spin systems 
at low temperatures $T \ll |J|$, 
a quantity which is not accessible with
pseudofermion FRG methods~\cite{Reuther10,Reuther11,Reuther11a,Buessen16,Thoenniss20,Kiese20,Niggemann20}.

For completeness it should be mentioned  
that the idea of working directly with
physical spin correlation functions is also central to the
equation of motion approach for quantum spin systems
pioneered by Bogolyubov, Tyablikov, and others \cite{Bogoliubov59,TahirKheli62,Akhiezer68}.
In this approach the infinite hierarchy of equations of motion for the
spin correlation functions is closed by some 
decoupling procedure for correlation functions involving more than two spins, resulting in a closed self-consistency equation for the spin-spin correlation function.
A notable example is given by the 
Tyablikov-decoupling \cite{Bogoliubov59,TahirKheli62,Akhiezer68}  
which for $S = 1/2$ Heisenberg ferromagnets amounts to approximating a  
mixed three-spin correlation function by a product of a transverse two-spin correlation 
function and the magnetization. While in the ordered phase 
this seems to be a reasonable 
approximation, it 
is only of limited use in the paramagnetic zero-field limit, especially when we are interested in the dynamics. 
An important difference between our SFRG approach and 
methods based on the decoupling of equations of motion for spin correlation functions
is that
SFRG is formulated  in terms of irreducible vertices, which provide a more compact parametrization of 
higher order spin correlations and allow for sophisticated truncation strategies
compatible with the constraints 
imposed by the Ward identities and the ergodicity of the system.

Experimentally, the dynamic structure factor  can be measured via inelastic neutron scattering.
Moreover, the nuclear spin-lattice relaxation rate in magnetic insulators measured in 
nuclear magnetic resonance (NMR) experiments 
is proportional to a weighted Brillouin zone average of $S ( \bd{k} , \omega_N )$,
where the NMR frequency $\omega_N$ is usually much smaller 
than the exchange couplings \cite{Beeman68}. Our results for $S ( \bd{k} , \omega )$ 
presented in Sec.~\ref{sec:dissip} can therefore be used to calculate the high-temperature behavior
of the NMR relaxation rate
in Heisenberg magnets.

\section*{Acknowledgments}
This work was financially supported by the Deutsche Forschungsgemeinschaft (DFG)
through project KO 1442/10-1.

\begin{appendix}

\section*{APPENDIX A:  
High-temperature spin diffusion on a bcc lattice}

\setcounter{equation}{0}
\renewcommand{\theequation}{A\arabic{equation}}

In this appendix we give some technical details of the solution of the
integral equation (\ref{eq:Deltaint}) for the dissipation energy
$\Delta ( \bd{k} , i \omega)$
on a body-centered cubic lattice at $T = \infty$ including next-nearest neighbor exchange. The geometry is shown
in Fig.~\ref{fig:bccgeometry}. The self-energy contribution to the relevant 
high-temperature limit of the kernel $V ( \bd{k} , \bd{q} )$ in
Eq.~(\ref{eq:Khigh}) can then be written as
\begin{eqnarray}
 & & 2\Sigma_2( \bd{q} ) - \Sigma_2( \bd{q + k} ) - \Sigma_2(\bm{q - k})  \nonumber 
 \\ 
 & = & \frac{2J^2_1}{3} 
 \left[ 2 \gamma^{\rm{bcc}}_{\bm{q}} - \gamma^{\rm bcc}_{\bm{q + k}} 
 - \gamma^{\rm bcc}_{\bm{q - k}} \right]
 \nonumber \\ &+ &
  \frac{J^2_2}{2} \left[ 2\gamma_{\bm{q}} - \gamma_{\bm{q + k}} - \gamma_{\bm{q - k}} \right],
\end{eqnarray}
where  the form factors $\gamma^{\rm bcc}_{\bd{k}}$ and $\gamma_{\bd{k}}$ are defined
in Eqs.~(\ref{eq: bccJ_1}) and (\ref{eq:gammak3def}).
Analogous to Eq.~(\ref{eq:tildeDelta}), it is convenient to
introduce again the dimensionless quantities
$\tilde \Delta(\bd{k}, i {\omega})  = \Delta(\bd{k}, i\omega) / (|J_1|\sqrt{b'_0}) $ and
 $\tilde \omega =  \omega / ( |J_1|\sqrt{b'_0})$.
The solution of our integral equation (\ref{eq:Deltaint})
can then be expressed in terms of six independent form factors,
\begin{eqnarray}
\tilde \Delta(\bd{k},i {\omega}) & = & (1 - \gamma^{\rm bcc}_{\bm{k}})  \tilde \Delta^{\rm bcc}_{1}(i {\omega}) + (1 - \gamma_{\bm{k}})\tilde \Delta^{\rm sc}_{1}(i {\omega})
\nonumber \\ & + & 
(1 - \gamma^{\rm bcc}_{2\bm{k}})\tilde \Delta^{{\rm bcc}, \parallel}_2(i {\omega})
+ (1 - \gamma^{\perp}_{\bm{k}})\tilde \Delta^{{\rm sc},\perp}_2(i {\omega})
\nonumber \\ & + &
 (1 - \gamma^{\rm bcc,sc}_{\bm{k}})\tilde \Delta^{\rm bcc, sc}_2(i {\omega})
+ (1 - \gamma_{2\bm{k}})\tilde \Delta^{{\rm sc},\parallel}_2(i {\omega}),\nonumber \\ & &
\end{eqnarray}
where  the off-diagonal form factor $\gamma_{\bd{k}}^{\bot}$
can be obtained by setting $d=3$ in the general definition (\ref{eq:gammabot}),
 \begin{eqnarray}
 \gamma_{\bd{k}}^{\bot} & = & \frac{1}{3}
 \bigl[ \cos ( k_x a ) \cos ( k_y a ) + \cos ( k_y a ) \cos ( k_z a )
 \nonumber
 \\
 & & \hspace{3mm}
 + \cos ( k_z a ) \cos ( k_x a ) \bigr],
 \label{eq:gammabot3}
 \end{eqnarray}
and the mixed  form factor $ \gamma^{\rm bcc,sc}_{\bm{k}} $ is given by
\begin{eqnarray}
& & \gamma^{\rm bcc,sc}_{\bm{k}} =  \frac{1}{3}\Big[\cos\Big(\frac{3k_x a}{2}\Big)\cos\Big(\frac{k_y a}{2}\Big) \cos\Big(\frac{k_z a}{2}\Big)
\nonumber \\ & &   \hspace{15mm} 
 + (x \leftrightarrow z) + (x \leftrightarrow y) \Big].
\end{eqnarray}
Introducing a short notation for the ratio of exchange couplings,
 \begin{equation}
 \mu = J_2 / J_1,
 \end{equation}
the system (\ref{eq:falpha}) of self-consistency equations then reduces to the following six coupled 
equations,
 \begin{subequations}
\begin{eqnarray}
\tilde{\Delta}^{\rm bcc}_{1}(i {\omega}) & = & \frac{4}{3b'_0}\int_{\bm{q}}
 \frac{\gamma^{\rm bcc}_{\bm{q}}}{|\tilde{\omega}| + \tilde \Delta(\bm{q},i {\omega})} 
 +8 \int_{\bm{q}}\frac{1 + 3\gamma_{\bm{q}} + 3\gamma^\perp_{\bm{q}}}{|\tilde{\omega}| 
 + \tilde \Delta(\bm{q},i {\omega})}  
  \nonumber \\ & &
- \tilde{\Delta}^{\rm bcc,sc}_{2}(i {\omega})   - 2 \tilde{\Delta}^{{\rm bcc},\parallel}_{2} (i {\omega}) ,
\end{eqnarray}
\begin{eqnarray}
\tilde{\Delta}^{{\rm bcc},\parallel}_{2}(i {\omega}) 
 & = &  -4\int_{\bm{q}}\frac{\gamma^{\rm bcc}_{2\bm{q}}}{|\tilde{\omega}| + \tilde 
 \Delta(\bm{q},i {\omega})},
\end{eqnarray}
\begin{eqnarray}
\tilde{\Delta}^{\rm bcc,sc}_{2}(i {\omega}) 
 =  -24\mu\int_{\bm{q}}\frac{\gamma^{\rm bcc,sc}_{\bm{q}}}{|\tilde{\omega}| 
 + \tilde \Delta(\bm{q},i {\omega})},
\end{eqnarray}
\begin{eqnarray}
\tilde{\Delta}^{\rm sc}_{1}(i {\omega}) & = &  
 -12\int_{\bm{q}}\frac{\gamma_{\bm{q}}}{|\tilde{\omega}|
 + \tilde \Delta(\bm{q},i {\omega})} 
 \nonumber
 \\
 & & + 24\mu \int_{\bm{q}}\frac{\gamma^{\rm bcc}_{\bm{q}}}{|\tilde{\omega}| 
 + \tilde \Delta(\bm{q},i {\omega})} - \tilde{\Delta}^{\rm bcc,sc}_{2}(i {\omega})
\nonumber \\ & &
 + \frac{\mu^2}{b'_0}\int_{\bm{q}}
 \frac{\gamma_{\bm{q}}}{|\tilde{\omega}| + \tilde \Delta(\bm{q},i {\omega})}
\nonumber \\  & &
+ 6\mu^2\int_{\bm{q}}\frac{1 +  \gamma_{2\bm{q}} + 4\gamma^{\bot}_{\bm{q}}}{|\tilde{\omega}| 
 + \tilde \Delta(\bm{q},i {\omega})},
\end{eqnarray}
\begin{eqnarray}
\tilde{\Delta}^{{\rm sc},\bot}_{2}(i {\omega}) = -12(1 + \mu^2)\int_{\bm{q}}\frac{\gamma^\perp_{\bm{q}}}{|\tilde{\omega}| + \tilde \Delta(\bm{q},i {\omega})},
\end{eqnarray}
\begin{eqnarray}
\tilde{\Delta}^{{\rm sc}, \parallel}_{2}(i {\omega}) = -3\mu^2\int_{\bm{q}}
 \frac{\gamma_{2\bm{q}}}{{|\tilde{\omega}| + \tilde \Delta(\bm{q},i {\omega})}}.
\end{eqnarray}
 \end{subequations}
According to Eq.~(\ref{eq:Ddifdef})
the spin-diffusion coefficient ${\cal{D}}$ can then be obtained from the 
term of order $k^2$  in the expansion 
of  $\Delta ( \bd{k} , 0 ) =  | J_1 | \sqrt{ b_0^{\prime}} \tilde{\Delta} ( \bd{k} , 0 )$
in powers of the momentum, so that we finally
arrive at the following expression for the spin-diffusion coefficient
in the limit of infinite temperature,
\begin{eqnarray}
 {\cal{D}} & = & \frac{|J_1|\sqrt{b'_0}a^2}{6} \Big[ \frac{3}{4}\tilde{\Delta}^{\rm bcc}_{1}(0) 
+ 3\tilde{\Delta}^{{\rm bcc},\parallel}_{2}(0) + \frac{11}{4}\tilde{\Delta}^{\rm bcc,sc}_{2}(0)
 \nonumber \\ & & 
  + \tilde{\Delta}^{\rm sc}_{1}(0)  + 4 \tilde{\Delta}_{2}^{{\rm sc}, \parallel}(0) 
 + 2 \tilde{\Delta}_{2}^{{\rm sc}, \bot}(0)
 \Big] .
\end{eqnarray}

\section*{APPENDIX B:  
High-temperature spin diffusion on hypercubic lattices}

\setcounter{equation}{0}
\renewcommand{\theequation}{B\arabic{equation}}

Here we give some technical details of the solution of the
integral equation~(\ref{eq:Deltaint}) on
hypercubic lattice in dimensions $d = 1,2,3$ for a Heisenberg model with nearest-neighbor exchange $J_1$ and next-nearest-neighbor exchange~$J_2$.

\subsection{One dimension}
Setting again $\mu = J_2 / J_1$, the Fourier transform of the 
exchange interaction in $d=1$ is 
\begin{equation}
J(\bm{k}) = 2J_1 [ \cos(k_x a)  + \mu \cos(2k_x a)].
\end{equation}
At high temperatures
the self-energy contribution to the
kernel $V ( \bd{k} , \bd{q} )$ in
Eq.~(\ref{eq:Khigh}) can then be written as
 \begin{eqnarray}
& &  2 \Sigma_2 ( \bd{q} ) - \Sigma_2 ( \bd{q} + \bd{k} ) -  \Sigma_2 ( \bd{q} - \bd{k} ) 
 \nonumber
 \\
 & = & \frac{J_1^2}{6}\Big[ 2\cos(q_x a) - \cos((q_x + k_x)a)  - \cos((q_x - k_x)a) 
 \nonumber \\ & &
   \hspace{5mm} + \mu^2\bigl[ 2\cos(2q_x a) - \cos(2(q_x + k_x)a)  
 \nonumber
 \\ & & \hspace{12mm} - \cos(2(q_x - k_x)a) \bigr] \Big]. 
 \end{eqnarray}
The solution of the integral equation (\ref{eq:Deltaint})
in $d=1$ can then be written as
$\Delta ( \bd{k} , i \omega ) = | J_1 | b_0^{\prime} \tilde{\Delta} ( {k}_x , i \omega )$,
where the dimensionless function $\tilde{\Delta} ( k_x , i \omega )$ can be expressed in terms of four different form factors,
\begin{eqnarray}
\tilde{\Delta}(k_x,i {\omega}) = \sum_{j = 1}^4 [ 1 - \cos(j k_x a)]
 \tilde{\Delta}_j(i {\omega}).
\label{eq: 1dnnn}
\end{eqnarray}
With the abbreviations $\int_{q_x} = a  \int_{ -\pi /a}^{\pi /a } \frac{d q_x}{2 \pi}$
and $\tilde{\omega} = \omega /( | J_1 | b_0^{\prime} )$
the self-consistency equations (\ref{eq:falpha}) for the amplitudes reduce to
 \begin{subequations}
 \label{eq:1damplitudes}
\begin{eqnarray}
\tilde{\Delta}_1(i{\omega}) & = & 
\frac{1}{3b'_0}\int_{q_x}\frac{\cos(q_x a)}{|\tilde{\omega}| 
 + \tilde \Delta( q_x,i {\omega})} 
 \nonumber
 \\
& + & 2 \int_{{q}_x}\frac{1 + \cos(2q_x a)}{{|\tilde{\omega}| + \tilde \Delta( q_x ,i{\omega})}} 
- \tilde{\Delta}_3(i {\omega}),
\end{eqnarray}
\begin{eqnarray}
 \tilde{\Delta}_2(i {\omega}) & = & \Big(\frac{\mu^2}{3b'_0} - 1 \Big)\int_{{q}_x}\frac{\cos(2q_x a)}{{|\tilde{\omega}| + \tilde \Delta( q_x ,i {\omega})}}  
\nonumber \\ & + &  2 \mu \int_{{q}_x}\frac{  \cos(q_x a) + \mu}{{|\tilde{\omega}| 
+ \tilde \Delta(q_x ,i {\omega})}} 
- \tilde{\Delta}_3(i {\omega})   - 2\tilde{\Delta}_4(i {\omega}),
 \nonumber
 \\
& &
\end{eqnarray}
\begin{eqnarray}
\tilde{\Delta}_3(i {\omega}) = -2\mu \int_{{q}_x}\frac{\cos(3q_x a)}{{|\tilde{\omega}| + \tilde \Delta( q_x ,i{\omega})}},
\end{eqnarray}
\begin{eqnarray}
\tilde{\Delta}_4(i {\omega}) = -\mu^2\int_{{q}_x}\frac{\cos(4q_x a)}{{|\tilde{\omega}| + \tilde \Delta( q_x ,i{\omega})}}.
\end{eqnarray}
 \end{subequations}
For small frequencies  $|\tilde \omega| \ll 1$ we obtain for the amplitudes to
leading order
 \begin{subequations}
 \begin{eqnarray}
\tilde{\Delta}_1(i {\omega}) & = &  \Big(4 + \frac{1}{3b'_0} + 2\mu\Big)
 \left[ \frac{3b'_0}{2 ( 1 + 4\mu^2) | \tilde{\omega} | }\right]^{1/3} ,
 \\
\tilde{\Delta}_2(i {\omega}) & = &  \Big(\frac{\mu^2}{3b'_0} -1 + 4\mu^2 + 4\mu\Big)
 \left[ \frac{3b'_0}{2 ( 1 + 4\mu^2) | \tilde{\omega} | }\right]^{1/3} ,
 \nonumber
 \\
 & &
\\
\tilde{\Delta}_3 (i {\omega}) & = &  - 2\mu
 \left[ \frac{3b'_0}{2(1 + 4\mu^2) | \tilde{\omega} | }\right]^{1/3} ,
\\
\tilde{\Delta}_4(i {\omega}) & = &  - \mu^2\left[ 
 \frac{3b'_0}{2(1 + 4\mu^2) | \tilde{\omega} | }\right]^{1/3} .
\end{eqnarray}
 \end{subequations}
Substituting these expressions into Eq.~(\ref{eq: 1dnnn}) and expanding
to second order in
$k_x$ we obtain the anomalous spin-diffusion coefficient in $d=1$,
 \begin{eqnarray}
{\cal{D}}(i\omega) 
 & = & \frac{|J_1|\sqrt{b'_0}a^2}{2} \Bigl[ \tilde{\Delta}_{1}(i\tilde{\omega}) + 
 4 \tilde{\Delta}_{2}(i\tilde{\omega}) 
 \nonumber
 \\
 & & \hspace{15mm}
+ 9 
 \tilde{\Delta}_{3}(i\tilde{\omega}) + 16\tilde{\Delta}_{4} (i\tilde{\omega})  \Bigr]
 \nonumber \\
& = & \left[ \frac{|J_1|(1 + 4\mu^2)^2}{144|\omega|}\right]^{1/3}|J_1|a^2
 \nonumber
 \\
 & = & \left[ \frac{  \sqrt{ J_1^2  + 4  J_2^2} }{144|\omega|}\right]^{1/3}  \sqrt{   J_1^2 + 4 J_2^2} a^2 . 
 \hspace{7mm}
 \label{eq:D1res2}
\end{eqnarray}
\subsection{Square lattice}

On a square lattice the Fourier transform of the exchange couplings 
with nearest-neighbor exchange $J_1$ and next-nearest-neighbor exchange $J_2 = \mu J_1$ is
\begin{equation}
J(\bm{k}) =  4J_1 [ \gamma_{\bd{k}} + \mu \gamma^{\bot}_{ \bd{k} } ],
\end{equation}
 where now
 \begin{eqnarray}
 \gamma_{\bd{k}} & = & \frac{1}{2}  \left[ \cos ( k_x a ) + \cos ( k_y a ) \right],
 \\
  \gamma^{\bot}_{ \bd{k} } & = & \cos ( k_x a ) \cos ( k_y a ).
 \end{eqnarray}
The self-energy contribution to the high temperature
kernel $V ( \bd{k} , \bd{q} )$ defined in
Eq.~(\ref{eq:Khigh}) can then be written as
 \begin{eqnarray}
& &  2 \Sigma_2 ( \bd{q} ) - \Sigma_2 ( \bd{q} + \bd{k} ) -  \Sigma_2 ( \bd{q} - \bd{k} ) 
 \nonumber
 \\
 & = & \frac{J_1^2}{3}\Big[  2\gamma_{\bm{q}} - \gamma_{\bm{q + k}} - \gamma_{\bm{q - k}}  
 \nonumber \\ & & \hspace{4mm}
 + \mu^2\left(  2\gamma^{\bot}_{\bm{q}} - \gamma^{\bot}_{\bm{q + k}} - \gamma^{\bot}_{\bm{q - k}} \right)\Big].
 \end{eqnarray}
The solution of the integral equation (\ref{eq:Deltaint})
can be expressed in terms of five different form factors,
 \begin{eqnarray}
  \tilde \Delta ( \bd{k}, i{\omega} ) & = & ( 1 - \gamma_{\bd{k}} )   \tilde{\Delta}_1(i {\omega}) + 
 ( 1 - \gamma_{2\bd{k}} ) \tilde{\Delta}_{2}^{\parallel}(i {\omega})
 \nonumber \\ & &
 +  ( 1 - \gamma^{\bot}_{\bd{k}} ) \tilde{\Delta}_{2}^{\bot}(i {\omega})  +  ( 1 - \gamma^{\bot}_{2\bd{k}} ) \tilde{\Delta}_{2,2}^{\parallel}(i {\omega}) 
 \nonumber \\  & &
 + ( 1 - \gamma^{(2,1)}_{\bd{k}} )\tilde{\Delta}_{2,1}(i {\omega}),
 \label{eq:2dnnn}
 \end{eqnarray}
where we have introduced the mixed form factor
 \begin{equation}
 \gamma^{(2,1)}_{\bm{k}} = \frac{1}{2}\Big[\cos(2 k_x a) \cos(k_y a) + \cos(k_x a) \cos(2 k_y a) \Big].
 \end{equation}
The self-consistency equations (\ref{eq:falpha}) for the amplitudes can then be written
in the following form
 \begin{subequations}
\begin{eqnarray}
\tilde{\Delta}_1(i {\omega}) & = & \frac{2}{3b'_0}\int_{\bm{q}}\frac{\gamma_{\bm{q}}}{|\tilde{\omega}| + \tilde \Delta(\bm{q},i {\omega})}
\nonumber \\ & & 
+ 4 \int_{\bm{q}}\frac{1 + \gamma_{2 \bm{q}} + 2 \gamma^{\bot}_{\bm{q}}}{|\tilde{\omega}| + \tilde \Delta(\bm{q},i {\omega})}  - \tilde \Delta_{2,1}(i {\omega}),
 \hspace{7mm}
\end{eqnarray}
\begin{equation}
\tilde{\Delta}^{\parallel}_2(i {\omega}) = -2(1 + 2\mu^2)\int_{\bm{q}}\frac{\gamma_{2\bm{q}}}{|\tilde{\omega}| + \tilde \Delta(\bm{q},i {\omega})},
\end{equation}
\begin{eqnarray}
& & \tilde{\Delta}^{\bot}_2(i {\omega}) = \left( \frac{2\mu^2}{3b'_0} - 4\right)
 \int_{\bm{q}}\frac{\gamma^{\bot}_{\bm{q}}}{|\tilde{\omega}| + \tilde \Delta(\bm{q},i {\omega})} 
\nonumber \\ & &
+ 8\mu \int_{\bm{q}}\frac{\gamma_{\bm{q}}}{|\tilde{\omega}| + \tilde \Delta(\bm{q},i {\omega})}
+ 4\mu^2  \int_{\bm{q}}\frac{1}{|\tilde{\omega}| + \tilde \Delta(\bm{q},i {\omega})} 
\nonumber \\ & &
+ 8\mu^2 \int_{\bm{q}}\frac{\gamma_{2\bm{q}}}{|\tilde{\omega}| + \tilde \Delta(\bm{q},i {\omega})}
- \tilde{\Delta}_{2,1}(i {\omega}) - 2 \tilde{\Delta}_{2,2}^{\parallel}(i {\omega}), 
\nonumber \\ & &
\end{eqnarray}
\begin{equation}
\tilde{\Delta}_{2,2}^{\parallel}(i {\omega})  = -2\mu^2\int_{\bm{q}}\frac{\gamma^{\bot}_{\bm{2q}}}{|\tilde{\omega}| + \tilde \Delta(\bm{q},i {\omega})},
\end{equation}
\begin{equation}
\tilde{\Delta}_{2,1}(i {\omega}) = - 8\mu\int_{\bm{q}}\frac{\gamma^{(2,1)}_{\bm{q}}}{|\tilde{\omega}| + \tilde \Delta(\bm{q},i {\omega})}.
\end{equation} 
 \end{subequations}
For small frequencies $ | \omega  | \ll | J_1|$ the solution of the 
above equations is to leading logarithmic order given by
 \begin{subequations}
  \label{eq:2damplitudes}
 \begin{eqnarray}
\tilde{\Delta}_1(i{\omega}) & = & 2 \Big(\frac{1}{3b'_0} + 8 + 4 \mu\Big)
 \sqrt{ 
\frac{ 3b'_0\ln\Big(\frac{|J_1|\sqrt{1 + 2\mu^2}}{\sqrt{24\pi}|\omega|}\Big) }{2\pi( 1 + 2 \mu^2)}},
 \hspace{7mm}
 \nonumber
 \\
 & &
 \\
\tilde{\Delta}^{\parallel}_2(i {\omega}) & = & - 2 ( 1 + 2 \mu^2 )
\sqrt{ 
\frac{ 3b'_0\ln\Big(\frac{|J_1|\sqrt{1 + 2\mu^2}}{\sqrt{24\pi}|\omega|}\Big) }{2\pi( 1 + 2 \mu^2)}},
 \\
\tilde{\Delta}^{\bot}_2(i {\omega}) & = & 2 \left(  8 \mu^2 +  8 \mu + \frac{\mu^2}{3b'_0} - 2\right)
 \nonumber
 \\
 & &  \times 
\sqrt{ 
\frac{ 3b'_0\ln\Big(\frac{|J_1|\sqrt{1 + 2\mu^2}}{\sqrt{24\pi}|\omega|}\Big) }{2\pi( 1 + 2 \mu^2)}},
 \\
\tilde{\Delta}_{2,2}^{\parallel}(i {\omega}) & = &  -2\mu^2
\sqrt{ 
\frac{ 3b'_0\ln\Big(\frac{|J_1|\sqrt{1 + 2\mu^2}}{\sqrt{24\pi}|\omega|}\Big) }{2\pi( 1 + 2 \mu^2)}},
 \\
\tilde{\Delta}_{2,1}(i {\omega})  & = & -8\mu
\sqrt{ 
\frac{ 3b'_0\ln\Big(\frac{|J_1|\sqrt{1 + 2\mu^2}}{\sqrt{24\pi}|\omega|}\Big) }{2\pi( 1 + 2 \mu^2)}}.
\end{eqnarray}
 \end{subequations}
The resulting  anomalous diffusion coefficient on a square lattice is
\begin{eqnarray}
{\cal{D}}(i\omega) & = & \frac{|J_1|\sqrt{b'_0}a^2}{4}\Big[ \tilde{\Delta}_{1}(i {\omega}) + 
 4 \tilde{\Delta}_{2}^{\parallel}(i {\omega}) + 2 
 \tilde{\Delta}_{2}^{\bot}(i {\omega}) 
 \nonumber \\ & & \hspace{15mm}
 + 8  \tilde{\Delta}_{2,2}^{\parallel}(i {\omega})  + 5\tilde{\Delta}_{2,1}(i {\omega})\Big]  
   \nonumber \\ & =  &
 \sqrt{\frac{\ln\Big( \frac{|J_1|\sqrt{1 + 2\mu^2}}{\sqrt{24\pi}|\omega|}\Big)}{24\pi}}  | J_1|\sqrt{1 + 2\mu^2}a^2.
 \label{eq:Dres2d}
\end{eqnarray}
Keeping in mind that  $ | J_1|\sqrt{1 + 2\mu^2} = \sqrt{ J_1^2 + 2 J_2^2 }$,
we see that in the expansion of $\Delta ( \bd{k} , i \omega )$ to  order $k^2$ 
the next-nearest-neighbor interaction $J_2$ can be taken into account via the
following replacement of the
nearest-neighbor interaction,  
$| J_1 | \rightarrow \sqrt{ J_1^2 + (a^{\prime} / a )^2 J_2^2 }$,
where $a^{\prime}$ is the distance between next-nearest neighbors.
From Eq.~(\ref{eq:D1res2}) it is clear that this is also true in one dimension.
Effects depending on the sign of  $ J_2 / J_1$  can be only seen 
by expanding $\Delta ( \bd{k} , i \omega )$ beyond the leading order,
implying that these effects are only visible for momenta $ka = \mathcal{O}(1)$.

\subsection{Simple cubic lattice}

For a simple cubic lattice with nearest-neighbor exchange $J_1$ and
next-nearest-neighbor exchange $J_2 = \mu J_1$  
the Fourier transform of the exchange interaction is
\begin{equation}
J(\bm{k}) = 6J_1 [ \gamma_{\bm{k}} + 2\mu \gamma^{\bot}_{\bm{k}} ],
\end{equation}
where the form factors $\gamma_{\bd{k}}$ and
$\gamma^{\bot}_{\bd{k}}$ are defined in Eqs.~(\ref{eq:gammak3def}) and (\ref{eq:gammabot3}),
respectively. 
The self-energy contribution to the high temperature
kernel $V ( \bd{k} , \bd{q} )$ defined in
Eq.~(\ref{eq:Khigh}) is then
 \begin{eqnarray}
& &  2 \Sigma_2 ( \bd{q} ) - \Sigma_2 ( \bd{q} + \bd{k} ) -  \Sigma_2 ( \bd{q} - \bd{k} ) 
 \nonumber
 \\
 & = & \frac{J_1^2}{2}\Big[  2\gamma_{\bm{q}} - \gamma_{\bm{q + k}} - \gamma_{\bm{q - k}}  
 \nonumber \\ & & \hspace{3mm}
 + 2\mu^2\left( 2\gamma^{\bot}_{\bm{q}} - \gamma^{\bot}_{\bm{q + k}} - \gamma^{\bot}_{\bm{q - k}} \right) \Big].
 \end{eqnarray}
   \begin{figure}[tb]
 \begin{center}
 \centering
 \includegraphics[width=0.45\textwidth]{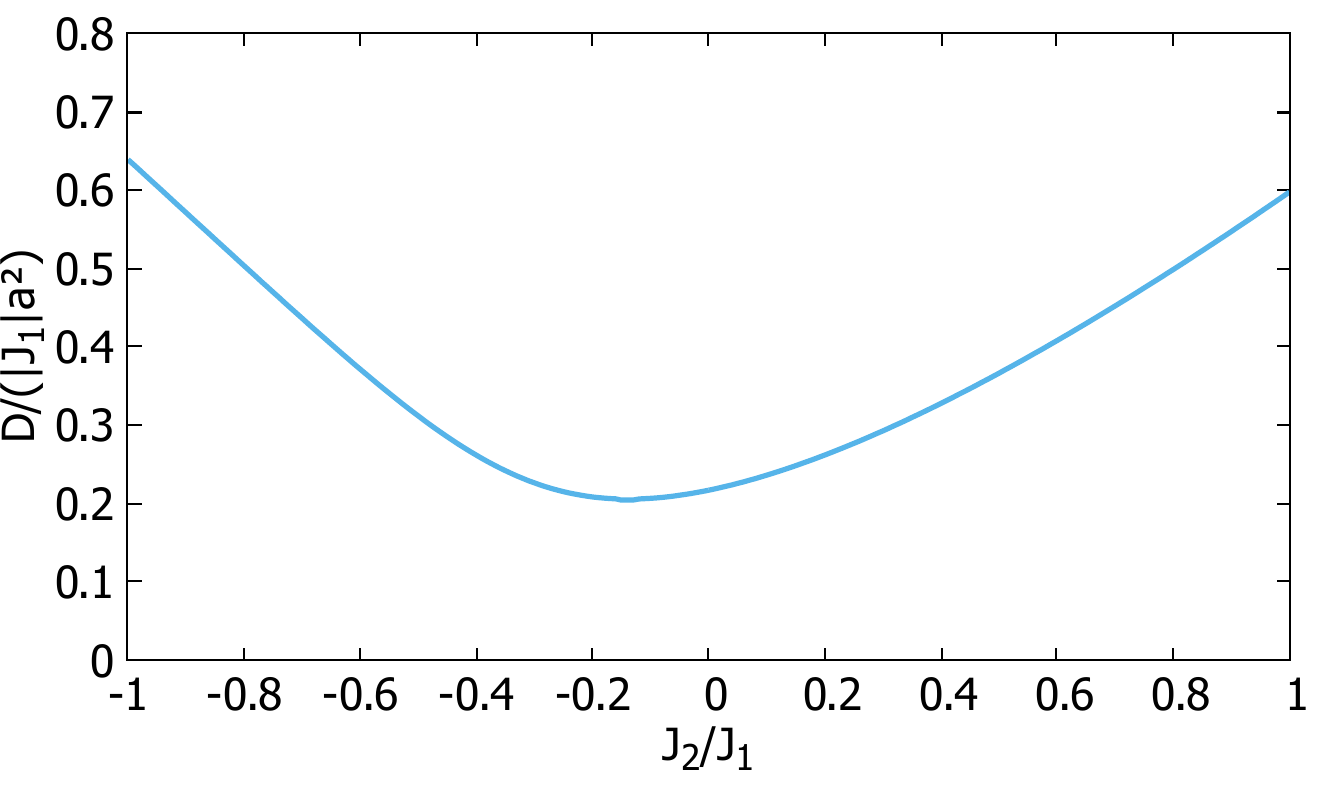}
   \end{center}
  \caption{%
Spin-diffusion coefficient $\mathcal{D}$ for a 
spin-$1/2$ Heisenberg model on a simple cubic lattice with nearest-neighbor interaction $J_1$and next-nearest-neighbor interaction $J_2$ as a function
of $J_2 / J_1$ for $T = \infty$.}
\label{fig:diffsc}
\end{figure}
At high temperatures, 
the solution $\Delta ( \bd{k} , i \omega )$ of the integral equation (\ref{eq:Deltaint})
can be expressed in terms of seven different form factors. 
Hence, the dimensionless dissipation energy
 $\tilde{\Delta} ( \bd{k} , i \omega ) = \Delta ( \bd{k} , i \omega ) / ( | J_1 | \sqrt{b_0^{\prime}} )$ can be written in the following form,
 \begin{eqnarray}
 & &  \tilde \Delta ( \bd{k},i {\omega} ) =  ( 1 - \gamma_{\bd{k}} )   \tilde{\Delta}_1(i {\omega}) + 
 ( 1 - \gamma_{2\bd{k}} ) \tilde{\Delta}_{2}^{\parallel}(i {\omega})
 \nonumber \\ & &
 +  ( 1 - \gamma^{\bot}_{\bd{k}} ) \tilde{\Delta}_{2}^{\bot}(i {\omega})
   +  ( 1 - \gamma^{\bot}_{2\bd{k}} ) \tilde{\Delta}_{2,2}^{\parallel}(i {\omega})
   \nonumber \\ & & 
    +  ( 1 - \gamma^{(2,1,0)}_{\bd{k}} ) \tilde{\Delta}_{2,1,0}(i {\omega})
 +  ( 1 - \gamma^{(2,1,1)}_{\bd{k}} ) \tilde{\Delta}_{2,1,1}(i {\omega})
    \nonumber \\ & & 
 +  ( 1 - \gamma^{(1,1,1)}_{\bd{k}} ) \tilde{\Delta}_{1,1,1}(i {\omega}) ,  
 \label{eq:fabcNNN}
 \end{eqnarray}
 where we have introduced three additional form factors
 \begin{subequations}
 \begin{eqnarray}
 \gamma^{(2,1,0)}_{\bd{k}} &=  & \frac{1}{6}
 \Big[\cos(2k_x a)\cos(k_y a) + \cos(k_x a)\cos(2k_y a) 
 \nonumber \\  & & \hspace{5mm}
 + (x\leftrightarrow z) + (y \leftrightarrow z)\Big],
 \\
\gamma^{(2,1,1)}_{\bd{k}} & = &  \frac{1}{3}\Big[\cos(2k_x a)\cos(k_y a)\cos(k_z a)
 \nonumber \\  & & \hspace{5mm}
 + (x\leftrightarrow z) + (y \leftrightarrow z)\Big],
 \\
 \gamma^{(1,1,1)}_{\bd{k}} & = &  \cos(k_x a)\cos(k_y a)\cos(k_za ).
 \end{eqnarray}
 \end{subequations}
The self-consistency equations (\ref{eq:falpha}) for the amplitudes are
 \begin{subequations}
   \label{eq:3damplitudes}
\begin{eqnarray}
 \tilde{\Delta}_1(i {\omega}) & = & \frac{1}{b'_0}\int_{\bm{q}}\frac{\gamma_{\bm{q}}}{|\tilde{\omega}| + \tilde \Delta(\bm{q},i {\omega})} + 6 \int_{\bm{q}}\frac{1 + \gamma_{2\bm{q}} + 4 \gamma^{\bot}_{\bm{q}}}{|\tilde{\omega}| + \tilde \Delta(\bm{q},i {\omega})} 
\nonumber \\ & &
- \tilde{\Delta}_{2,1,0}(i {\omega}) - \tilde{\Delta}_{1,1,1}(i {\omega})  ,
 \\
\tilde{\Delta}^{\parallel}_{2}(i {\omega}) & = &  
 - 3(1 + 4\mu^2)\int_{\bm{q}}\frac{\gamma_{2\bm{q}}}{|\tilde{\omega}| + 
 \tilde \Delta(\bm{q},i{\omega})},
 \nonumber
 \\
 & &
 \\
\tilde{\Delta}^{\bot}_{2}(i {\omega})  & = &  \left(
 \frac{2\mu^2}{b'_0} + 24 \mu^2 - 12 \right) \int_{\bm{q}}\frac{\gamma^{\bot}_{\bm{q}}}{|\tilde{\omega}| + \tilde \Delta(\bm{q},i {\omega})} 
\nonumber \\  
 &  &  \hspace{-16mm} + 24\mu \int_{\bm{q}}\frac{\gamma_{\bm{q}}}{|\tilde{\omega}| + \tilde \Delta(\bm{q},i {\omega})} 
+  12\mu^2  \int_{\bm{q}}\frac{1 + 2\gamma_{2\bm{q}}}{|\tilde{\omega}| + \tilde \Delta(\bm{q},i {\omega})}
 \nonumber \\  & & \hspace{-16mm}
 - \tilde{\Delta}_{2,1,0}(i {\omega})  - \tilde{\Delta}_{1,1,1}(i {\omega})  
  - 2 \tilde{\Delta}^{\parallel}_{2,2}(i {\omega}) - 2\tilde{\Delta}_{2,1,1}(i {\omega}) ,
 \nonumber
 \\
 & &
 \\
\tilde{\Delta}^{\parallel}_{2,2}(i {\omega}) & = & 
 -6\mu^2\int_{\bm{q}}\frac{\gamma^{\bot}_{2\bm{q}}}{|\tilde{\omega}| + \tilde \Delta(\bm{q},i {\omega})} ,
 \\
\tilde{\Delta}_{2,1,0}(i {\omega})  & = &  - 24\mu\int_{\bm{q}}\frac{\gamma^{(2,1,0)}_{\bm{q}}}{|\tilde{\omega}| + \tilde \Delta(\bm{q},i {\omega})},
\\
\tilde{\Delta}_{2,1,1}(i {\omega})  &  = &  -24\mu^2\int_{\bm{q}}\frac{\gamma^{(2,1,1)}_{\bm{q}}}{|\tilde{\omega}| + \tilde \Delta(\bm{q},i {\omega})},
 \\
\tilde{\Delta}_{1,1,1}(i {\omega}) & = & - 24\mu\int_{\bm{q}}\frac{\gamma^{(1,1,1)}_{\bm{q}}}{|\tilde{\omega}| + \tilde \Delta(\bm{q},i {\omega})}.
\end{eqnarray}
 \end{subequations}
The spin-diffusion coefficient at infinite temperature is then given by
\begin{eqnarray}
 {\cal{D}} & = & \frac{|J_1|\sqrt{b'_0}a^2}{6}\Big[ \tilde{\Delta}_{1}(0) + 4 \tilde{\Delta}_{2}^{\parallel}(0) + 2 
 \tilde{\Delta}_{2}^{\bot}(0) + 8  \tilde{\Delta}_{2,2}^{\parallel}(0) 
 \nonumber \\ & &
  + 5\tilde{\Delta}_{2,1,0}(0) + 6\tilde{\Delta}_{2,1,1}(0) + 3\tilde{\Delta}_{1,1,1}(0)\Big].
\end{eqnarray}
In Fig.~\ref{fig:diffsc} we show a graph of ${\cal{D}}$ for spin $S=1/2$ 
as a function of $\mu = J_2 / J_1$. 
The  asymmetry with respect to $\mu \rightarrow - \mu$ 
has also been found on a bcc lattice in Fig.~\ref{fig:diffbcc}.
In contrast, in reduced dimensions the anomalous 
spin-diffusion coefficient ${\cal{D}} ( i \omega )$
in Eqs.~(\ref{eq:D1res2}) and (\ref{eq:Dres2d})
is symmetric with respect to $\mu \rightarrow - \mu$.

\end{appendix}






\begin{thebibliography}{99}
\bibitem{footnoteCor}
The full solution of the integral equation implies non-analytic corrections to diffusion which appear if we expand the dissipation function beyond leading order in momentum and frequency.
\end{thebibliography}
\end{document}